\documentclass[prb,twocolumn,amsmath,amsfonts,amssymb]{revtex4-1}
\usepackage[position=top,singlelinecheck=false,]{subcaption}
\usepackage{color}
\usepackage{setspace}
\usepackage{xstring}
\usepackage{braket}
\usepackage{nicefrac}
\usepackage[export]{adjustbox}
\usepackage{placeins}
\usepackage{csquotes}
\usepackage{siunitx}
\usepackage{float}
\usepackage{graphicx}
\usepackage{changepage}
\usepackage{booktabs}
\usepackage{multirow}

\newcommand{\bilayer}[1]{%
    \IfEqCase{#1}{%
        {1}{MoS$_2$/MoS$_2$}%
        {2}{MoSe$_2$/MoSe$_2$}%
        {3}{WS$_2$/WS$_2$}%
        {4}{WSe$_2$/WSe$_2$}%
        {5}{WS$_2$/MoS$_2$}%
        {6}{WSe$_2$/MoSe$_2$}%
        {7}{MoSe$_2$/MoS$_2$}%
        {8}{MoSe$_2$/WS$_2$}%
        {9}{WSe$_2$/MoS$_2$}%
        {10}{WSe$_2$/WS$_2$}%
    }[\PackageError{\bilayer}{Undefined option to bilayer:#1}{}]%
}

\newcommand{\monolayer}[1]{%
    \IfEqCase{#1}{%
        {1}{MoS$_2$}%
        {2}{MoSe$_2$}%
        {3}{WS$_2$}%
        {4}{WSe$_2$}%
    }[\PackageError{\bilayer}{Undefined option to bilayer:#1}{}]%
}

\newcommand{\region}[1]{%
    \IfEqCase{#1}{%
    {AA}{AA}%
    {AB}{B$^{\mathrm{M/X}}$}%
    {BA}{B$^{\mathrm{X/M}}$}%
    }[\PackageError{\region}{Undefined option to region:#1}{}]%
}

\newcommand{\homoangle}[1]{%
    \IfEqCase{#1}{%
        {1}{~$1.6$\si{\degree}}
        {2}{~$2.6$\si{\degree}}
        {3}{~$3.1$\si{\degree}}
        {4}{~$3.9$\si{\degree}}
        {5}{~$5.1$\si{\degree}}
        {6}{~$7.3$\si{\degree}}
    }[\PackageError{\bilayer}{Undefined option to homoangle:#1}{}]%
}

\newcommand{\heteroangle}[1]{%
    \IfEqCase{#1}{%
        {1}{~$4.5$\si{\degree}}
        {2}{~$5.4$\si{\degree}}
        {3}{~$7.9$\si{\degree}}
    }[\PackageError{\bilayer}{Undefined option to homoangle:#1}{}]%
}

\begin{document}

\title{Flat band properties of twisted transition metal dichalcogenide homo- and heterobilayers of \protect\monolayer{1}, \protect\monolayer{2}, \protect\monolayer{3} and \protect\monolayer{4}}

\author{Valerio Vitale$^1$}
\affiliation{$^1$Departments of Materials and Physics and the Thomas Young Centre for Theory and Simulation of Materials, Imperial College London, London SW7 2AZ, UK\\}
\author{Kemal Atalar$^1$}
\affiliation{$^1$Departments of Materials and Physics and the Thomas Young Centre for Theory and Simulation of Materials, Imperial College London, London SW7 2AZ, UK\\}
\author{Arash A. Mostofi$^1$}
\affiliation{$^1$Departments of Materials and Physics and the Thomas Young Centre for Theory and Simulation of Materials, Imperial College London, London SW7 2AZ, UK\\}
\author{Johannes Lischner$^1$}
\affiliation{$^1$Departments of Materials and Physics and the Thomas Young Centre for Theory and Simulation of Materials, Imperial College London, London SW7 2AZ, UK\\}

\begin{abstract}
Twisted bilayers of two-dimensional materials, such as twisted bilayer graphene, often feature flat electronic bands that enable the observation of electron correlation effects. In this work, we study the electronic structure of twisted transition metal dichalcogenide (TMD) homo- and heterobilayers that are obtained by combining \monolayer{1}, \monolayer{3}, \monolayer{2} and \monolayer{4} monolayers, and show how flat band properties depend on the chemical composition of the bilayer as well as its twist angle. We determine the relaxed atomic structure of the twisted bilayers using classical force fields and calculate the electronic band structure using a tight-binding model parametrized from first-principles density-functional theory. We find that the highest valence bands in these systems can derive either from $\Gamma$-point or $K$/$K'$-point states of the constituent monolayers. For homobilayers, the two highest valence bands are composed of monolayer $\Gamma$-point states, exhibit a graphene-like dispersion and become flat as the twist angle is reduced. The situation is more complicated for heterobilayers where the ordering of $\Gamma$-derived and $K$/$K'$-derived states depends both on the material composition and also the twist angle. In all systems, qualitatively different band structures are obtained when atomic relaxations are neglected. 
\end{abstract}

\keywords{twisted bilayers, transition metal dichalcogenides, TMDs, DFT, tight-binding}

\maketitle

\section{Introduction}\label{sec:intro}
Introducing a twist between two van der Waals stacked two-dimensional materials creates a moir\'e pattern which results in novel emergent properties. For example, a graphene bilayer with a twist of $\sim$1.1~degree exhibits flat bands, strong electron correlations and superconductivity which are absent in the constituent monolayers\cite{Cao_nat1,Cao_nat2,efetov_natphys2,Yankowitz1059,Hartree}. These findings have generated significant interest and established the new field of twistronics\cite{Carr_PRB_95}. 

Besides graphene, there exist many two-dimensional materials that can be used as building blocks of moir\'e materials\cite{Mounet2018}. In particular, the transition metal dichalcogenides (TMDs) with chemical formula MX$_2$ [with M being a transition metal atom such as tungsten (W), molybdenum (Mo), niobium (Nb) or tantalum (Ta) and X denoting a chalcogen atom such as sulphur (S), selenium (Se) or tellurium (Te)] are a promising class of candidate materials. In contrast to graphene, many monolayer TMDs are semiconductors with band gaps in the range of 1-2~eV which makes these materials promising for applications in nano- and optoelectronics\cite{Mueller2018,2d_TMDs_review,Manzeli2017,Schaibley2016}. Moreover, monolayer TMDs exhibit strong spin-orbit coupling and spin-valley locking as a consequence of their crystal structure and the presence of heavy transition metal atoms\cite{Manzeli2017,Schaibley2016}.

Recently, several experimental groups have started to explore the properties of twisted TMD bilayers. 
For example, Wang and coworkers\cite{Wang2020} fabricated bilayers of \monolayer{4} with different twist angles and observed a correlated insulator state when the lowest valence band was half filled with holes. In the same system, Huang {\it et al.}\cite{Huang2020}~measured a giant nonlinear Hall effect at small twist angles, and control of optical properties through twisting has been reported in \monolayer{1} bilayers\cite{Gao2020}.

In addition to homobilayers consisting of two identical TMD monolayers, it is also possible to create heterobilayers consisting of two different TMD monolayers. 
For heterobilayers, a moir\'e pattern emerges even without a twist between the layers, as a consequence of the different lattice constants of the constituent monolayers. 
Tran and coworkers\cite{Tran2019,Tran2020} studied the optical properties of twisted \bilayer{6} bilayers and observed signatures of interlayer excitons that are trapped by the moir\'e potential. A similar experiment but with an untwisted \bilayer{6} bilayer was performed by Gerardot and coworkers\cite{Kremser2020}, who also observed spin-layer locking of interlayer excitons in a 2H-\bilayer{2}/\monolayer{4} trilayer\cite{Brotons-Gisbert2020}. 
Tang {\it et al.}\cite{Tang2020} detected interactions between excitons and magnetically ordered holes in angle-aligned \bilayer{10} structures indicating that this system can be used to simulate the phase diagram of the triangular Hubbard model. The existence of stripe phases over a large doping range has recently been reported in untwisted \bilayer{10} bilayers by Mak {\it et al.}\cite{jin2020stripe}. The same system also shows an abundance of correlated insulating states across a range of electron and hole doping levels\cite{xu2020abundance}.

To understand these experimental findings, detailed knowledge of the electronic structure of twisted TMD bilayers is required. Several groups have carried out density-functional theory (DFT) calculations of twisted homobilayers. For example, Naik and Jain\cite{Naik_PRL_121} have calculated the band structure of several homobilayers (neglecting the effect of spin-orbit coupling) at a twist angle of $3.5$~degree and found flat valence bands. However, accessing smaller twist angles is challenging because of the unfavorable scaling of standard first-principles techniques with system size. To access smaller twist angles, Zhan {\it et al.}\cite{zhan2020multiultraflatbands} used the \emph{ab initio} tight-binding model developed by Fang and coworkers\cite{Kaxiras_11bands} for untwisted homobilayers and calculated the band structure of \monolayer{1} homobilayers for twist angles as small as $1.6$~degree, including the effect of spin-orbit coupling. In a similar work, de Laissard\`iere {\it et al.}\cite{deLaissardiere_PRB} used a Slater-Koster based tight-binding approach to study the evolution of flat bands in twisted homobilayer \monolayer{1}.
As an alternative to atomistic methods, Wu and coworkers\cite{Wu_PRL_121} employed a continuum effective mass approach to study the electronic structure of twisted heterobilayers. Similar work was carried out by Zhang, Yuan and Fu\cite{zhang2020moire,Zhang_PRB_2021} and Vogl {\it et al.}\cite{Vogl_PRB_2021}. A different approach based on generalised Wigner crystals, has been proposed by Phillips and coworkers\cite{Phillips_PRB_2021} to explain the emergence of insulating states at fractional filling. 

In this work, we systematically study the atomic and electronic structure of all 3R stacked ($\theta\sim0^\circ$) twisted homo- and heterobilayers that can be constructed by combining \monolayer{1}, \monolayer{2}, \monolayer{3} and \monolayer{4} monolayers. Specifically, we use classical force fields to calculate the relaxed atomic structure of these systems, which display significant in-plane and 
out-of-plane relaxations. For the relaxed structures, we use an atomistic tight-binding model derived from first-principles DFT calculations to calculate the electronic band structure including the effect of spin-orbit coupling. In all homobilayers, we find that for relatively small angles ($\theta<4^\circ$), the two highest valence bands are composed of $\Gamma$-valley states of the constituent monolayers and become extremely flat as the twist angle approaches zero, reaching bandwidths of a few meV for twist angles near $1.5^\circ$. In contrast, not all heterobilayers exhibit such $\Gamma$-derived valence bands. In some heterobilayers (most notably those containing a \monolayer{4} layer), the top valence bands derive from monolayer states at $K$ and $K'$. Such $K$/$K'$-derived valence states are less affected by interlayer coupling and are found to be more dispersive compared to $\Gamma$-derived states. The different ordering of $\Gamma$-derived and $K$/$K'$-derived valence states in the various twisted bilayer systems can be understood by comparing the energy scale for interlayer hopping with the energy difference between the valence band $K$- and $\Gamma$-states of the constituent monolayers. Importantly, the neglect of atomic relaxations leads to qualitatively different electronic properties.

\section{Methods}\label{sec:methods}

\subsection{Atomic structure}\label{subsec:methods_atomic_structure}
As a first step, we generate structures of flat (i.e. unrelaxed) twisted TMD homo- and heterobilayers (tBL-TMDs). 
We start from 3R stacked bilayers, where metal and chalcogen atoms of the top layer are directly above corresponding metal and chalcogen atoms of the bottom layer, and rotate the top layer by an angle $\theta$ around the axis perpendicular to the plane of the bilayer and going through the metal atoms.
For homobilayers, a commensurate structure is obtained when the moir\'e cell vectors $\mathbf{t}_1$ and $\mathbf{t}_2$ can be expressed as\cite{deLaissardiere_PRB}
\begin{equation}
\mathbf{t}_1 = n\mathbf{a}_1 + m\mathbf{a}_2, \quad \mathbf{t}_2 = -m\mathbf{a}_1 + (n+m)\mathbf{a}_2,
\label{eq1:moire_homos}
\end{equation}
where  $\mathbf{a}_1=\frac{a}{2}(\sqrt{3}, 1, 0)$ and $\mathbf{a}_2=\frac{a}{2}(\sqrt{3}, -1, 0)$ are primitive lattice vectors of the monolayer (with $a$ being the lattice constant) and $m$ and $n$ are integers. The twist angle is given by $\cos\theta=\frac{n^2 + 4nm + m^2}{2(n^2 + nm + m^2)}$ and the number of atoms in the cell is $N_{\mathrm{at}}=6(n^2 + nm + m^2)$. 

For heterobilayers, we first consider systems whose constituent monolayers contain the same species of chalcogen atom. In this case, the lattice constants of both monolayers differ by less than $1\%$ and we generate a commensurate moir\'e cell for the twisted heterobilayers by
increasing the lattice constant of the monolayer with the smaller lattice constant to the value of the larger lattice constant and then use the same approach described above for homobilayers. 

In contrast, for heterobilayers whose constituent monolayers contain different species of chalcogen atom, the lattice constants of the monolayers differ by several percent. To generate moir\'e cells for these systems, we follow the approach of Zeller and G\"unther\cite{Zeller_2014}. In their work the moir\'e vectors $\mathbf{t}_1'$ and $\mathbf{t}_2'$ are defined as
\begin{equation}
\mathbf{t}_1' = n\mathbf{a}_1' + m\mathbf{a}_2', \quad \mathbf{t}_2' = -m\mathbf{a}_1' + (n-m)\mathbf{a}_2',
\end{equation}
where $\mathbf{a}_1'=a(1, 0, 0)$ and $\mathbf{a}_2'=\frac{a}{2}(-1,\sqrt{3}, 0)$ are the primitive lattice vectors of the monolayer with smaller equilibrium lattice constant $a$ ($a'$ denoting the lattice constant of the other layer). We use DFT equilibrium lattice constants from Ref.~\cite{C7CP00012J}. The integers $n$ and $m$ are determined from the numerical solution of a diophantine equation (see Appendix of Ref.\cite{Zeller_2014} for details). Here, we only consider so-called first-order moir\'e structures\cite{Zeller_2014}. Importantly, to generate a commensurate moir\'e cell near a desired target twist angle a certain level of strain must be applied. In this work, we only study systems with an overall strain of less than $3\%$. The strain (which can be either tensile or compressive) is always applied to the layer with the larger equilibrium lattice constant.

For both homo- and hetero-bilayers, using the flat twisted bilayers as starting points, we determine the relaxed equilibrium atomic structure via classical force fields as implemented in the LAMMPS software package\cite{PLIMPTON19951,LAMMPs}. In particular, we employ the force fields developed by Naik and coworkers, based on the Kolmogorov-Crespi potential for the interlayer interaction\cite{Naik_KC_JPC2019}. For intralayer interactions, a Stillinger-Weber type force field is used\cite{Jiang2017HandbookOS}. The relaxed structures of most twisted TMD bilayers are of a breathing-mode type, i.e., the two layers have out-of-plane displacements in opposite directions (see Sec.~\ref{subsec:results_atomic_structure}). The only exception are heterobilayers with different chalcogen atoms. These systems also exhibit breathing-mode relaxations for twist angles $\geq$ \heteroangle{1}, but for smaller twist angles qualitatively different relaxed structures are found in which the two layers have out-of-plane displacements in the same direction and the amplitude of these displacements is larger compared to those in breathing mode structures. In an experimental setting, we expect such structures to be less likely to occur because the twisted TMD bilayers are placed on a substrate. Therefore, we do not present results for these systems and focus our attention on breathing mode structures.

\subsection{Electronic structure}\label{subsec:methods_electronic_structure}
To calculate the electronic properties of twisted TMD homo- and heterobilayers, we use an atomistic tight-binding approach based on the work of Fang and coworkers\cite{Kaxiras_11bands}, who studied \emph{untwisted} homobilayers. The atomic orbital basis consists of 5 d-like orbitals for the metal atoms and 3 p-like orbitals for each chalcogen atom (which doubles to 10 d-like orbitals and 6 p-like orbitals, respectively, if spin-orbit coupling is included). In a first step, we construct a symmetry-adapted tight-binding model for the monolayers including on-site, first, second and selected third nearest neighbor hoppings. The required hopping parameters are determined from a Wannier transformation\cite{Marzari_RMP84,Pizzi_2020} of the DFT Hamiltonian. To model bilayers, Fang and coworkers describe interlayer hoppings between the p-orbitals of the chalcogen atoms at the interface between the two layers, which we refer to as \emph{inner} chalcogens, using the Slater-Koster approach\cite{Kaxiras_11bands}. The Slater-Koster parameters are fitted to a set of DFT calculations of untwisted bilayers in which the top layer is translated horizontally while the bottom layer is kept fixed. Finally, spin-orbit coupling is introduced via an on-site atomic term $\lambda_{\mathrm{M/X}}^{\mathrm{SO}} \mathbf{L} \cdot \mathbf{S}$ (with $\mathbf{L}$ and $\mathbf{S}$ denoting orbital and spin angular momentum operators, respectively, and $\lambda_{\mathrm{M/X}}^{\mathrm{SO}}$ is the spin-orbit coupling strength of M or X atoms, whose value for each atom is given in Ref.~\cite{Kaxiras_11bands}.) 

To model \emph{twisted} homo- and heterobilayers, we have extended the tight-binding model of Fang {\it et al.}\cite{Kaxiras_11bands}~in several ways. In particular, we have included interlayer hoppings from inner chalcogen p$_z$-like orbitals on one layer to metal d$_{z^2}$-like orbitals on the other layer using a Slater-Koster approach. Moreover, to better capture the effect of out-of-plane displacements of the atoms, we improve the description of interlayer hoppings (both p-p and p$_z$-d$_{z^2}$) by using a different set of Slater-Koster parameters for different values of the interlayer separation. All Slater-Koster parameters for the interlayer interactions as well as all intralayer hoppings were obtained from a Wannier transformation\cite{Marzari_RMP84,Pizzi_2020} of the DFT Hamiltonian. For heterobilayers, additional care must be taken to ensure that the on-site energies are referenced to the vacuum level. 

To determine the interlayer hoppings in a twisted bilayer, the orbital basis of the rotated monolayer must be transformed. As described above, only p$_x$-, p$_y$-, p$_z$- and d$_{z^2}$-like orbitals are involved in interlayer hoppings. Since d$_{z^2}$-like and p$_z$-like orbitals are unaffected by rotations around the $z$-axis, we only need to transform the p$_x$-like and p$_y$-like orbitals and the rotated orbitals are given by $(p'_x,p'_y)^{T} = R(\theta)(p_x,p_y)^{T}$ with $R(\theta)$ denoting a two-dimensional rotation matrix. Of course, interlayer hoppings involving p$_x$-like and p$_y$-like orbitals transform in a similar fashion when a twist is introduced.

Additional details about the interlayer tight-binding model, the determination of the hopping parameters and a full list of the parameters for all systems can be found in the Appendix A and in the Supplementary Information (Sec.~S5). 
We have compared the band structures (without spin-orbit coupling) from this tight-binding model to results from explicit DFT calculations for different twisted bilayers, see Sec.~S2 in the Supplementary Information, and find good agreement between the two methods, in particular for the valence bands. 

Besides modulating the interlayer hopping, the introduction of a twist also gives rise to significant in-plane atomic relaxations which in turn induce changes in the intralayer hoppings. Such changes, however, are not captured by our model as we assume that intralayer hoppings of the twisted bilayer are the same as those in a monolayer. Recently, it was shown that such twist-induced changes to the intralayer hoppings are responsible for the flattening of $K$/$K$'-derived valence band states in \bilayer{10} superlattices~\cite{Li2021}. To capture this effect, a fully position-dependent intralayer tight-binding Hamiltonian for the TMDs should be developed in the future. 

\begin{figure*}[t]
    \subcaptionbox{}{
    \centering
        \includegraphics[width=0.5\textwidth,trim={0px 0px 0px -80px},clip]{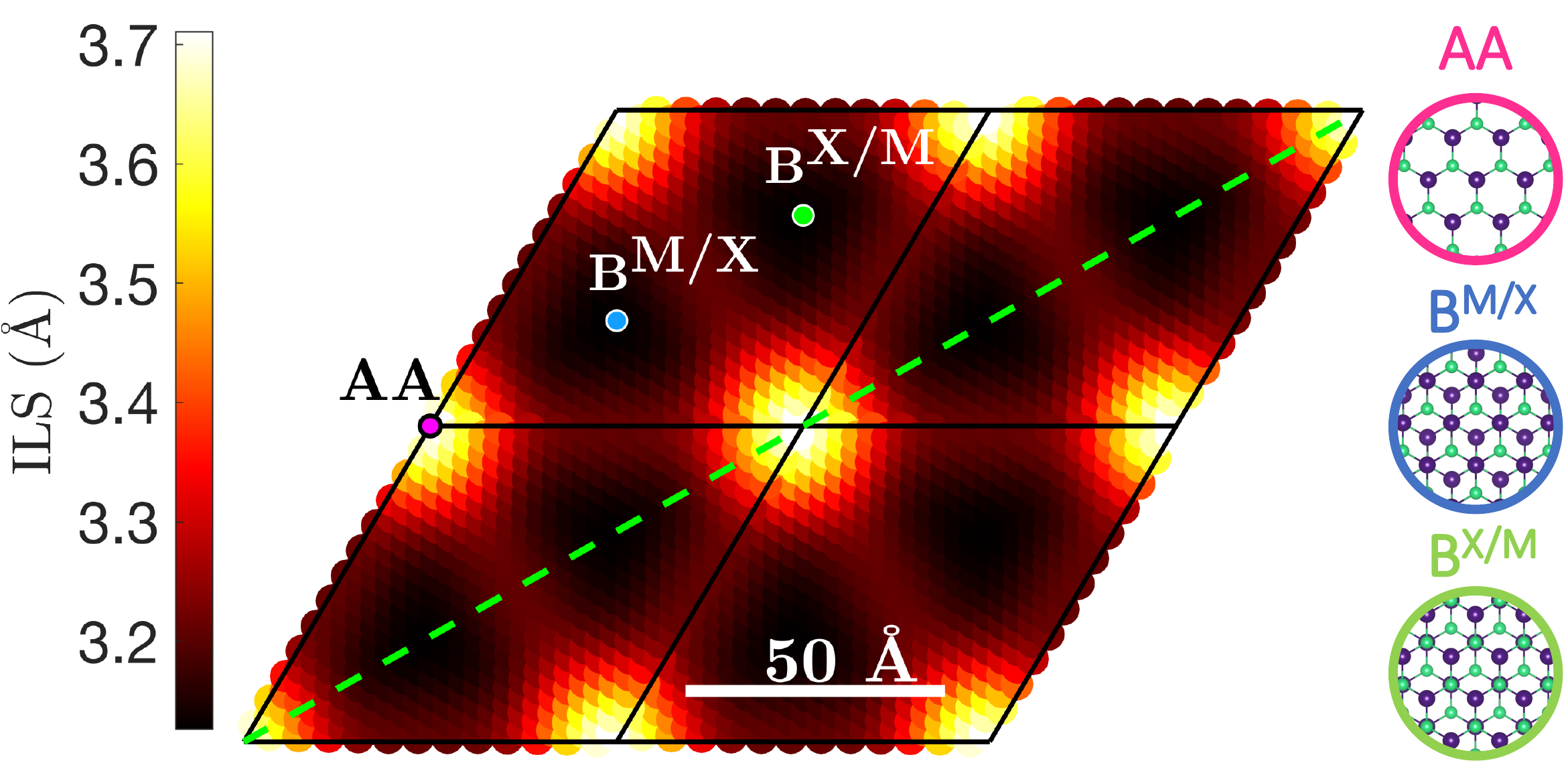}}
    \subcaptionbox{}{
            \centering
        \includegraphics[width=0.45\textwidth]{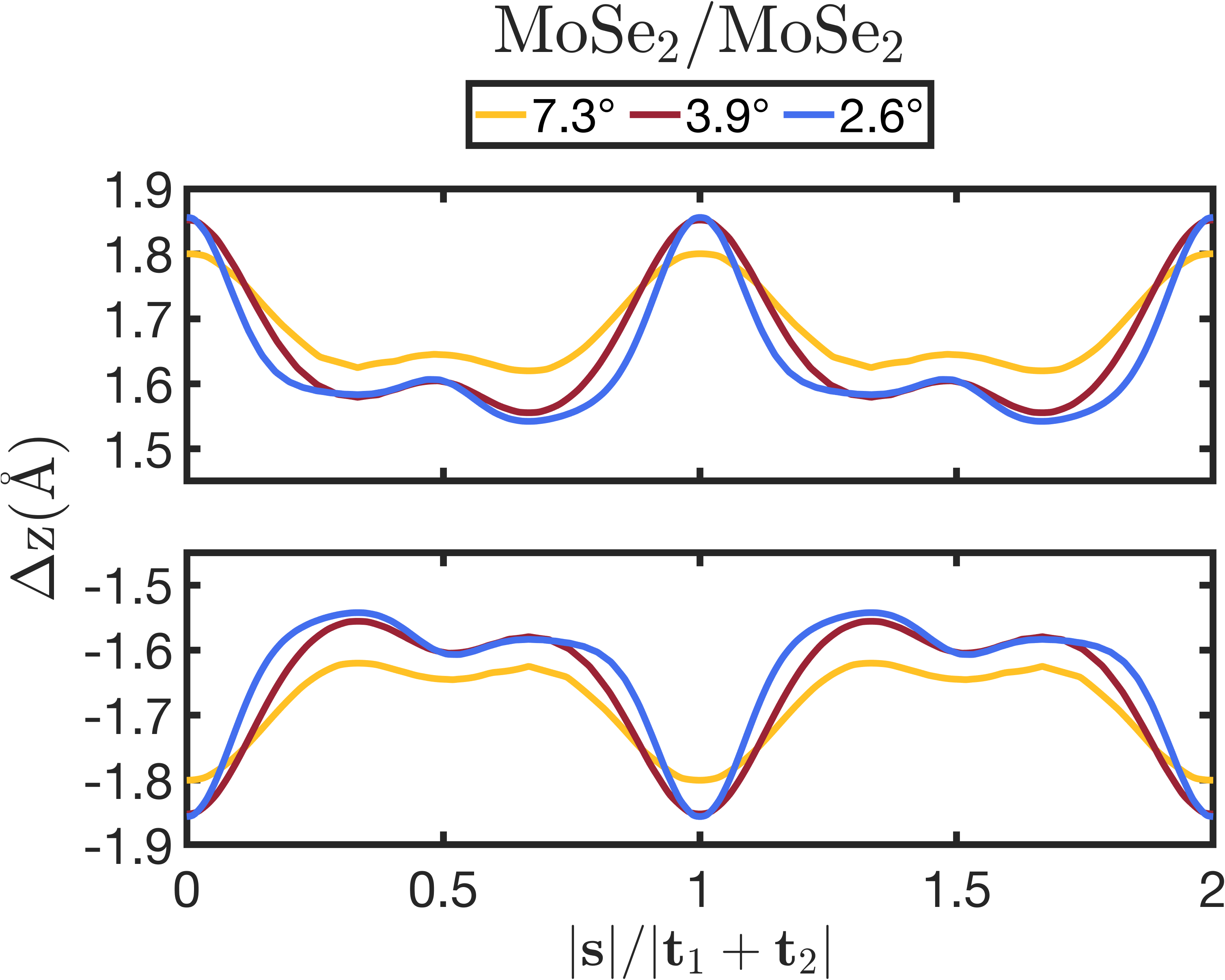}}
    \\
    \vspace{0.5cm}
    \subcaptionbox{\protect\monolayer{2} top layer}{
    \centering    
    \includegraphics[width=0.45\textwidth]{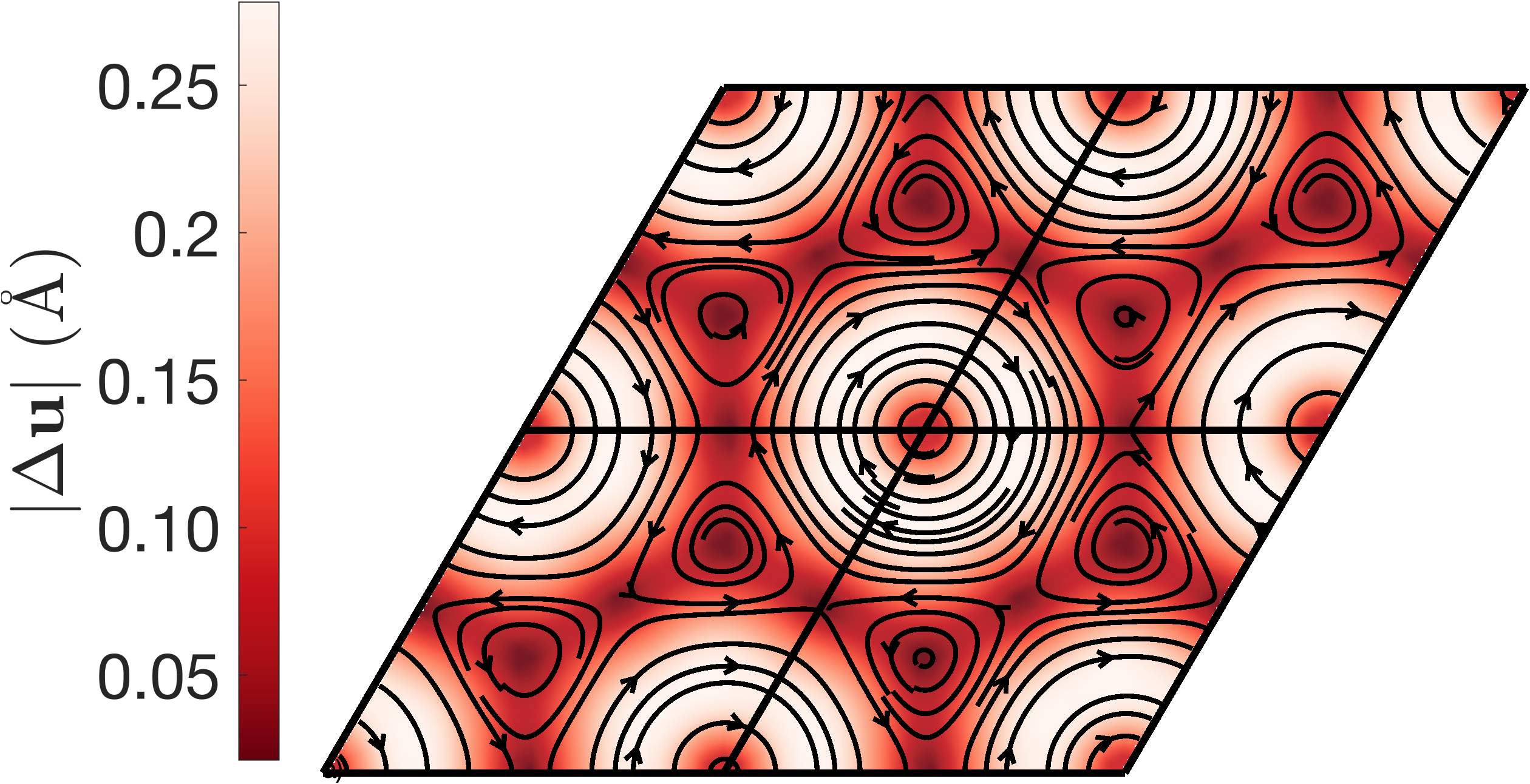}}
    \subcaptionbox{\protect\monolayer{2} bottom layer}{
    \centering
    \includegraphics[width=0.45\textwidth]{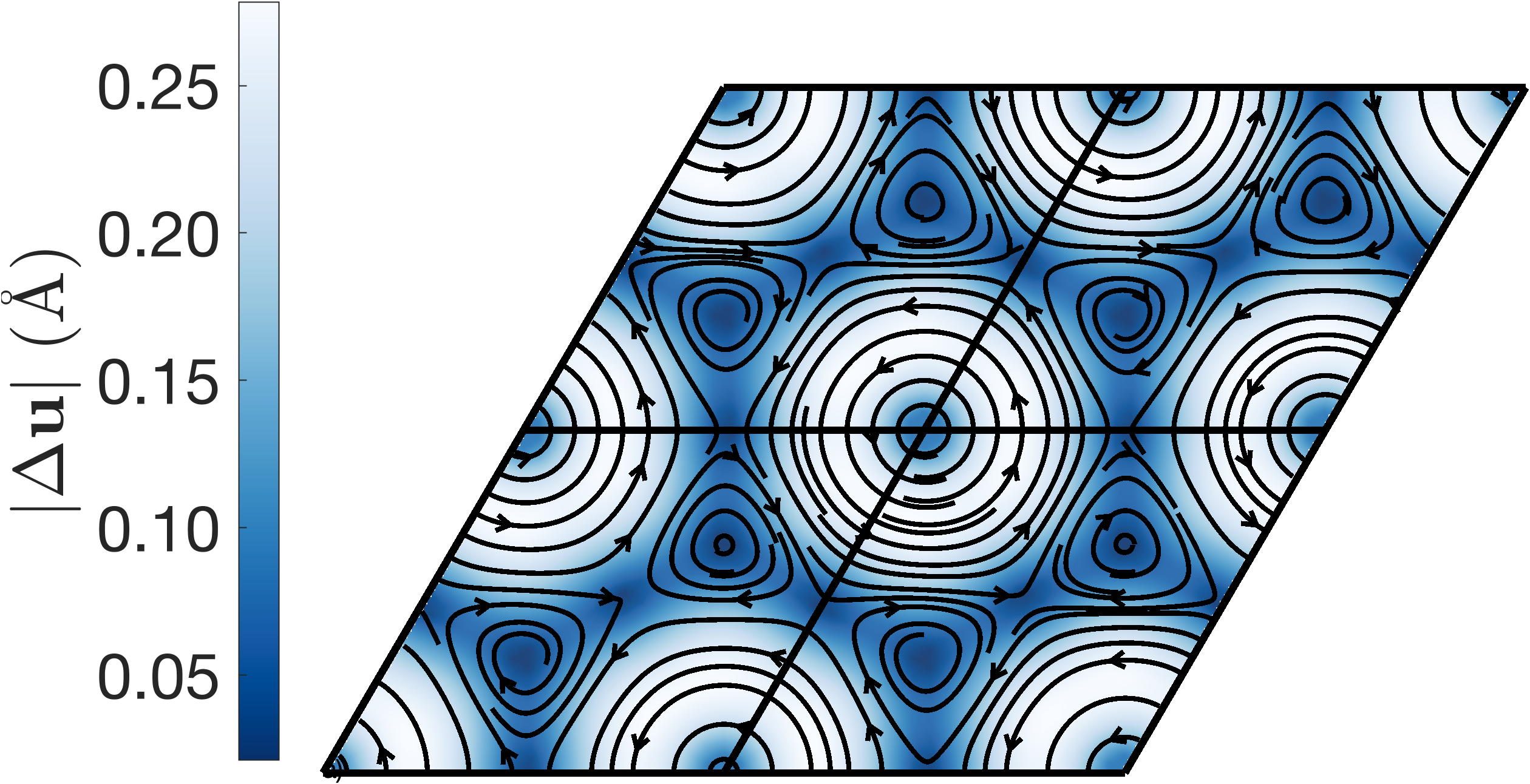}}
    \caption{Atomic relaxations in twisted \protect\bilayer{2} homobilayer. (a) Left: Inter-layer separation (ILS), defined as the distance between the two surfaces generated by the inner chalcogen atoms, for $\theta=$\protect\homoangle{2}. Right: atomic stacking arrangements in \protect\region{AA}, \protect\region{AB} and \protect\region{BA} regions of the moir\'e cell. (b) Out-of-plane displacement $\Delta z$ along the diagonal of a $2\times2$ moir\'e supercell $\mathbf{s}=\alpha(\mathbf{t}_1+\mathbf{t}_2)$ with $\alpha$ ranging from 0 to 2, shown as green dashed line in panel (a), for three different twist angles. (c) and (d) show the in-plane displacements $|\Delta \mathbf{u}|$ of the top and bottom layer, respectively, for $\theta=$\protect\homoangle{2}. Arrows indicate the direction of the in-plane displacements, with the magnitude given by the color map.}
    \label{fig1:corr_hobl}
\end{figure*}

\section{Results}
\subsection{Atomic structure}\label{subsec:results_atomic_structure}

\begin{figure*}[t]
    \subcaptionbox{}{
    \centering
    \includegraphics[width=0.5\textwidth,,trim={0px 0px 0px -50px},clip]{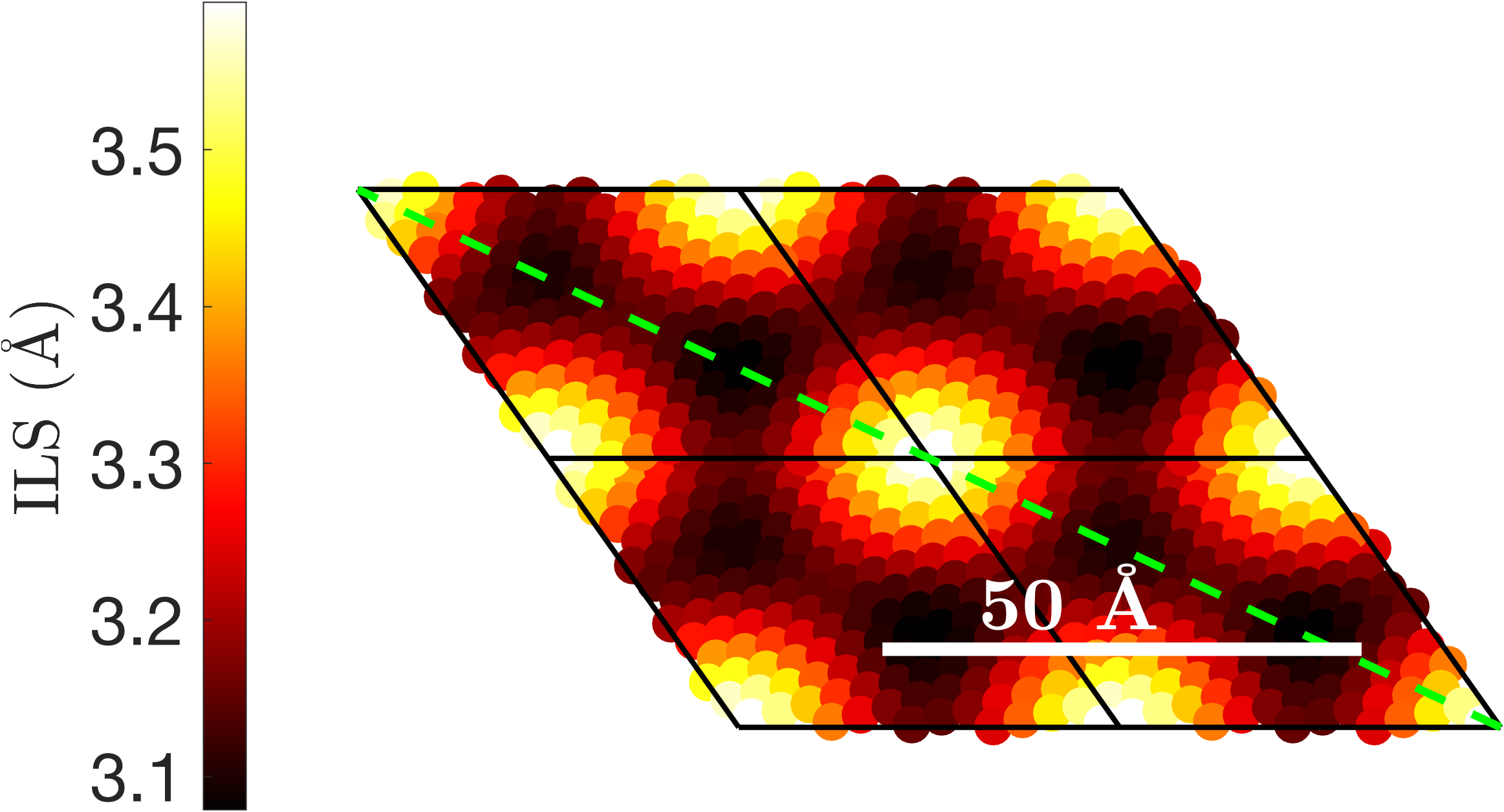}}
	\subcaptionbox{}{
    \centering
    \includegraphics[width=0.45\textwidth]{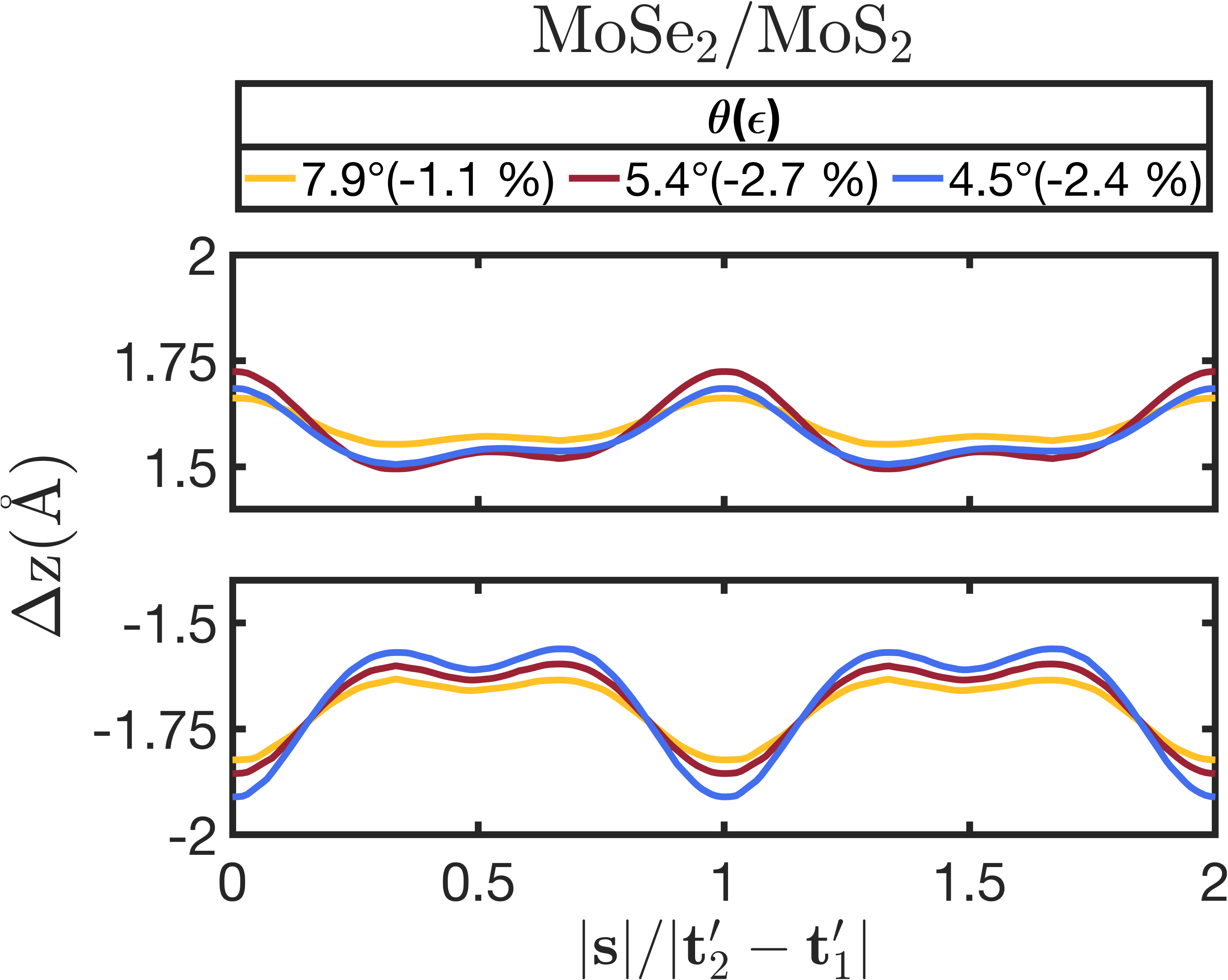}
    }\\
    \centering
    \subcaptionbox{\monolayer{2} layer}{
    \centering
    \includegraphics[width=0.45\textwidth]{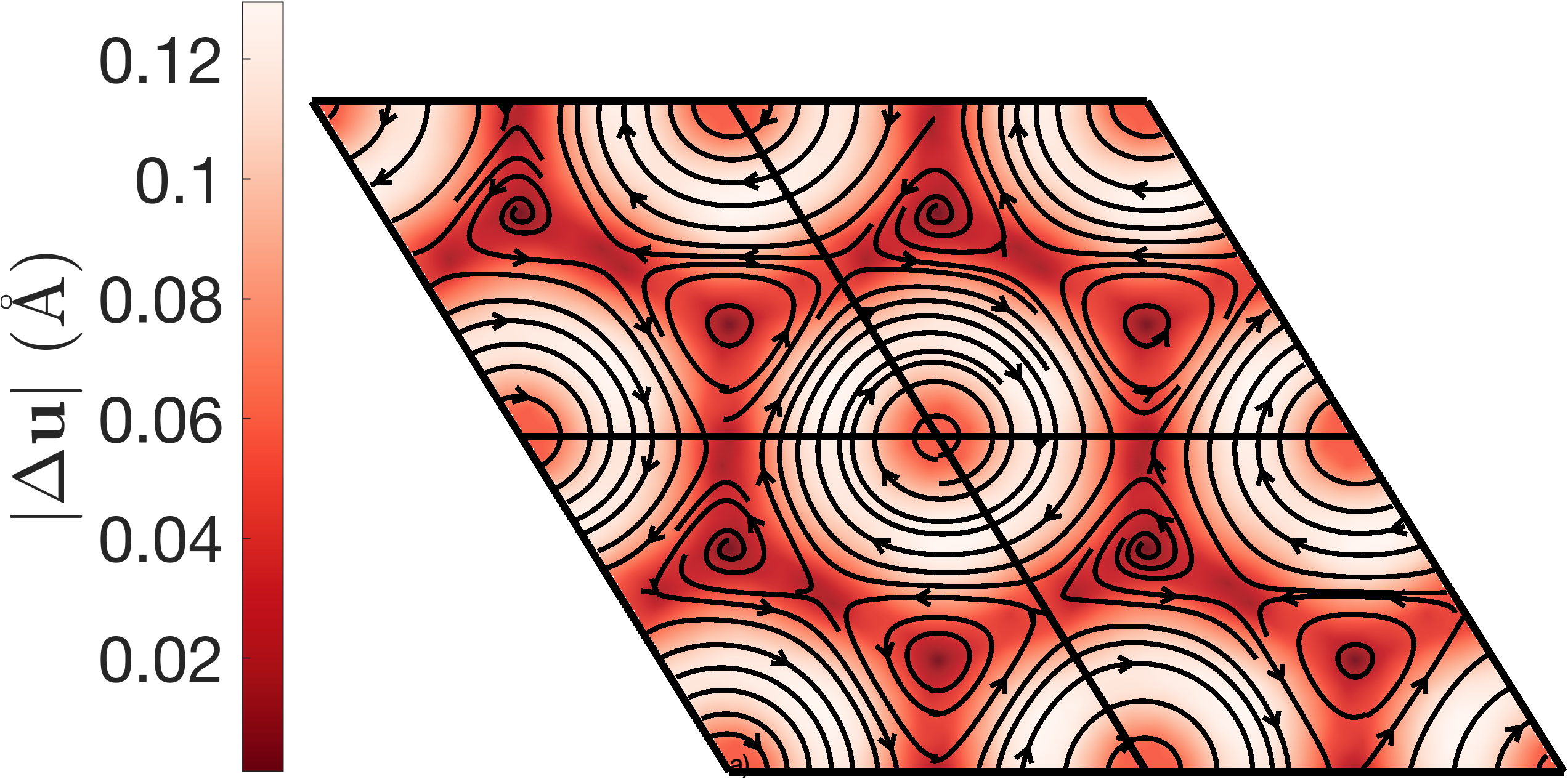}
    }
    \subcaptionbox{\monolayer{1} layer}{    
    \centering
    \includegraphics[width=0.45\textwidth]{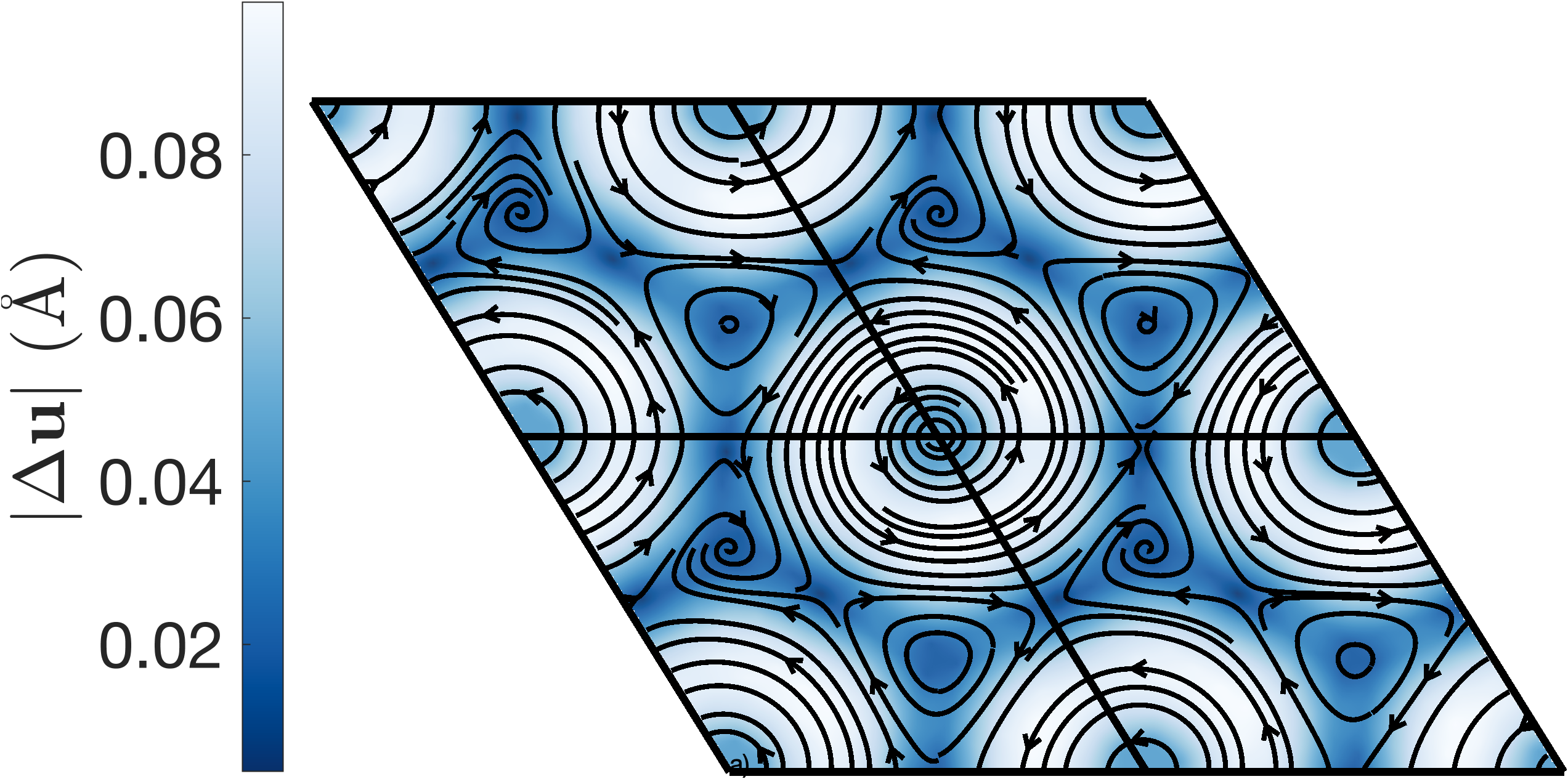}
    }
    \caption{Atomic relaxations in twisted \protect\bilayer{7} heterobilayer. (a) Inter-layer separation (ILS), defined as the distance between the two surfaces generated by the inner chalcogen atoms, for $\theta=$\protect\heteroangle{1}. Colored dots refer to different stacking regions as described in Fig.~\ref{fig1:corr_hobl}(a). (b) Out-of-plane displacement $\Delta z$ along the diagonal of a $2\times2$ moir\'e supercell $\mathbf{s}=\alpha(\mathbf{t}_2'-\mathbf{t}_1')$ with $\alpha$ ranging from 0 to 2, shown as green dashed line in (a), for three different twist angles. $\epsilon$ denotes the compressive strain in the \protect\monolayer{2} layer which is needed to generate commensurate moir\'e cells. (c) and (d) show the in-plane displacements $|\Delta \mathbf{u}|$ of the top and bottom layer, respectively, for $\theta=$\protect\heteroangle{1}. Arrows indicate the direction of the in-plane displacements, with the magnitude given by the color map.}
    \label{fig2:corr_hebl}
\end{figure*}

Introducing a twist between two 3R aligned TMD layers results in the creation of a moir\'e pattern consisting of regions with different stacking arrangements. High-symmetry stackings include \region{AA} regions, where the metal (chalcogen) atoms of the one layer are directly above the metal (chalcogen) atoms of the other layer, as well as two types of Bernal-like regions, where in one case (denoted \region{AB}) the chalcogen atom (X) in one layer lies directly above the metal atom (M) in the other layer, or vice versa (denoted \region{BA}). The high-symmetry stackings are shown in Fig.~\ref{fig1:corr_hobl}(a).

\subsubsection{Homobilayers.} 
All twisted homobilayers exhibit similar out-of-plane and in-plane displacement patterns upon relaxation. For example, Figs.~\ref{fig1:corr_hobl}(a)-(d) show results for twisted \bilayer{2} at a twist angle of $\theta=$\homoangle{2}. The interlayer separation (ILS), defined as the distance between the two surfaces on which the inner chalcogen atoms lie, is large in the \region{AA} regions (which form a triangular lattice), but smaller in the triangle-shaped \region{AB} and \region{BA} regions (which form a honeycomb lattice), see Fig.~\ref{fig1:corr_hobl}(a). 

Fig.~\ref{fig1:corr_hobl}(b) shows the out-of-plane displacement along the diagonal of the moir\'e unit cell for three different twist angles. As the twist angle decreases, the size of the \region{AA} regions shrinks, whereas \region{AB} and \region{BA} regions expand. This allows the system to reduce its energy as \region{AA} regions are energetically unfavorable because of their large steric repulsion. It can further be observed that the maximum ILS increases, while the minimum ILS decreases as the twist angle is reduced. Again, this reduces the energy cost associated with steric repulsion. 

Figures~\ref{fig1:corr_hobl}(c) and (d) show the in-plane displacements of twisted \bilayer{2}. Similar to twisted bilayer graphene\cite{Liang_PRB_2020}, the in-plane displacements in tBL-TMDs form vortices around the \region{AA} regions, with the atoms in the top and bottom layers rotating in opposite directions. Atoms in the \region{AB} and \region{BA} regions are almost unaffected by in-plane relaxations. The magnitude of in-plane atomic displacements around \region{AA} regions increases for small angles. This allows the system to reduce the size of the energetically unfavorable \region{AA} regions.

\subsubsection{Heterobilayers.}
Figure~\ref{fig2:corr_hebl} shows the in-plane and out-of-plane relaxations of twisted \bilayer{7} at a twist angle of $\theta=$\heteroangle{1}. For the set of angles studied in this work, we find that heterobilayers exhibit similar relaxation patterns as homobilayers: large ILSs are found in the \region{AA} regions, which form a triangular lattice. The relative size of the \region{AA} regions shrinks as the twist angle is decreased while \region{AB} and \region{BA} regions grow. In contrast to the homobilayers, the in-plane and out-of-plane displacements of the two layers are not symmetric, as can be seen in Figs.~\ref{fig2:corr_hebl}(b),(c) and (d). The difference of the out-of-plane displacements in the \region{AA} and \region{AB} regions is about four times larger in the \monolayer{1} layer than in the \monolayer{2} layer (Fig.~\ref{fig2:corr_hebl}(b)). As we show in the next section, this asymmetry is less pronounced in heterobilayers that have the same chalcogens. Similar to the out-of-plane displacements, the in-plane displacements are also larger in the \monolayer{1} layer compared to the \monolayer{2} layer (Figs.~\ref{fig2:corr_hebl}(c) and \ref{fig2:corr_hebl}(d)).

\begin{figure*}[t]
\centering
    \subcaptionbox{}{    
    \centering
    \includegraphics[width=0.5\textwidth]{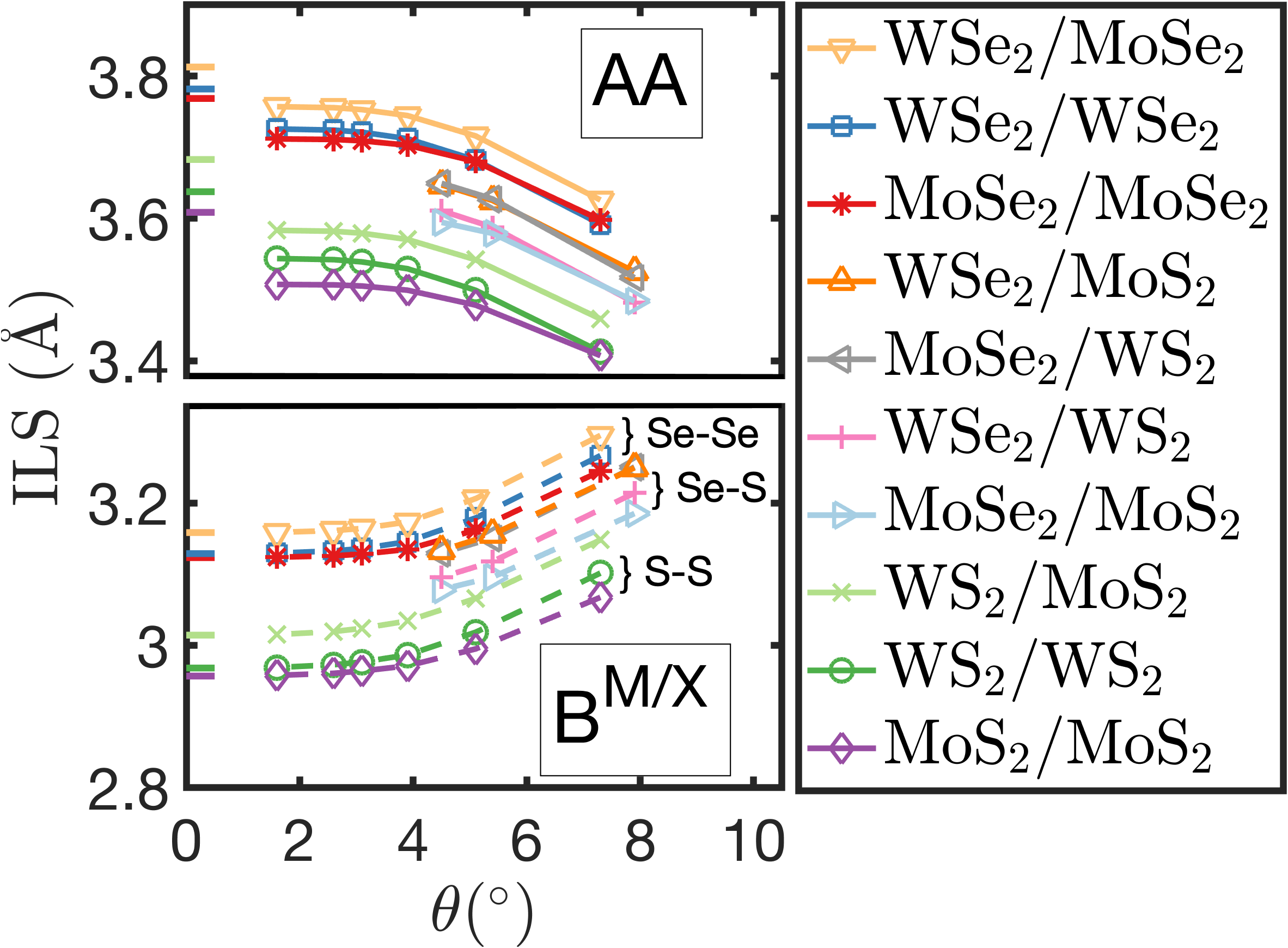}}
    \\
    \subcaptionbox{}{
    \centering
        \includegraphics[width=0.47\textwidth]{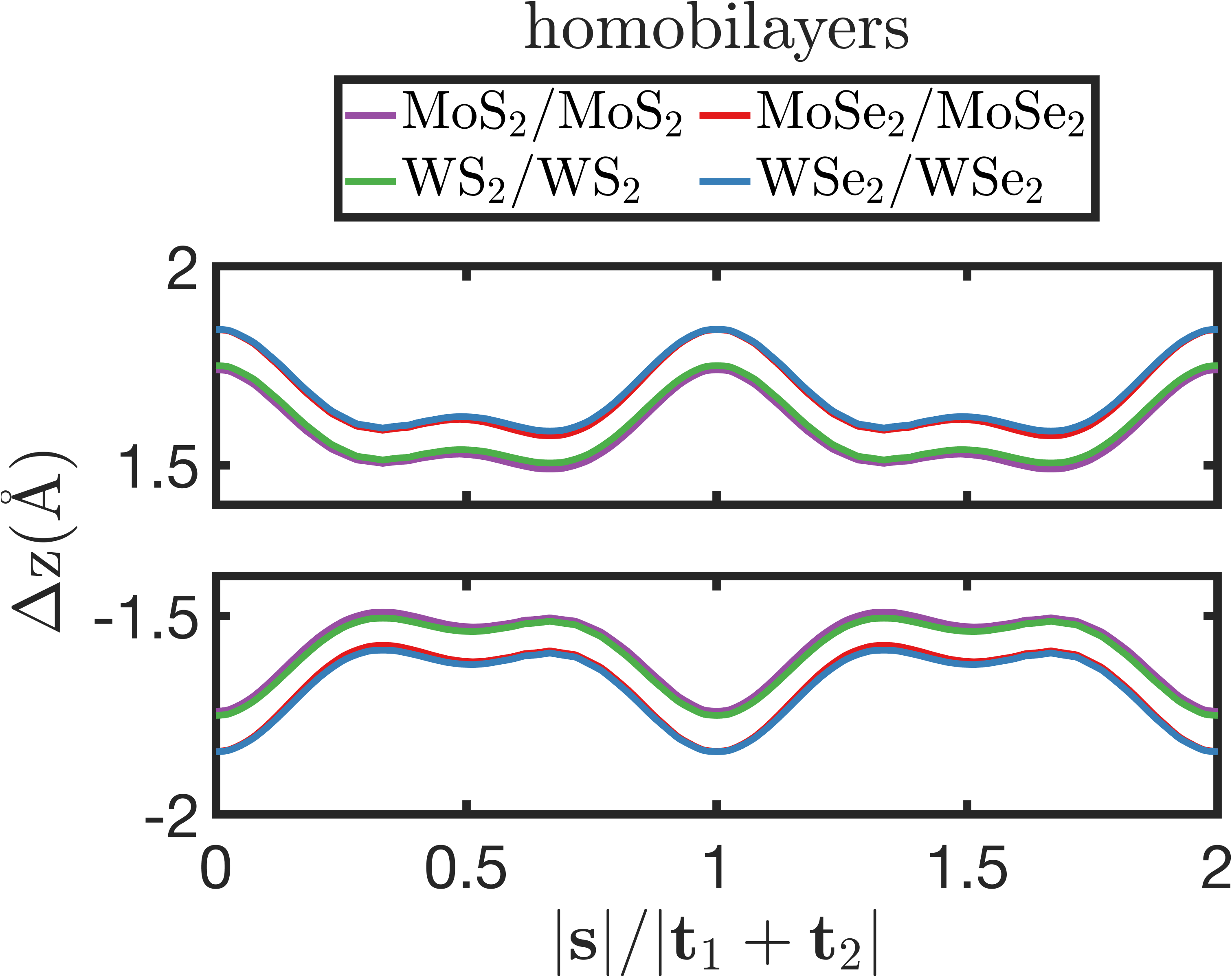}
        \includegraphics[width=0.47\textwidth]{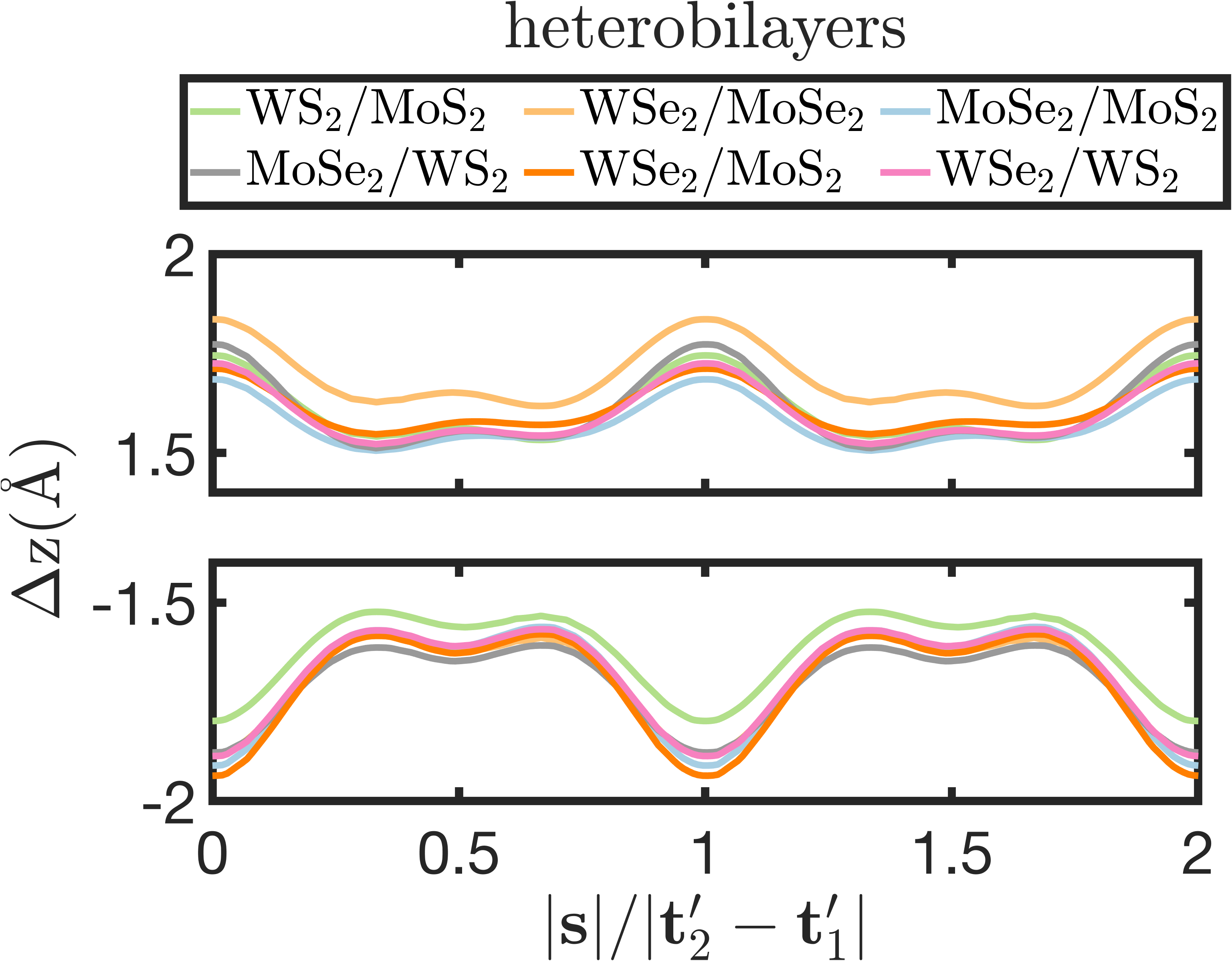}}
    \caption{(a) Minimum (bottom panel) and maximum (top panel) interlayer separation (ILS), corresponding to \protect\region{AB} and \protect\region{AA} regions, respectively, for all tBL-TMDs as function of twist angle. 
    For heterobilayers with different chalcogen atoms three twist angles are shown and these are \protect\heteroangle{1}, \protect\heteroangle{2} and \protect\heteroangle{3}. For reference, the ILS of untwisted \protect\region{AA} bilayers (top panel) and \protect\region{AB} bilayers (bottom panel) are also shown by short horizontal lines on the left hand side of the plots.  
    (b) Out-of-plane displacements $\Delta z$ for all homobilayers (left) and all heterobilayers (right) along the diagonal of a $2\times2$ moir\'e supercell (with $\mathbf{s}=\alpha(\mathbf{t}_1+\mathbf{t}_2)$ for homobilayers, and $\mathbf{s}=\alpha(\mathbf{t}_2'-\mathbf{t}_1')$ for heterobilayers, $\alpha$ ranges from 0 to 2). For all homobilayers and heterobilayers with same chalcogen atoms the twist angle is \protect\homoangle{5}; for all heterobilayers with different chalcogens  $\theta=$\protect\heteroangle{2}.}
    \label{fig3:intdist_all}
\end{figure*}

\subsubsection{Chemical trends.} 
Figure~\ref{fig3:intdist_all}(a) shows the maximum and minimum ILS, corresponding to the ILS value in the center of the \region{AA} and \region{AB}/
\region{BA} regions, respectively, as function of twist angle for the entire set of TMD homo- and hetero-bilayers. At large twist angles, the ILSs in these regions differ significantly from the values in the untwisted \region{AA} and \region{AB} bilayers. The ILS in the \region{AB}/\region{BA} regions (bottom panel of Fig.~\ref{fig3:intdist_all}(a)) decreases monotonically as the twist angle is reduced and converges to the ILS of the untwisted bilayers. In contrast, the ILS in the \region{AA} regions (top panel of Fig.~\ref{fig3:intdist_all}(a)) increases with decreasing twist angle, but does not converge to the value of the untwisted bilayer in the case of homobilayers and heterobilayers with same chalcogen atoms. This discontinuity of the maximum ILS at $\theta=0^\circ$ is a consequence of the structural relaxations which result in a growth of the \region{AB} and \region{BA} regions and a shrinkage of the \region{AA} regions at small twist angles. At the center of the large \region{AB}/\region{BA} regions the twisted bilayer has a similar structure as the untwisted \region{AB}/\region{BA} bilayer while the small size of the \region{AA} restricts the atoms from reaching the same structure as the untwisted \region{AA} bilayer.

\begin{figure*}[t]
    \centering
    \subcaptionbox{}{    
    \centering
    \includegraphics[width=0.25\textwidth,trim={0px 0px 0px -110px},clip]{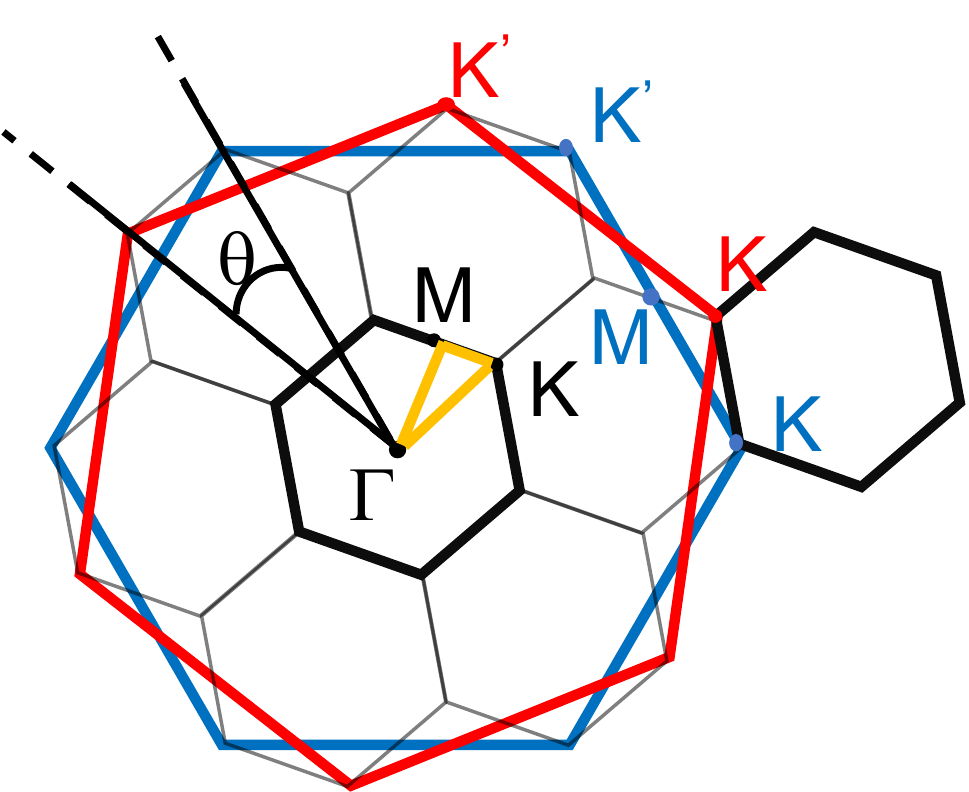}}
    \subcaptionbox{}{    
    \centering
    \includegraphics[width=0.6\textwidth,trim={-60px 0px 60px 20px},clip]{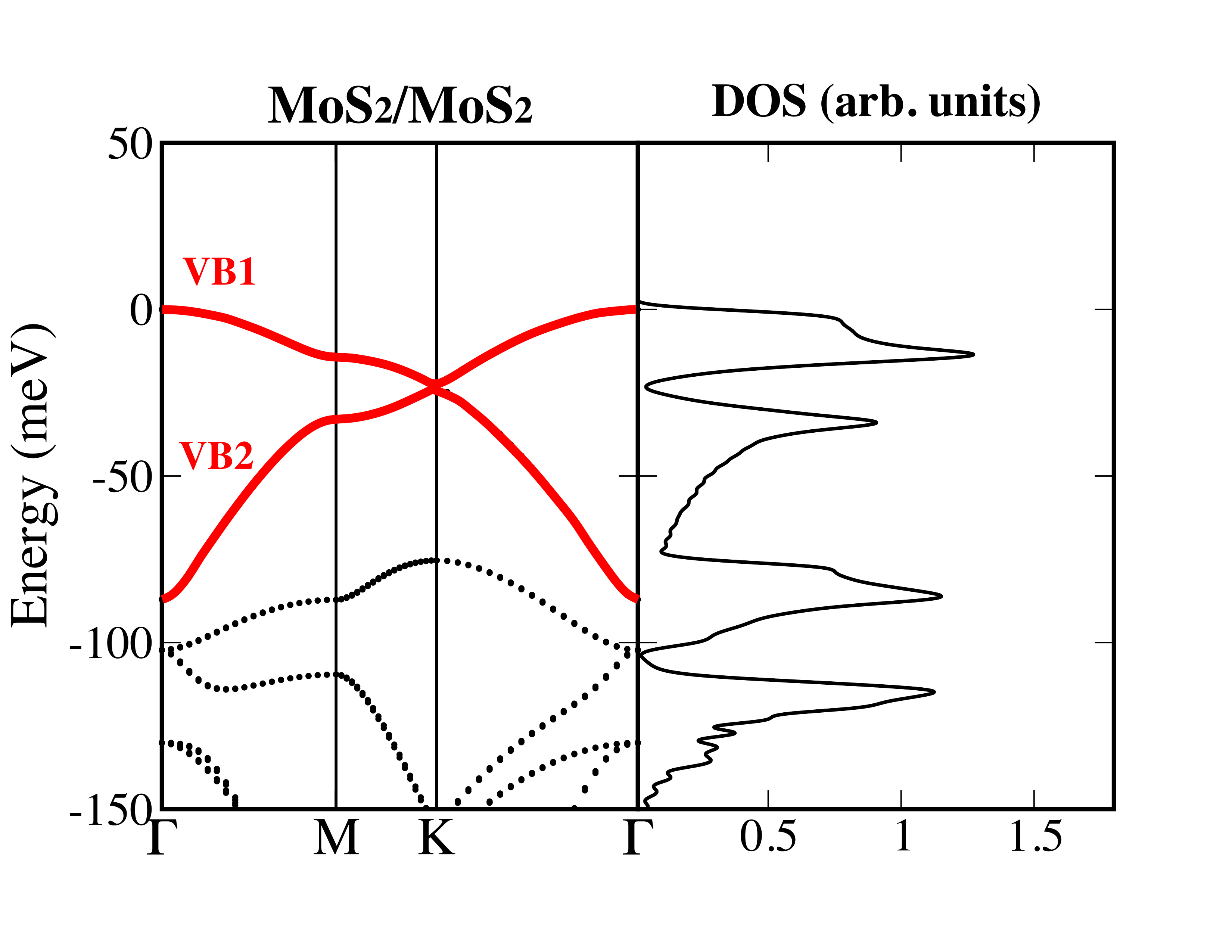}}\\
    \subcaptionbox{}{
    \centering
    \includegraphics[width=0.25\textwidth,trim={0px 0px 0px -100px},clip]{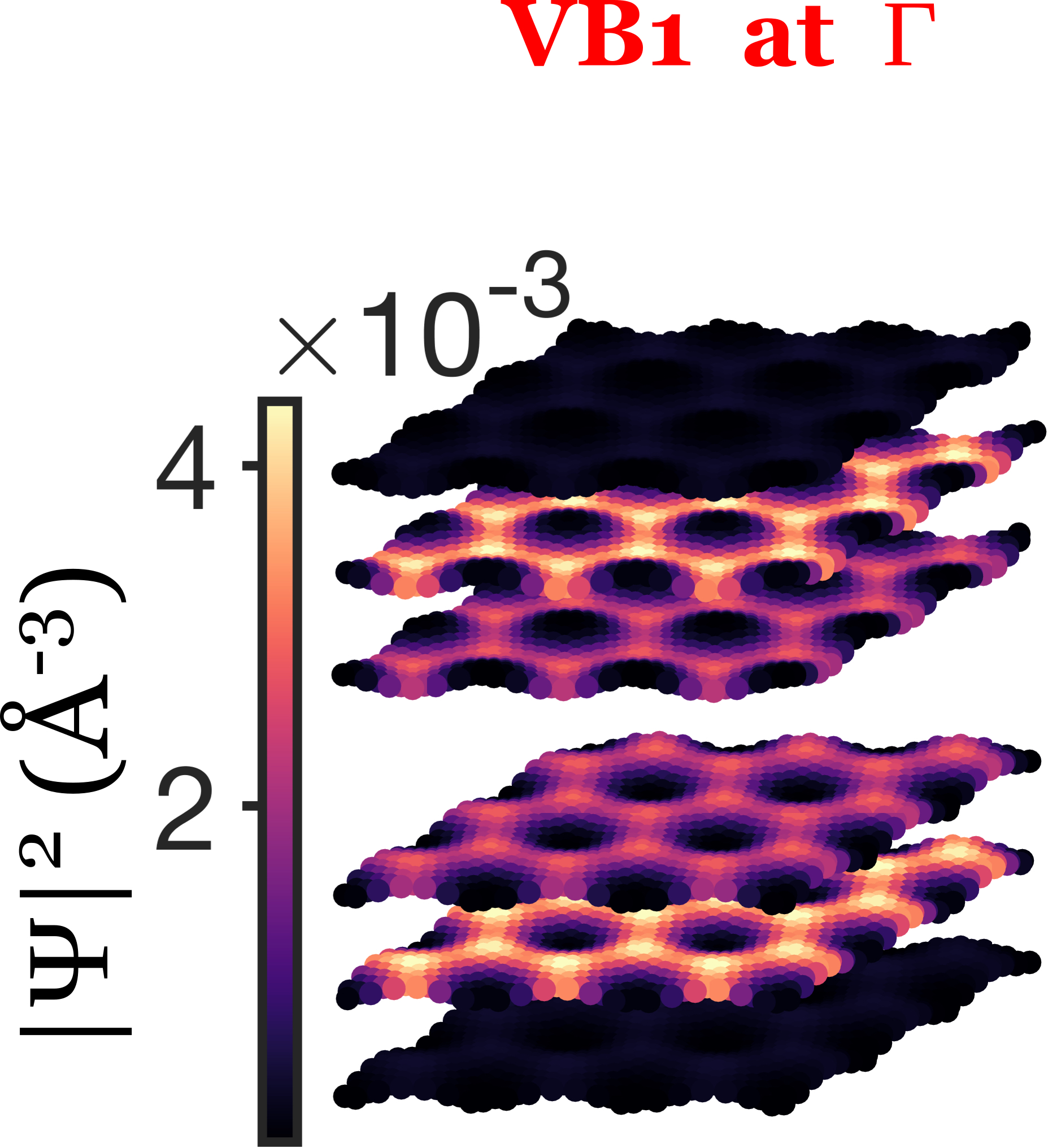}}
    \subcaptionbox{}{    
    \centering
    \includegraphics[width=0.63\textwidth]{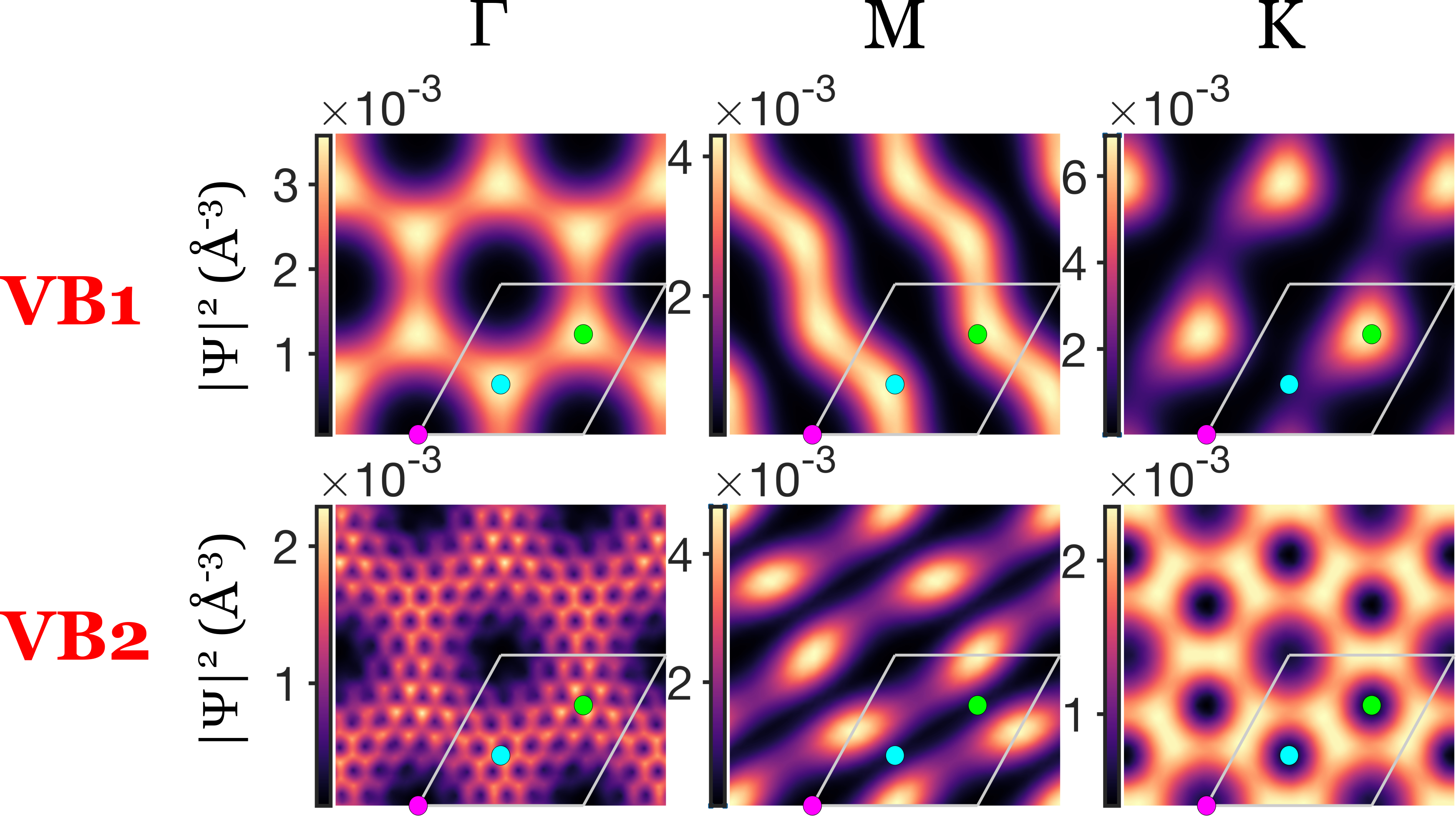}}
    \caption{Electronic structure of twisted \protect\bilayer{1}. (a) Moir\'e Brillouin zone (black hexagon) obtained by twisting two monolayers (whose Brillouin zones are indicated by blue and red hexagons) by $~\theta=22$\si{\degree}. The $\Gamma$-$M$-$K$-$\Gamma$ path used for computing band structures is also shown (yellow line). (b) Left: Band structure for $\theta=$\protect\homoangle{4} near the valence band edge. The two highest valence bands (denoted VB1 and VB2) are shown in red. Right: Density of states per \protect\monolayer{1} formula unit. (c) Layer-resolved $|\psi_\Gamma(\mathbf{r})|^2$ of VB1.
    (d) Layer-averaged squared wavefunctions of VB1 (top panels) and VB2 (bottom panels) at $\Gamma$, $M$ and $K$. Colored dots refer to different stacking regions as described in Fig.~\ref{fig1:corr_hobl}(a) and the moir\'e unit cell is indicated by grey lines.}
    \label{fig4:tBLWSe2}
\end{figure*}

Comparing the ILS of different bilayers, we observe that bilayers where both constituent monolayers contain S atoms (\bilayer{1}, \bilayer{3} and \bilayer{5}) exhibit the smallest interlayer distances (both in \region{AA} and \region{AB}/\region{BA} regions), whereas bilayers containing Se atoms (\bilayer{2}, \bilayer{4} and \bilayer{6}) in both layers exhibit the largest ILSs. Bilayers with S atoms in one layer and Se atoms in the other (\bilayer{7}, \bilayer{8}, \bilayer{9} and \bilayer{10}) show intermediate values of the ILS. These trends can be explained by the different van der Waals radii of S and Se atoms, which are $\sim 1.8$~\AA~ and $\sim 1.9$~\AA, respectively\cite{Batsanov2001}. 

The out-of-plane displacements for all homobilayers and heterobilayers with same chalcogen species at $\theta=$\homoangle{5} and for all heterobilayers with different chalcogens at $\theta=$\protect\heteroangle{2} are shown in Fig.~\ref{fig3:intdist_all}(b). As shown in the left panel of Fig.~\ref{fig3:intdist_all}(b), out-of-plane displacements in  homobilayers are layer-symmetric and the shape of the displacement patterns is similar for all systems. In contrast, out-of-plane displacements in heterobilayers (Fig.~\ref{fig3:intdist_all}(b) right panel) are layer-asymmetric. In these systems, the amplitude of the displacement pattern of the bottom layers (which are unstrained) is similar to that found in the homobilayers, while the amplitudes of the strained top layer are somewhat smaller. 

\subsection{Electronic structure}\label{subsec:results_electronic_structure}
\subsubsection{Homobilayers.}

In this section we study the evolution of the band structure of the twisted homobilayers \bilayer{1}, \bilayer{2}, \bilayer{3} and \bilayer{4} as function of twist angle. All calculations were carried out for the relaxed structures and include spin-orbit coupling. For all twist angles, the homobilayers exhibit a semiconducting band structure with a band gap separating the valence and conduction bands. Moreover, the two highest valence bands (each of which is spin degenerate) are separated from all other ``remote'' valence bands by energy gaps when $\theta< 4^\circ$. We refer to these two highest valence bands as VB1 and VB2, respectively.

\begin{figure}[t]
\centering
    \subcaptionbox{\bilayer{1} FLAT}{
        \includegraphics[width=0.22\textwidth,trim={70px 40px 0px 70px},clip]{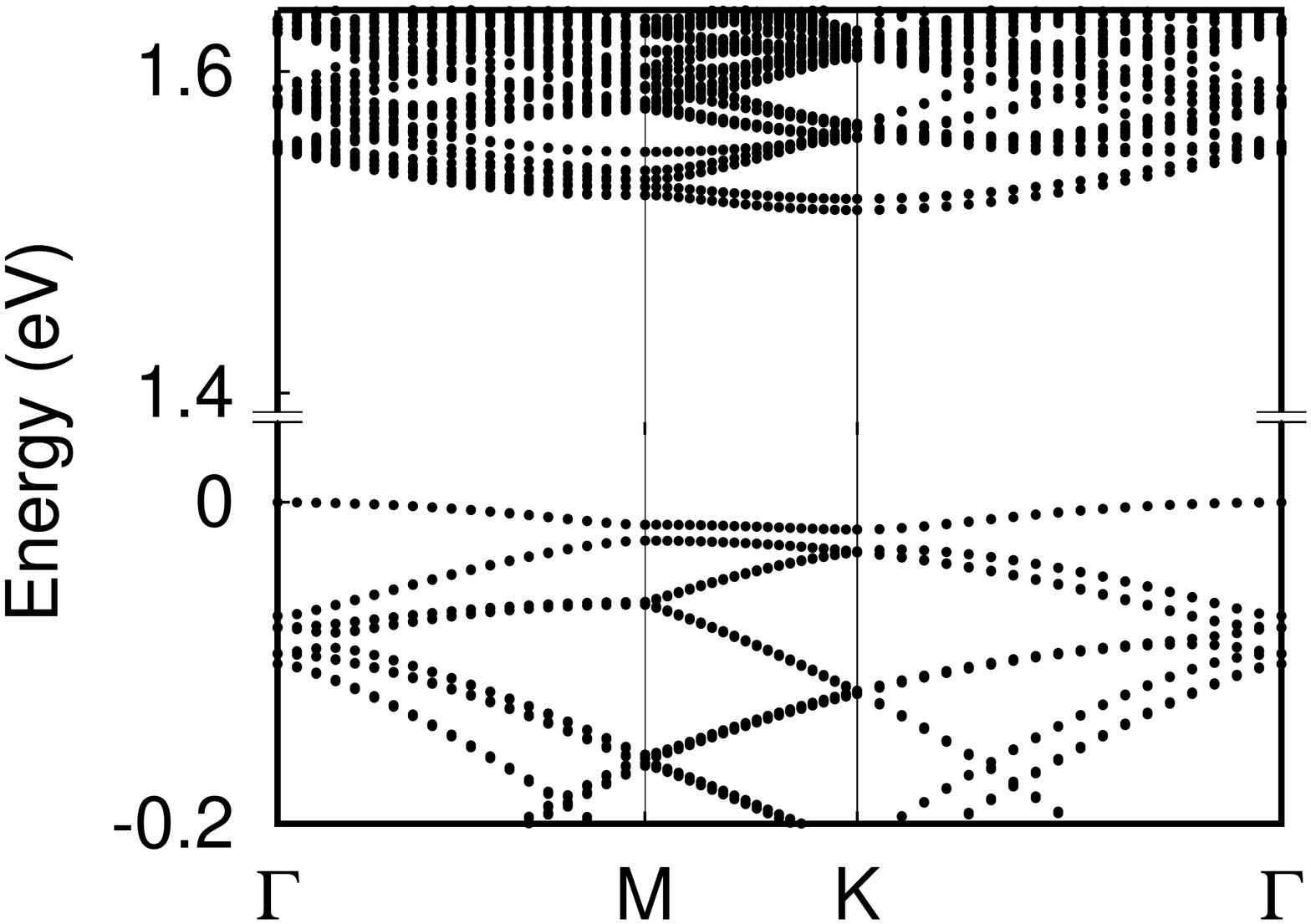}}
    \subcaptionbox{\bilayer{1} RELAXED}{
        \includegraphics[width=0.22\textwidth,trim={70px 40px 0px 70px},clip]{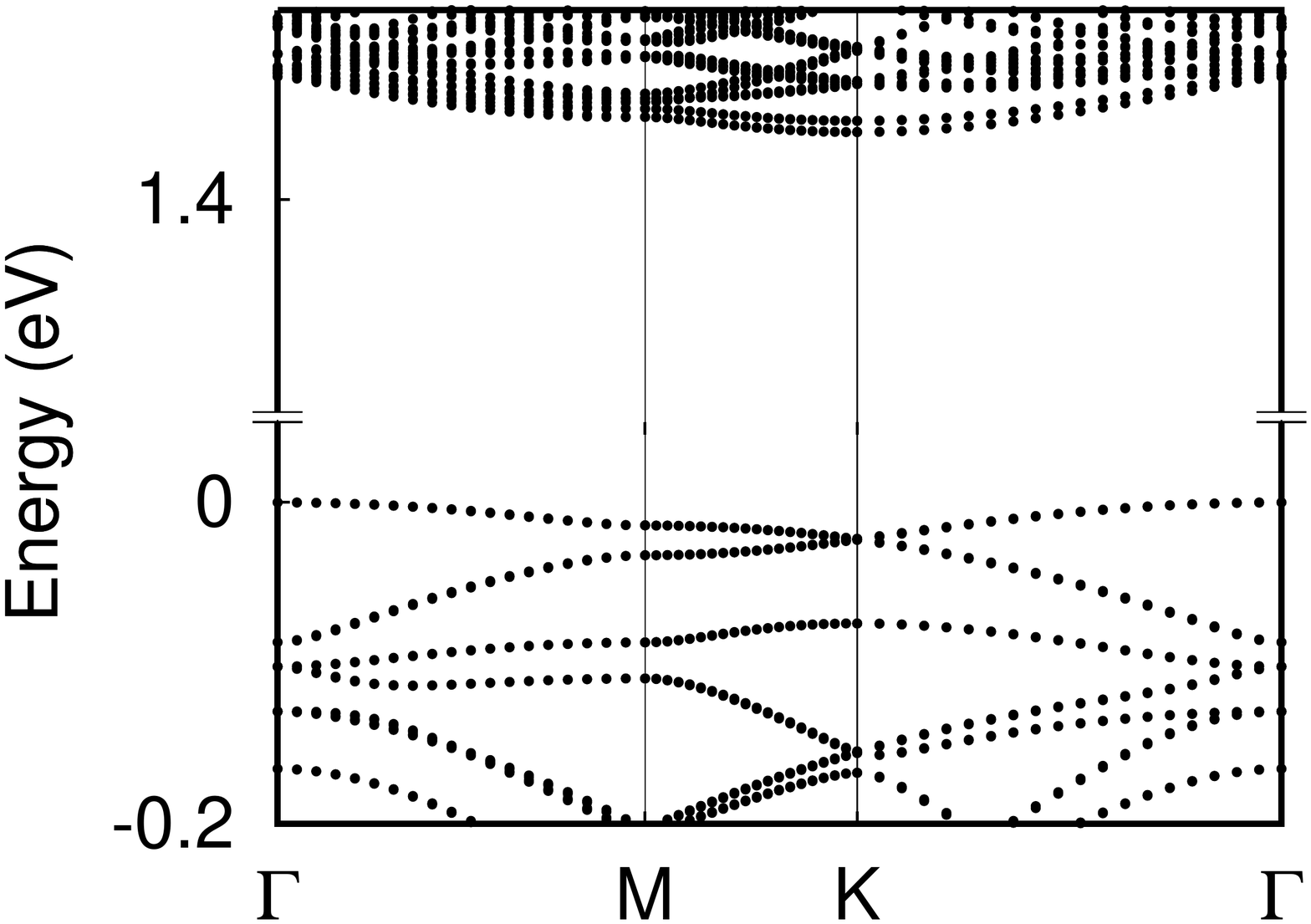}}\\
    \subcaptionbox{}{
        \includegraphics[width=0.22\textwidth]{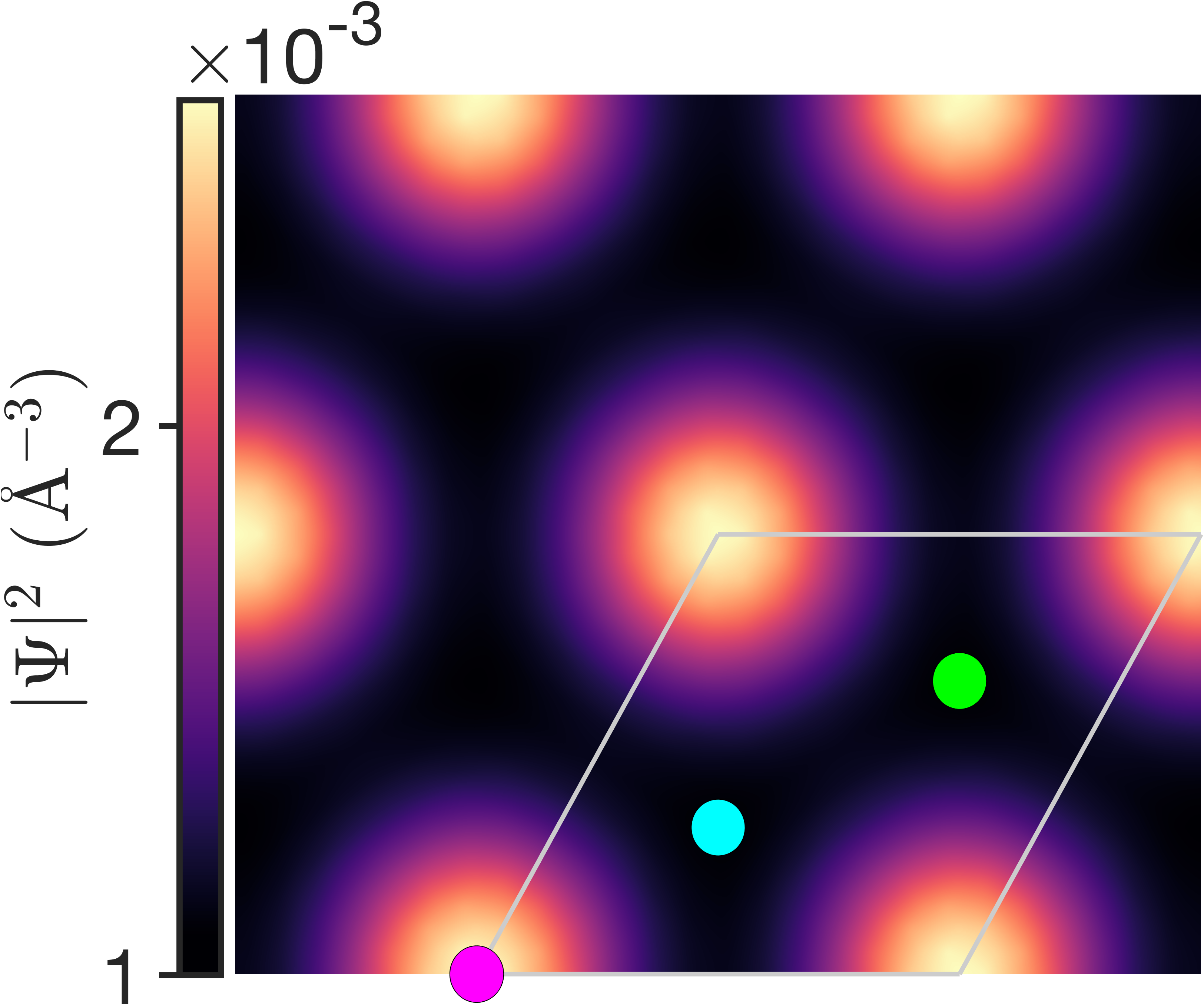}}
    \subcaptionbox{}{
        \includegraphics[width=0.22\textwidth]{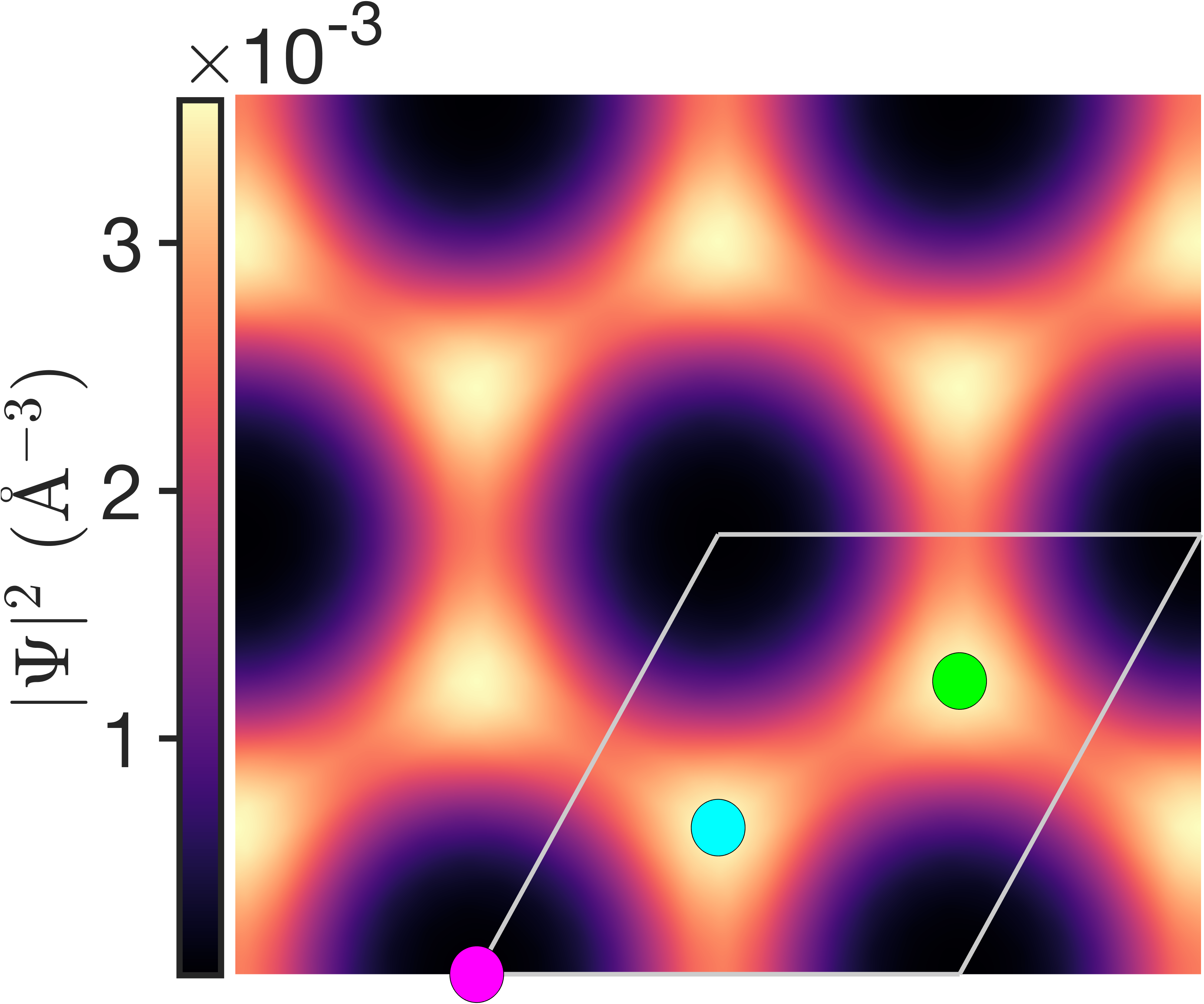}}
 \caption{Band structure of twisted \protect\bilayer{1} at $\theta=$\protect\homoangle{5} using (a) the unrelaxed flat atomic structure and (b) the relaxed atomic structure. Layer-averaged $|\psi_\Gamma(\mathbf{r})|^2$ of the highest valence band for (c) flat and (d) relaxed structures. Colored dots refer to different stacking regions as described in Fig.~\ref{fig1:corr_hobl}(a).}\label{fig5:hobls_flat_vs_relaxed}
\end{figure}

As an example, we focus on the valence band structure of twisted \bilayer{1}. Fig.~\ref{fig4:tBLWSe2} (b) shows the band structure at $\theta=$\homoangle{5}. It can be seen that the valence band maximum occurs at the $\Gamma$-point of the moir\'e Brillouin zone and that VB1 and VB2 touch at the $K$-point forming a Dirac cone. This is also reflected in the V-shaped density of states (Fig.~\ref{fig4:tBLWSe2} (b)). The graphene-like dispersion of VB1 and VB2 can be understood by analyzing the wavefunctions of these states. Fig.~\ref{fig4:tBLWSe2}(d) shows that the wavefunctions at $\Gamma$ are localized in the \region{AB} and \region{BA} regions which form a honeycomb lattice. Importantly, the total bandwidth of the two highest valence bands is less than 30~meV demonstrating the formation of flat bands upon twisting.

To understand the chemical origin of the flat bands, we analyze their projections onto atomic orbitals. Fig.~\ref{fig4:tBLWSe2}(c) shows that these states are localized symmetrically on the inner layers of chalcogen atoms and also on the two metal layers. These states are mostly composed of inner chalcogen p$_z$-like orbitals and metal d$_{z^2}$-like orbitals, as show in Fig.~S5 of the Supplementary Information. This suggests that VB1 and VB2 originate from $\Gamma$-states of the constituent monolayers: in all TMD monolayers, the top valence band states at $\Gamma$ have large projections onto chalcogen p$_z$-like orbitals and metal d$_{z^2}$-like orbitals, whereas the top valence band states at $K$ and $K'$ have large projections onto metal d$_{xy}$-like and d$_{x^2-y^2}$-like orbitals\cite{Kaxiras_11bands}. 

When atomic relaxations are not taken into account, a qualitatively different valence band structure is obtained. Figs.~\ref{fig5:hobls_flat_vs_relaxed}(a) and \ref{fig5:hobls_flat_vs_relaxed}(b) compare the band structures of unrelaxed and relaxed \bilayer{1} at $\theta=5.1^\circ$. In contrast to the relaxed system, the unrelaxed system does not exhibit a Dirac-like dispersion and exhibits an energy gap between VB1 and VB2. This is a consequence of the different spatial structure structure of the corresponding wavefunction, see Figs.~\ref{fig5:hobls_flat_vs_relaxed}(c) and ~\ref{fig5:hobls_flat_vs_relaxed}(d): in the relaxed system VB1 and VB2 localize in the \region{AB} and \region{BA} regions, while in the unrelaxed system the top valence band states are localized in the \region{AA} regions which form a triangular lattice\cite{Wu_PRL_121}. We find a similar effect of atomic relaxations in all homobilayers, see Supplementary Information (S1).  

Figure~\ref{fig6:homobilayers} compares the band structures of all homobilayers at three twist angles ($\theta=$\homoangle{5}, \homoangle{2} and \homoangle{1}). At $\theta=$\homoangle{5}, the top valence band in twisted \bilayer{1} and \bilayer{2} are $\Gamma$-derived and exhibit a Dirac-like dispersion with a valence band maximum at $\Gamma$, as discussed above. In contrast, for \bilayer{3} and \bilayer{4} the $\Gamma$-derived valence states are intersected by dispersive bands which are derived from monolayer $K$/$K'$-states (see discussion below). In \bilayer{4}, the highest valence band is $K$/$K'$-derived and the valence band maximum is found at the $K$-point. 

When the twist angle is reduced to $\theta=$\homoangle{2}, the $\Gamma$-derived bands become flatter and the $K$/$K'$-derived bands are shifted to lower energies such that they no longer intersect the flat $\Gamma$-derived top valence bands. Decreasing the twist angle further to $\theta=$\homoangle{1}, we observe that the highest four remote $\Gamma$-derived valence bands become isolated in energy from all other remote bands, see Figs.~\ref{fig6:homobilayers}(c),(f),(i),(l). The two middle bands of this set also exhibit a Dirac-like dispersion near the $K$-point, while the highest and lowest bands are very flat. Our results are in good agreement with DFT calculations performed by Xian and coworkers\cite{xian2020realization}, who also analyzed the origin of these bands and proposed that they can be described by a set of p$_x$-like and p$_y$-like orbitals on a honeycomb lattice. 

Figure~\ref{fig7:hobls_properties}(a) shows the bandwidth $w$, computed as the energy difference between states at $\Gamma$ and $K$, of the top valence band (denoted as VB1 as in Fig.~\ref{fig4:tBLWSe2}(b)) as function of twist angle for relaxed and unrelaxed (flat) homobilayers. As the twist angle decreases, the bandwidths approach zero. For relaxed homobilayers, the magnitude of the bandwidths in the different systems are relatively similar with \bilayer{2} exhibiting the smallest one (reaching $\approx 0.5$~meV at $\theta=$\homoangle{1}). When relaxations are neglected, the bandwidths are smaller. For example, a bandwidth of only $0.2$~meV is found in unrelaxed \bilayer{2} at $\theta=$\homoangle{2}.

\begin{figure*}[t]
    \centering 
    \includegraphics[width=0.9\textwidth]{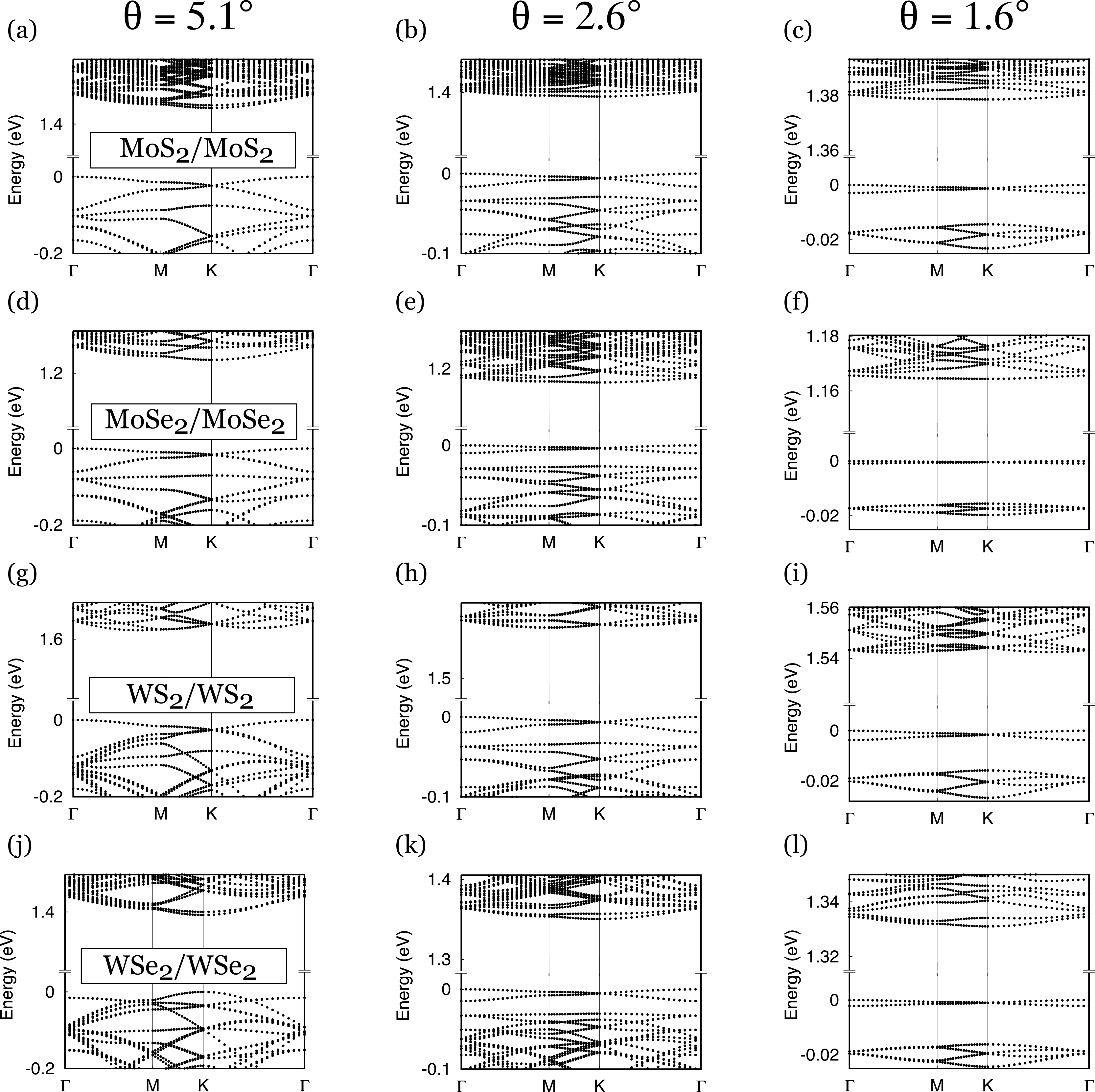}
    \caption{Band structures of twisted TMD homobilayers (\protect\bilayer{1}, \protect\bilayer{2}, \protect\bilayer{3} and \protect\bilayer{4}) for three twist angles $\theta=$\protect\homoangle{5}, \protect\homoangle{2} and \protect\homoangle{1}. The high-symmetry path $\Gamma$-$M$-$K$-$\Gamma$ is shown in Fig.~\ref{fig4:tBLWSe2}(a).}
    \label{fig6:homobilayers}
\end{figure*}

\begin{figure*}[t]
    \centering
    \subcaptionbox{ }{
        \includegraphics[width=0.47\textwidth,trim={0px 0px 0px 5px},clip]{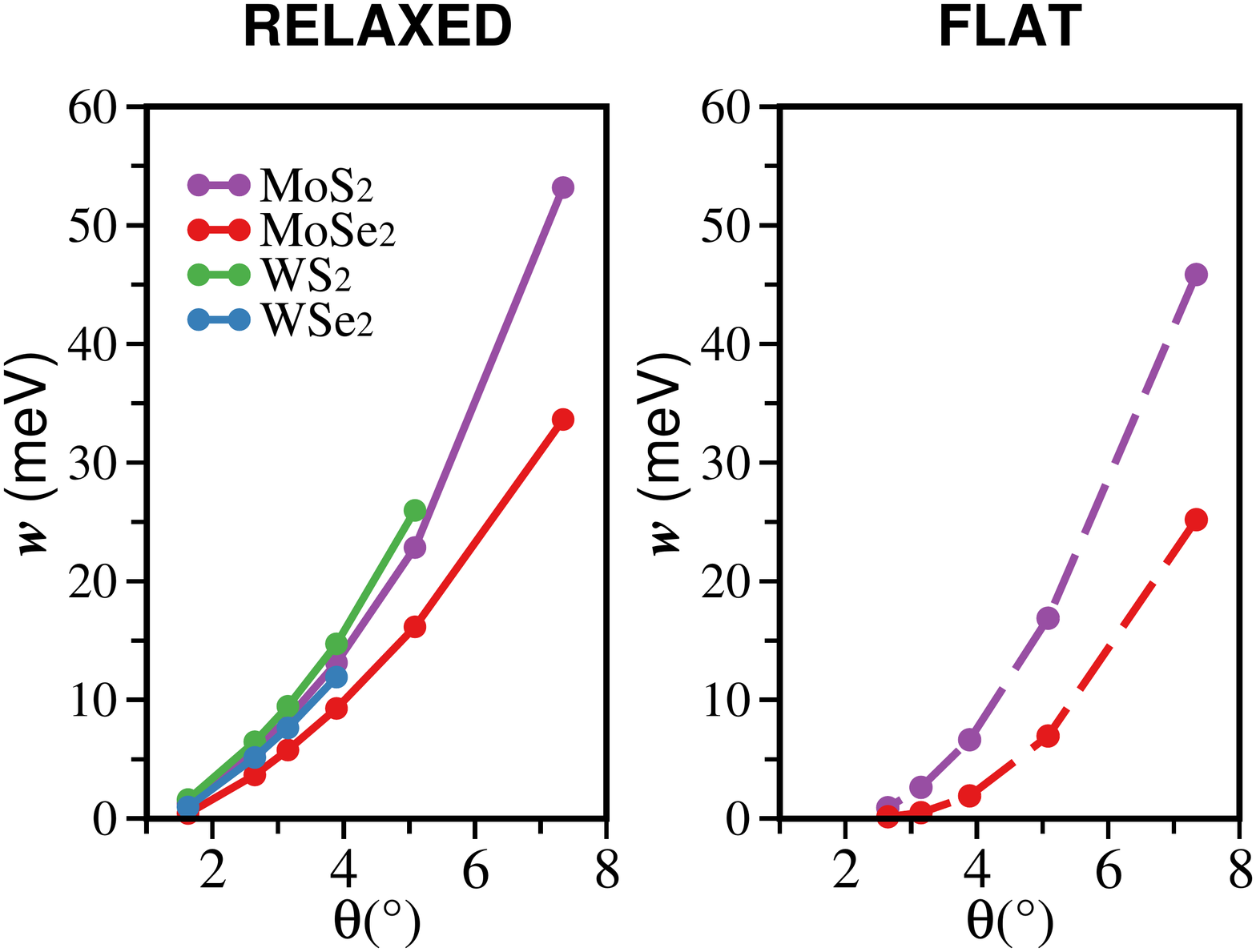}}
        \centering
    \subcaptionbox{ }{
        \includegraphics[width=0.47\textwidth,trim={0px 0px 0px 0px},clip]{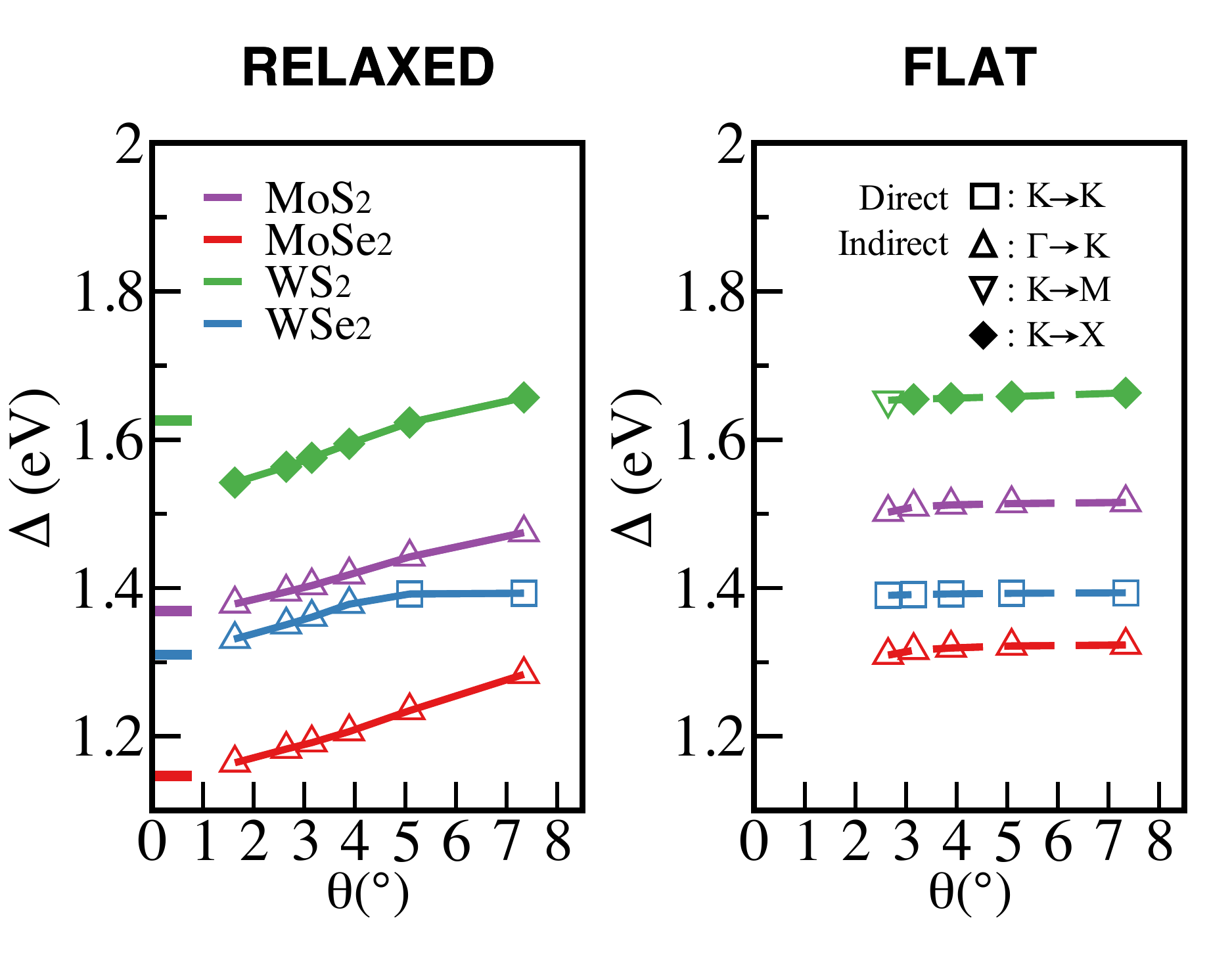}}\\
        \centering
    \caption{(a) Band width $w$ and (b) band gap $\Delta$ of the top $\Gamma$-derived valence band (VB1) as function of twist angle $\theta$ for relaxed and unrelaxed (flat) homobilayers. We only show band widths for systems in which the $\Gamma$-derived states are not intersected by $K$/$K'$-derived valence states.}
    \label{fig7:hobls_properties}
\end{figure*}
\begin{figure*}[t]
    \centering
    \includegraphics[width=0.9\textwidth]{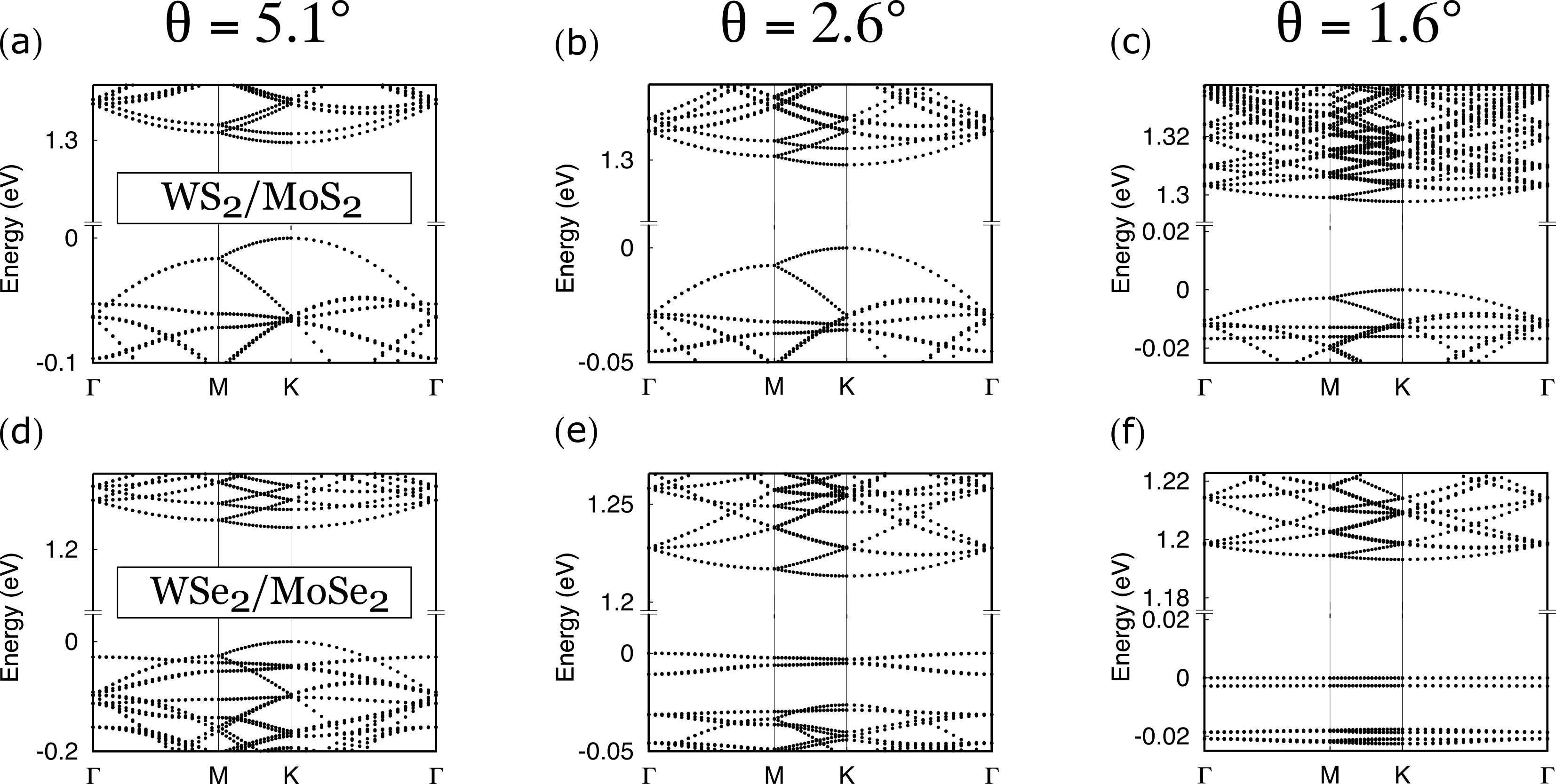}
    \caption{Band structures of twisted TMD heterobilayers with the same type of chalcogen in both monolayers. Results are presented for three twist angles: $\theta=$\protect\homoangle{5},\protect\homoangle{2} and \protect\homoangle{1}.}
    \label{fig9:heterobilayers1}
\end{figure*}

Figure~\ref{fig7:hobls_properties}(b) shows the band gap $\Delta$ between valence and conduction states as function of twist angle for both relaxed and unrelaxed homobilayers. For all relaxed homobilayers $\Delta$ decreases linearly with a slope of $\approx20$~meV/degree as the twist angle is reduced. \bilayer{3} has the largest band gap ($1.55-1.65$~eV) and \bilayer{2} the smallest ($1.15-1.3$~eV). The same ordering is found for the untwisted bilayers (independent of the stacking arrangement). With only the exception of \bilayer{3}, at small angles the band gaps of twisted bilayers approach the values of the untwisted \region{AB}(or \region{BA}) bilayers, shown on the left panel of Fig.~\ref{fig7:hobls_properties}(b) at $\theta=0$\si{\degree}. This is expected as the \region{AB} and \region{BA} regions are energetically favorable (compared to \region{AA} regions) and their relative size grows as the twist angle is reduced (see Sec.~\ref{subsec:methods_atomic_structure}). 

Interestingly, the nature of the band gap of \bilayer{4} changes from direct \mbox{($K \rightarrow K$)} to indirect ($\Gamma \rightarrow K$) around $\theta = $\homoangle{5}. This is a consequence of the change in ordering of $\Gamma$-derived and $K$/$K'$-derived valence states, see Fig.~\ref{fig6:homobilayers}. All other systems exhibit indirect band gaps. In particular, for \bilayer{1} and \bilayer{2} the valence band maximum is at $\Gamma$ and the conduction band minimum at $K$ as in the untwisted case, while for \bilayer{3} the conduction band minimum is half-way between the $\Gamma$-point and the $M$-point (referred to as the $X$-point), which explains the deviation from the untwisted case.

Without relaxations, the band gaps are almost constant and do not depend sensitively on the twist angle, see Fig.~\ref{fig7:hobls_properties}(b). Also, the nature of the band gap for \bilayer{4} is different compared to the relaxed systems for $\theta <$\homoangle{5}, as is for \bilayer{3} at $\theta=$\homoangle{2}.

\subsubsection{Heterobilayers.}\label{subsec:heterobilayers}
\begin{figure*}[t!]
    \centering
    \includegraphics[height=0.9\textwidth]{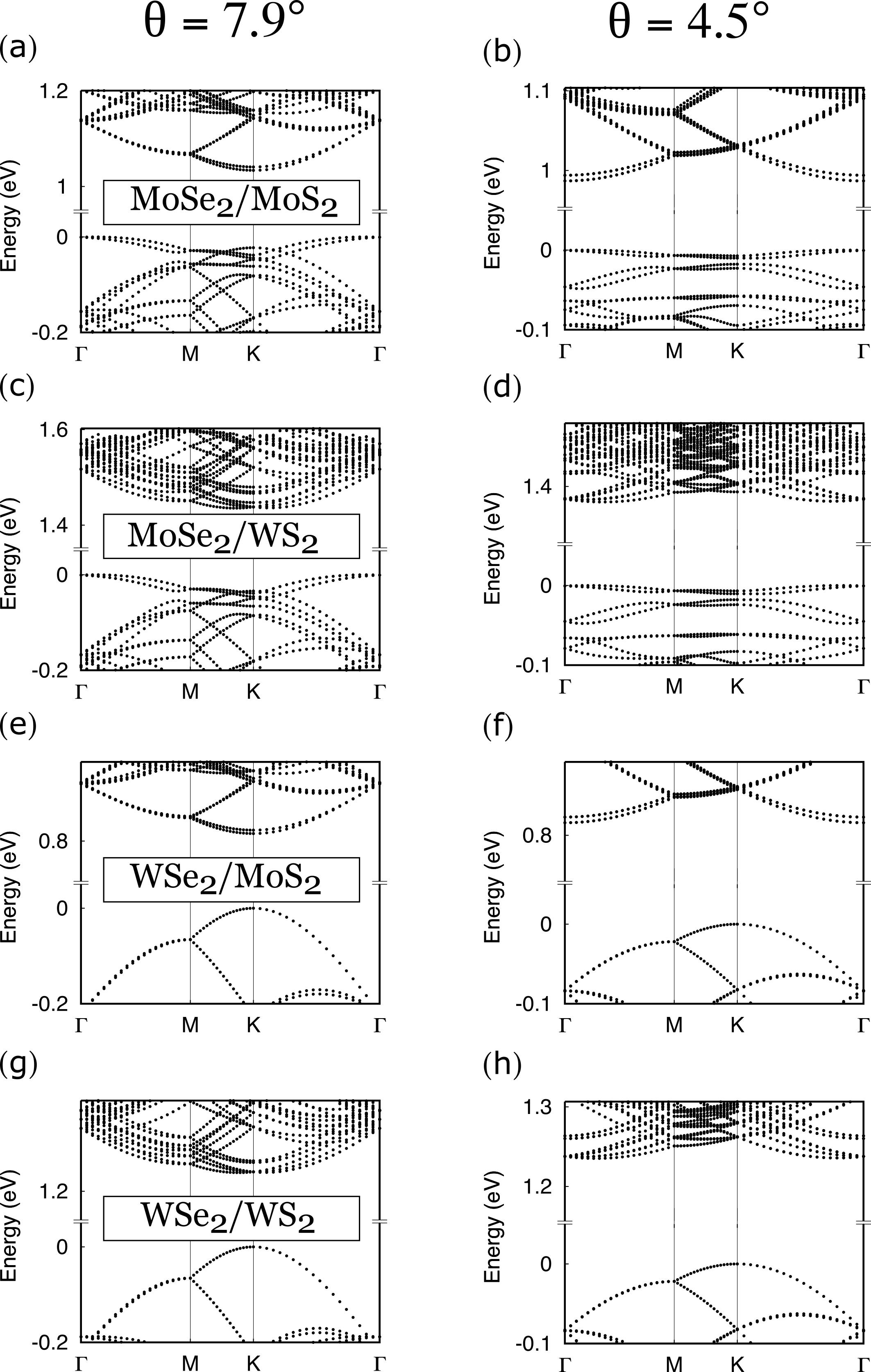}
    \caption{Band structures of twisted TMD heterobilayers with different types of chalcogen atoms in the constituent monolayers as function of twist angle. Results are presented for $\theta=$\protect\heteroangle{3} and \protect\heteroangle{1}.}
    \label{fig10:heterobilayers2}
\end{figure*}
We first consider heterobilayers with the same chalcogen species in each layer, i.e., \bilayer{5} and \bilayer{6}. As discussed in Sec.~\ref{sec:methods}, it is possible to generate commensurate moir\'e structures with very little strain for these systems. The band structures of these systems at three different twist angles are shown in Fig.~\ref{fig9:heterobilayers1}. Again, we find that these systems feature both flat $\Gamma$-derived valence bands as well as dispersive $K$/$K'$-derived states.

In \bilayer{6} at $\theta=5.1^\circ$, the valence band maximum corresponds to a $K$/$K$'-derived state. At smaller twist angles, the $K$/$K'$-derived states are shifted to lower energies and the top valence bands are $\Gamma$-derived and very flat. Similar to the homobilayers, the $\Gamma$-derived top two valence bands are separated from all other remote valence bands at small twist angles. However, these bands no longer have a Dirac-like dispersion, but are separated by an energy gap at $K$. This energy gap caused by the ``chemical asymmetry" of the two constituent layers.

In contrast, the highest valence bands in \bilayer{5} are derived from monolayer $K$/$K'$-states at all twist angles.

Next, we first consider the heterobilayers \bilayer{7} and \bilayer{8}, i.e. hetero-bilayers containing different species of chalcogens but no \monolayer{4}. Figs.~\ref{fig10:heterobilayers2}(a)-(d) show the band structures of these systems at two twist angles ($\theta=$\heteroangle{3} and $\theta=$\heteroangle{1}). For \bilayer{7} the top valence bands have large projections onto p$_z$ orbitals of the inner chalcogen atoms and metal d$_{z^2}$-like orbitals, and the projections are layer-asymmetric with significantly more weight on the \monolayer{2} layer, see Fig.~S7 of Supplementary Information. 
\begin{figure}[t]
\centering
\subcaptionbox{ }{
\includegraphics[width=0.33\textwidth,trim={0px 0px 0px 0px},clip]{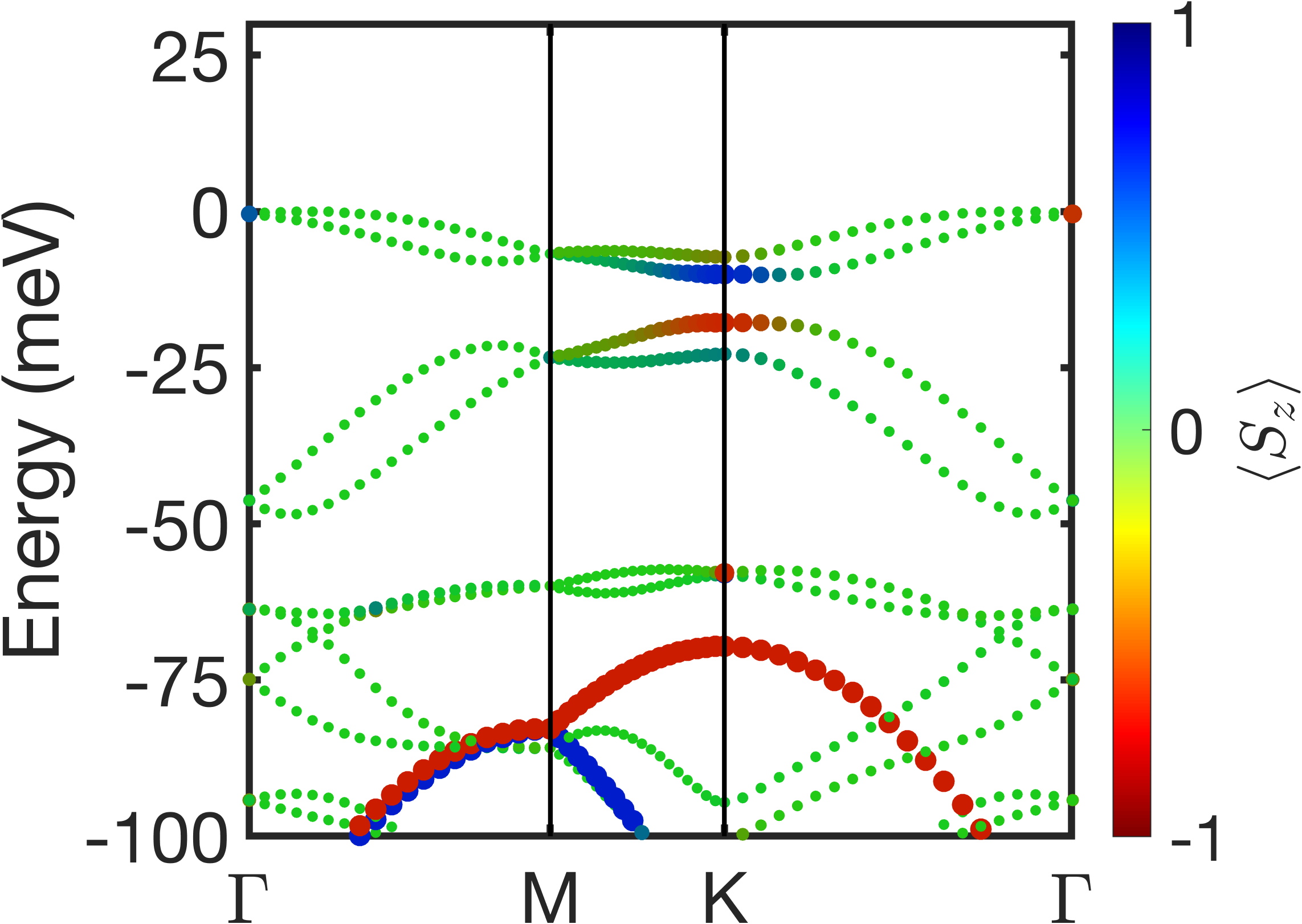}}\\
\subcaptionbox{ }{
\includegraphics[width=0.33\textwidth,trim={0px 0px 0px 0px},clip]{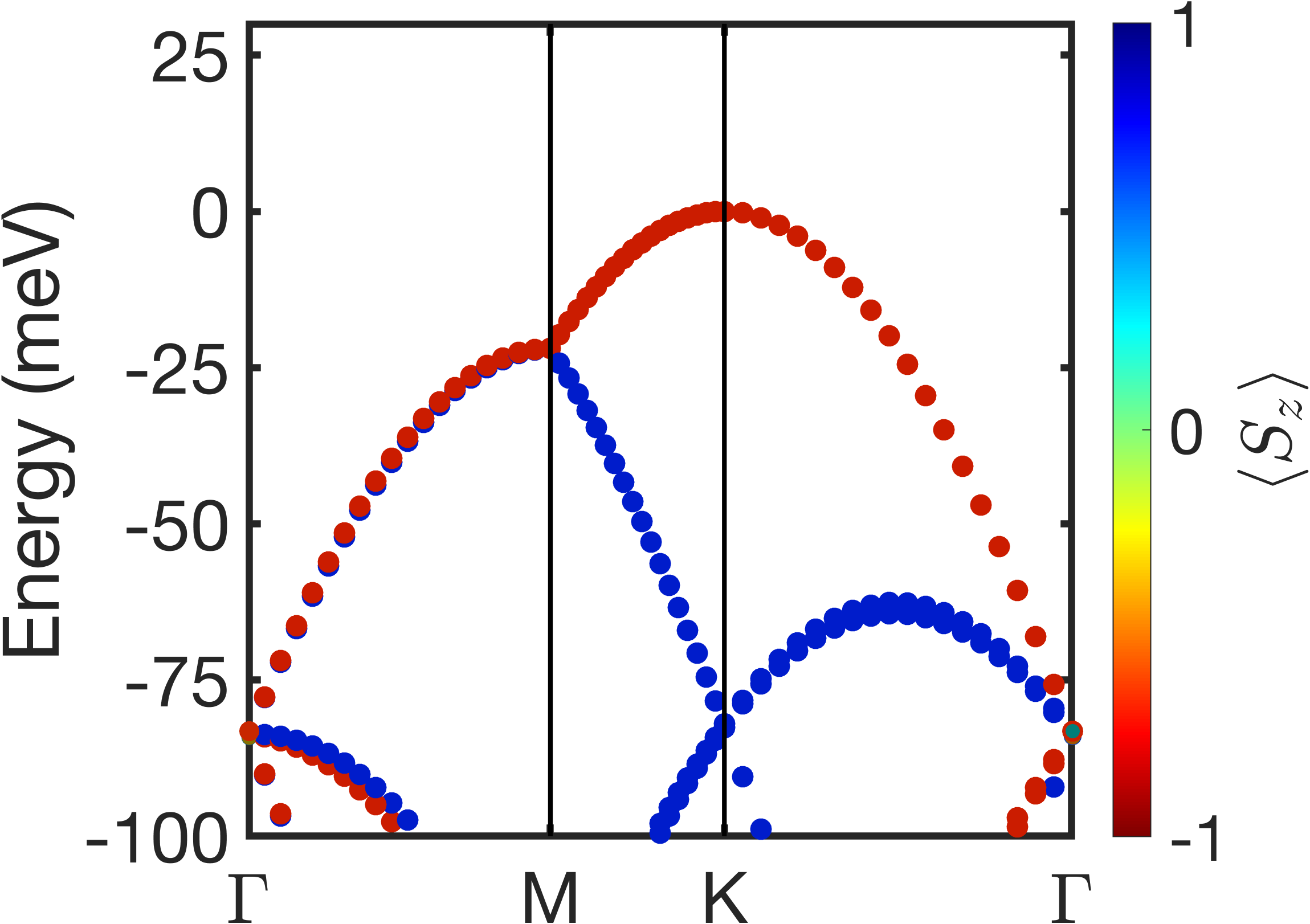}}
\caption{Spin-resolved valence band structure of (a) \protect\bilayer{7} at $\theta=$\protect\heteroangle{1} and (b) \protect\bilayer{9} at $\theta=$\protect\heteroangle{1}.} 
\label{fig11:spin_projection}
\end{figure}
In contrast to the homobilayers and the heterobilayers containing a single chalcogen species, the top valence bands in these systems are not spin-degenerate. Fig.~\ref{fig11:spin_projection}(a) shows the spin-resolved dispersion of top valence bands in \bilayer{7} at $\theta=$\heteroangle{1} which exhibits spin splittings with magnitudes up to 13~meV. Also, an energy gap between VB1 and VB2 of 8~meV is found at the $K$-point. Interestingly, these bands are partially spin-polarized: 
it can be observed that the top valence bands are only spin-polarized in the vicinity of the $K$-point even though spin splitting occurs along the whole band structure path. This scenario was recently discussed by Liu and coworkers\cite{Liu2019} who demonstrated that spin-orbit coupling can lead to spin splittings without spin polarization in non-magnetic materials without inversion symmetry.

Finally, Figs.~\ref{fig10:heterobilayers2}(e)-(h) show the band structures of twisted \bilayer{9} and \bilayer{10} bilayers, i.e., the heterobilayers with different chalcogen atoms that contain \monolayer{4}. The top valence bands in these systems are dispersive and the flat bands are observed at lower energies, see discussion in Sec.~\ref{sec:flat_vs_dispersive}.

Figure~\ref{fig13:hebls_properties} compares the bandwidths and band gaps of the different heterobilayers as function of twist angle. Note that we only show results for twisted bilayers whose top valence bands are flat, i.e., no results are shown for \bilayer{5}, \bilayer{9} and \bilayer{10}. For \bilayer{6} flat bands are only found for $\theta < $\homoangle{5}. Similar to the case of the homobilayers, the bandwidths of the heterobilayers decrease monotonically as the twist angle approaches zero and the value of the bandwidth at a fixed twist angle is roughly the same for the different homo- and heterobilayers.
\begin{figure}[h]
\centering
    \subcaptionbox{\protect\bilayer{6} UNRELAXED}{
        \includegraphics[width=0.22\textwidth,trim={70px 40px 0px 70px},clip]{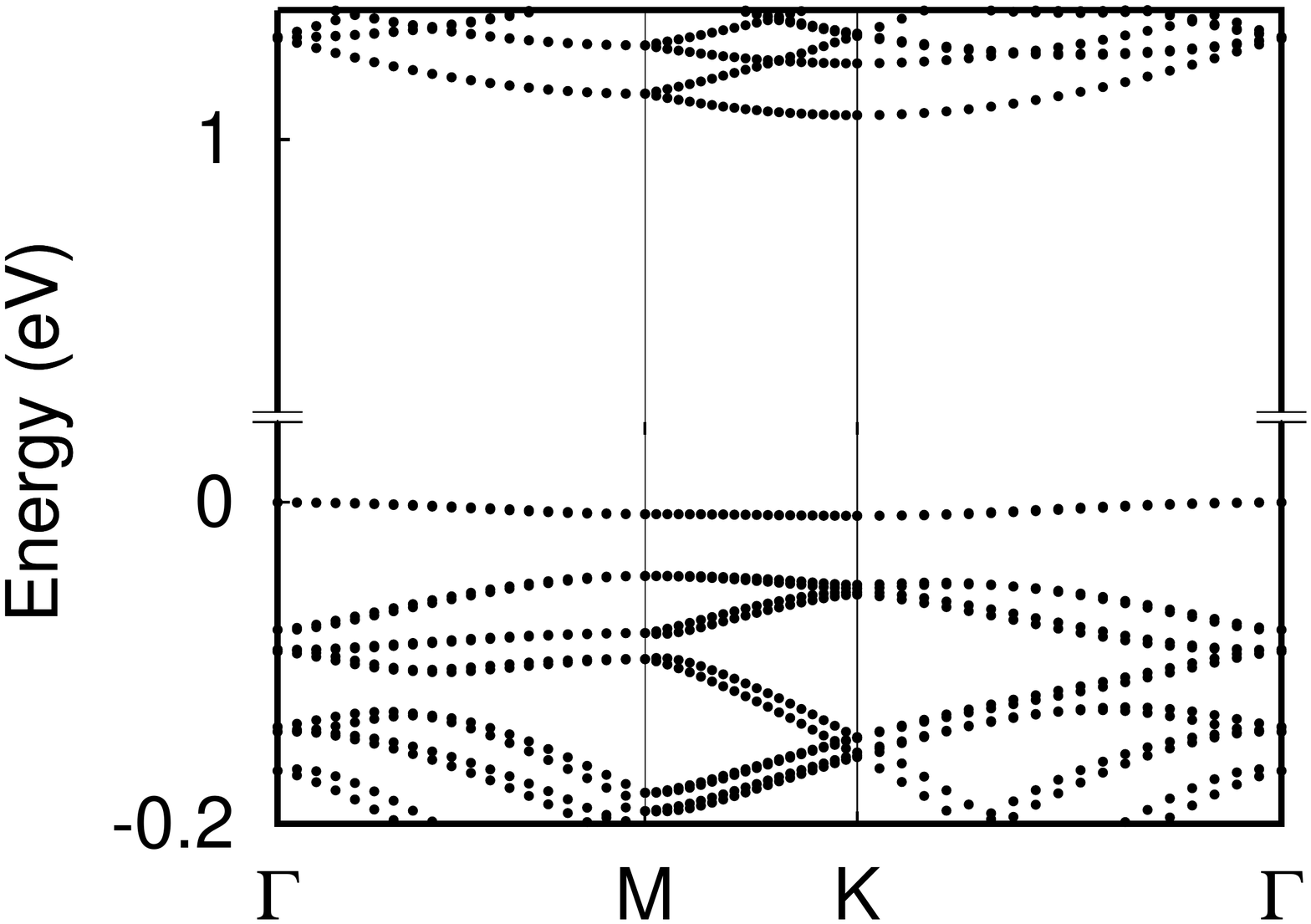}}
    \subcaptionbox{\protect\bilayer{6} RELAXED}{
        \includegraphics[width=0.22\textwidth,trim={70px 40px 0px 70px},clip]{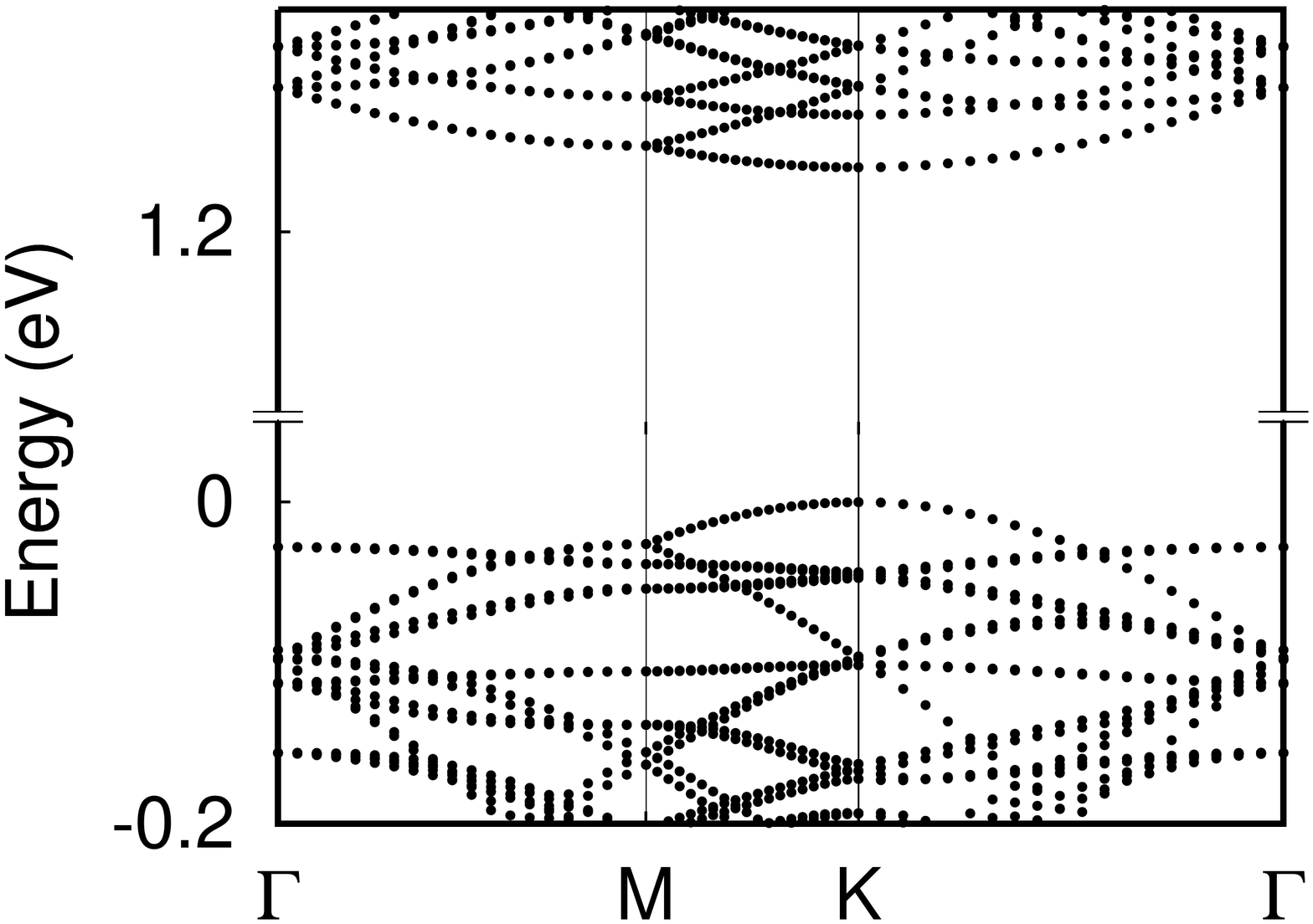}}

        \caption{Band structure of twisted \protect\bilayer{6} at $\theta=$\protect\homoangle{5} using (a) the unrelaxed (flat) atomic structure and (b) relaxed atomic structure.}\label{fig12:hebls_flat_vs_relaxed}
\end{figure}

The band gap of the twisted heterobilayers does not depend sensitively on twist angle, see Fig.~\ref{fig13:hebls_properties}(b). For most systems, a mild reduction of the gap can be observed as the twist angle decreases. The smallest bands gaps ($\approx0.8$~eV) are found for \bilayer{9}, while \bilayer{8} and \bilayer{5} exhibit the largest gaps ($\approx1.3-1.4$~eV). Interestingly, the nature of the band gap depends sensitively on the twist angle and many systems exhibit a change from a direct to an indirect gap as the twist angle is reduced. For example, the gap goes from indirect ($\Gamma$-$K$) to direct ($\Gamma$-$\Gamma$) in \bilayer{7}, and from indirect ($\Gamma$-X) to direct ($\Gamma$-$\Gamma$) in \bilayer{8}.

\begin{figure*}[t!]
    \centering
    \subcaptionbox{}{
        \includegraphics[width=0.47\textwidth,trim={0px 0px 0px 20px},clip]{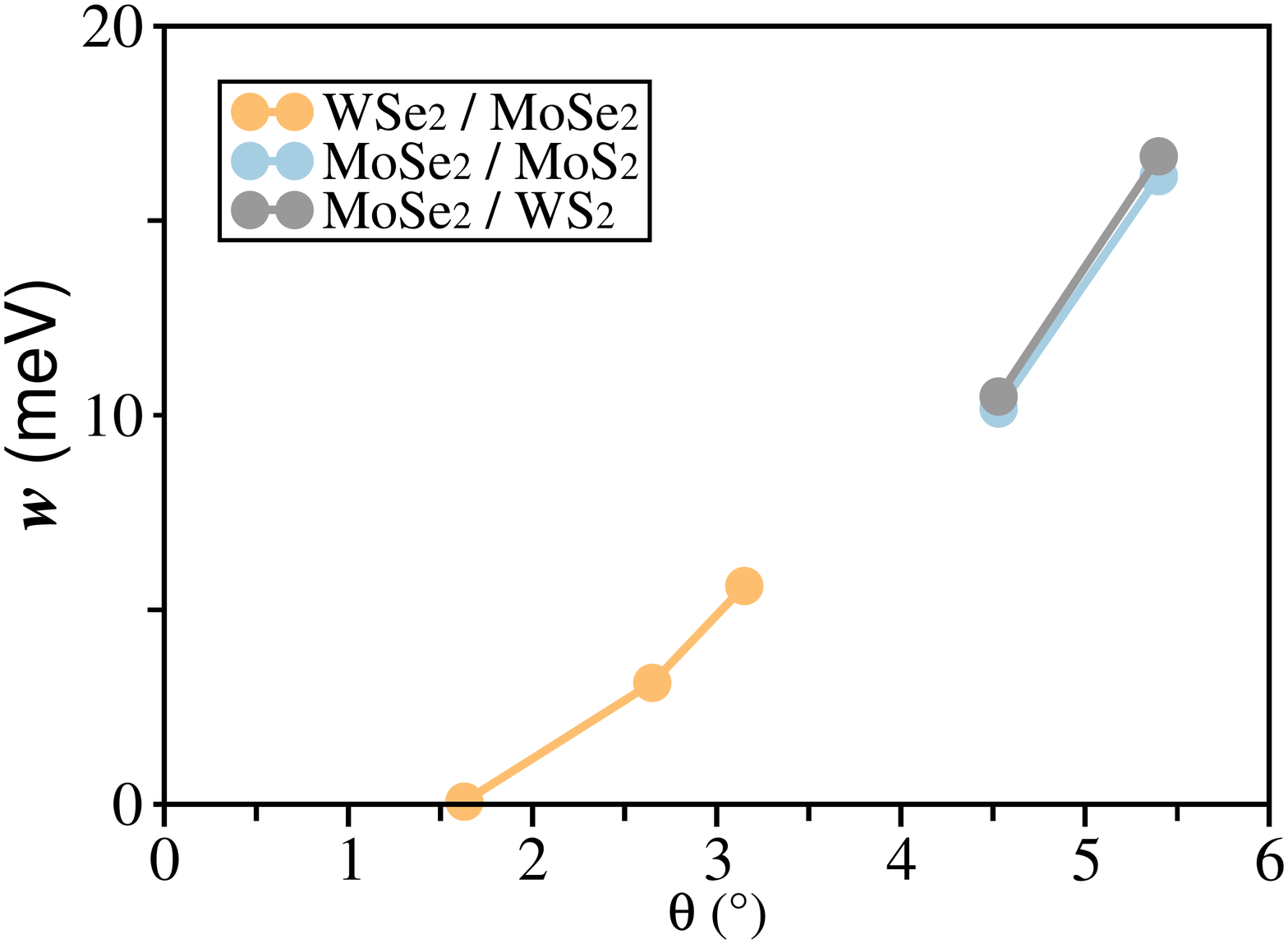}}
        \centering
    \subcaptionbox{}{
        \includegraphics[width=0.47\textwidth,trim={0px 0px 0px 10px},clip]{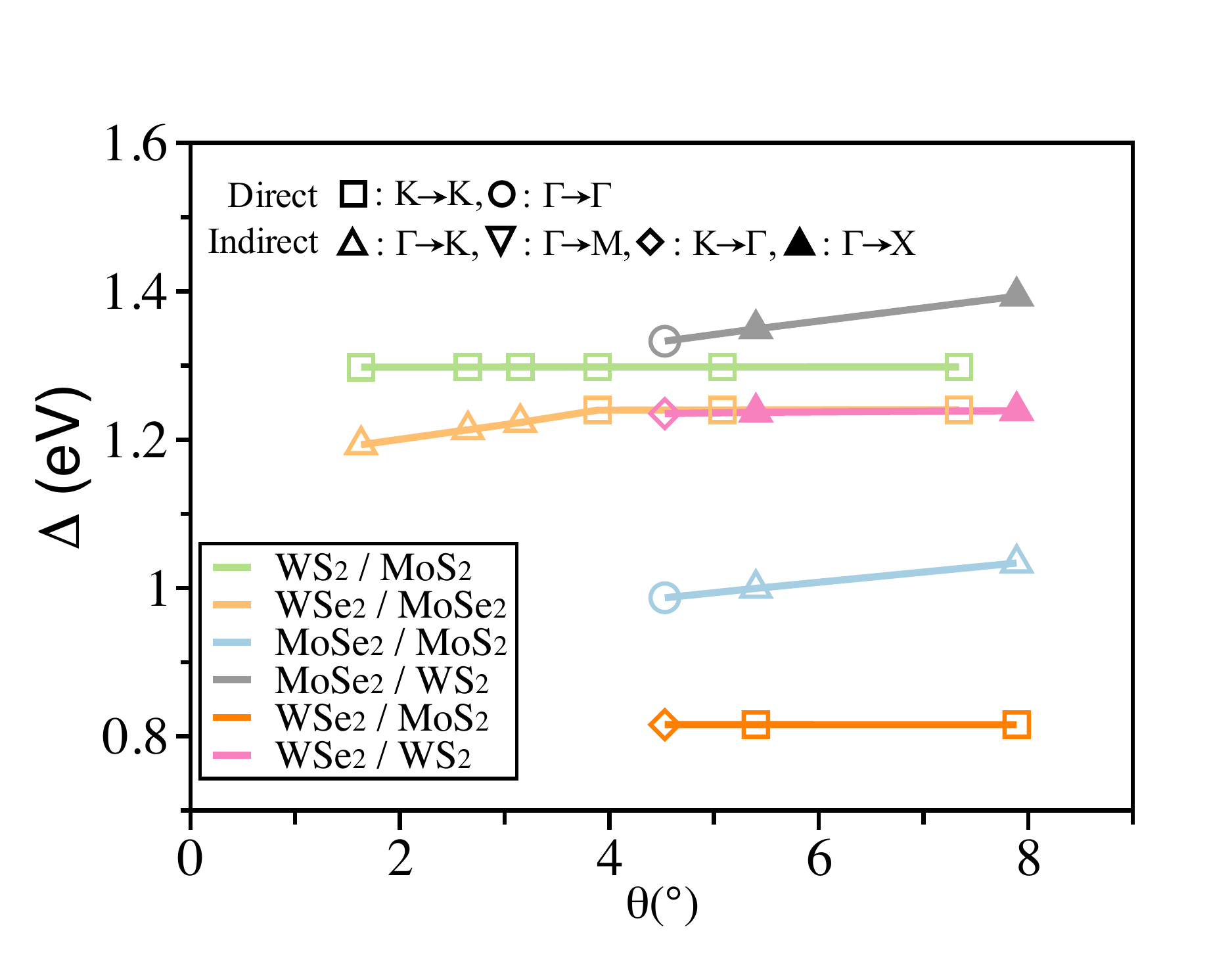}}\\
    \caption{(a) Bandwidth $w$ and (b) band gap $\Delta$ of the top valence band for different heterobilayers as a function of twist angle $\theta$. In panel (a) we only show results for twisted bilayers with flat valence bands (see discussion in the text). }
    \label{fig13:hebls_properties}
\end{figure*}

\subsection{Physical origin of the dispersive valence bands}\label{sec:flat_vs_dispersive}

We now focus on the set of heterobilayers that exhibit dispersive bands at the valence band edge and discuss the origin of these bands. As described above, such bands are observed near the top of the valence band in \bilayer{3} and \bilayer{4} homobilayers at large twist angles and form the top valence states at all twist angles in \bilayer{5}, \bilayer{9} and \bilayer{10}, see Figs.~\ref{fig10:heterobilayers2}(e)-(h). Compared to the flat bands, the width of these bands decreases much less as the twist angle is reduced. Projecting the corresponding states onto atomic orbitals reveals large contributions from W d$_{xy}$-like and d$_{x^2-y^2}$-like orbitals, see Fig.~S8 of Supplementary Information, suggesting that they originate from states at the $K$-point of \monolayer{3} monolayer in the case of \bilayer{5} and similarly from $K$-point states of the \monolayer{4} monolayer in the \bilayer{9} and \bilayer{10} systems\cite{Kaxiras_11bands}. As these states have very small projections onto the inner chalcogen atoms (see Fig.~S9 of the Supplementary Information), they are only weakly affected by interlayer coupling which explains why introduction of a twist does not result in a significant reduction of their band width in our model. However, it has been established that such $K$/$K'$-derived bands can become flat as a consequence of modulations in the intralayer hopping induced by in-plane relaxations~\cite{Li2021}. As explained in Sec.~\ref{sec:methods}, our current tight-binding approach does not capture such variations of the intralayer hopping and therefore does not capture this additional band flattening mechanism. 

The top valence band at the $K$-point of the \monolayer{3} and \monolayer{4} monolayers is spin-polarized and the dispersive valence bands of the twisted bilayers inherit this property as shown for \bilayer{9} in Fig.~\ref{fig11:spin_projection}(b). The valence band maximum in bilayers with dispersive bands is located at the $K$-point. In the \bilayer{5} bilayer the band gap is direct, whereas it is indirect in the \bilayer{9} and \bilayer{10} bilayers as the conduction band minimum is at $\Gamma$. 

We have found that the ordering of the flat and dispersive valence bands depends sensitively on the atomic structure and the twist angle. For example, Fig.~\ref{fig12:hebls_flat_vs_relaxed}(a) shows that neglecting atomic relaxations results in a different ordering of dispersive and flat valence bands. Moreover, we have found that it is possible to switch the order of flat and dispersive bands when the interlayer separation of the relaxed structures is reduced suggesting that the electronic properties of these materials can be easily tuned by applying pressure, as show in Fig.~S10 of the Supplementary Information.

To understand the ordering of flat and dispersive valence bands in the different heterobilayers, we propose a simple model in which the bilayer states originating from $K$/$K'$- or $\Gamma$-valleys of the constituent monolayers are obtained from a two-level system. Specifically, the Hamiltonians for the coupled valleys are given by 
\begin{equation}
\mathcal{H}_{\Gamma} = \left(\begin{array}{c c} 
                              \varepsilon^{(1)}_\Gamma & \Delta_\Gamma \\ 
                              \Delta_\Gamma & \varepsilon^{(2)}_\Gamma 
                              \end{array}
                        \right), \quad \mathcal{H}_{K} = \left(\begin{array}{c c} 
                              \varepsilon^{(1)}_{K} & \Delta_{K} \\ 
                              \Delta_{K} & \varepsilon^{(2)}_{K} 
                              \end{array}
                              \right),
 \label{eq:two-levelH}
\end{equation}
where $\epsilon_{\Gamma (K)}^{(1)}$ denotes the energy at $\Gamma$ ($K$) of the monolayer with the higher-lying valence band (and $\epsilon_{\Gamma (K)}^{(2)}$ denotes the corresponding energies for the monolayer with the lower-lying valence band). Also, $\Delta_{\Gamma (K)}$ describes the interlayer coupling. As the wavefunctions in the $K$/$K'$-valley are predominantly localized on the metal atoms, we assume that they are not affected by interlayer coupling and use $\Delta_{{K}}=0$. In contrast, the wavefunctions of $\Gamma$-valley states have projections onto chalcogen atoms and these states are pushed to higher energies by interlayer coupling\cite{C7CP00012J}. To calculate $\Delta_\Gamma$, we carry out tight-binding calculations for untwisted bilayers with \region{AB} stacking. Then, $\Delta_\Gamma$ is chosen such that the largest eigenvalue $\epsilon^\text{max}_\Gamma$ of $\mathcal{H}_\Gamma$ is equal to the energy  of the highest tight-binding valence band state at $\Gamma$ of the bilayer. For heterobilayers with different chalcogen atoms, we compute the value of $\epsilon^\text{max}_\Gamma$ and $\Delta_\Gamma$ by averaging the results from two calculations: one in which the lattice constants of both layers are set equal to $\max(a_1,a_2)$, and one in which the lattice constants of both layers are set to $\min(a_1,a_2)$, where $a_1$ and $a_2$ are the equilibrium lattice constants of the two monolayers, respectively. 

\begin{table*}[t!]
\centering
\caption{Parameters of $\mathcal{H}_\Gamma$ and $\mathcal{H}_{K}$ in Eq.~\ref{eq:two-levelH}. $\varepsilon_{\Gamma ({K})}^{(1)}$ denotes the energy at $\Gamma$ ($K$) of the monolayer with the higher-lying valence band, and $\varepsilon_{\Gamma ({K})}^{(2)}$ denotes the energy at $\Gamma$ ($K$) of the monolayer with the lower-lying valence band. $\Delta_{\Gamma}$ is the interlayer coupling. $\varepsilon^\text{max}_\Gamma$ denotes the largest eigenvalue of $\mathcal{H}_\Gamma$. All values are in eV.}
\sisetup{round-mode=places}
\begin{tabular}{@{}l *{6}{S[round-precision=2]} l @{}}
\toprule[1pt]
 & \multicolumn{1}{c}{$\varepsilon_\Gamma^{(1)}$} & \multicolumn{1}{c}{$\varepsilon_\Gamma^{(2)}$} & \multicolumn{1}{c}{$\varepsilon_{{K}}^{(1)}$} & \multicolumn{1}{c}{$\Delta_{\Gamma}$} & \multicolumn{1}{c}{$\varepsilon^\text{max}_\Gamma$}  & \multicolumn{1}{c}{$\varepsilon^{\text{max}}_\Gamma-\varepsilon^{(1)}_{{K}}$} & Fig.\\
\midrule[0.5pt]
 \bilayer{1}  &  -1.09528769  &  -2.305466512   &  -0.965323149   &  1.002254204  &  -0.5343	  &  0.4357   & \ref{fig6:homobilayers}(c)  \\
 \bilayer{2}  &  -0.44112537  &  -1.356268282   &  -0.252884315   &  0.856148124  &  0.0719	  &  0.3219   & \ref{fig6:homobilayers}(f) \\
 \bilayer{3}  &  -0.90632325  &  -2.137237419   &  -0.495155578   &  0.978359014  &  -0.3694	  &  0.1306   & \ref{fig6:homobilayers}(i) \\
 \bilayer{4}  &  -0.43890619  &  -1.362770759   &  -0.011798444   &  0.837712654  &  0.0557	  &  0.0657   & \ref{fig6:homobilayers}(l) \\ 
\midrule
 \bilayer{10} &  -0.43890619  &  -0.906323251   &  -0.011798444   &  0.542155222  &  -0.08275     &  -0.07275 & \ref{fig10:heterobilayers2}(h) \\ 
 \bilayer{9}  &  -0.43890619  &  -1.09528769    &  -0.011798444   &  0.561630431  &  -0.1161	  &  -0.1061  & \ref{fig10:heterobilayers2}(f) \\    
 \bilayer{5}  &  -0.90632325  &  -1.09528769    &  -0.495155578   &  0.026771066  &  -0.9063	  &  -0.4063  &  \ref{fig9:heterobilayers1}(c) \\
 \bilayer{6}  &  -0.43890619  &  -0.441125371   &  -0.011798444   &  0.5073       &   0.0673	  &  0.0773   & \ref{fig9:heterobilayers1}(f) \\ 
 \bilayer{7}  &  -0.44112537  &  -1.09528769    &  -0.252884315   &  0.564062555  &  -0.114	  &  0.136    & \ref{fig10:heterobilayers2}(b) \\
 \bilayer{8}  &  -0.44112537  &  -0.906323251   &  -0.252884315   &  0.542710796  &  -0.0823	  &  0.1677   & \ref{fig10:heterobilayers2}(d) \\
\bottomrule
 \bottomrule[1pt]
\end{tabular}
\label{tab:delta}
\end{table*}

Table~\ref{tab:delta} shows the results from this analysis. For the \bilayer{1} and \bilayer{2} homobilayers, we find that $\epsilon^\text{max}_\Gamma$ is significantly larger than $\epsilon^{(1)}_{{K}}$ indicating that the flat bands have higher energies than the dispersive bands. This is in agreement with our explicit band structure calculations, see Fig~\ref{fig6:homobilayers}. For the \bilayer{3} and \bilayer{4} homobilayers, $\epsilon^\text{max}_\Gamma$ is only slightly larger than $\epsilon^{(1)}_{{K}}$ and we expect that flat and dispersive bands have similar energies. Again, this is consistent with our explicit calculations which show that the ordering of flat and $K$/$K'$-derived bands can depend on the twist angle for these systems, see Fig.~\ref{fig6:homobilayers}. Considering the heterobilayers, we find that $\Gamma$-states are predicted to lie above the $K$/$K'$-derived states in \bilayer{6}, \bilayer{7} and \bilayer{8}, while the opposite ordering is predicted for \bilayer{5}, \bilayer{9} and \bilayer{10}. These predictions are again in agreement with our explicit band structure calculations.

\section{Conclusions}

We have studied the electronic band structures of all twisted transition metal dichalcogenide (TMD) bilayers that can be obtained by combining \monolayer{1}, \monolayer{2}, \monolayer{3} and \monolayer{4} monolayers. Specifically, we have carried out tight-binding calculations taking into account the effect of atomic relaxations and also spin-orbit coupling. In all twisted homobilayers, the top valence bands are derived from monolayer states at $\Gamma$ and become flat when the twist angle decreases. For twisted heterobilayers, we find two scenarios: either the highest valence band derives from $\Gamma$-states of the monolayer and becomes flat upon twisting or it derives from $K$- and $K'$-states of the monolayer and remains dispersive even at small twist angles. Interestingly, the ordering of flat and dispersive bands depends sensitively on the atomic structure of the bilayer and can be changed by applying pressure. Our findings reveal that the chemical complexity of the twisted TMD bilayers can be harnessed to design flat band properties and pave the way to understanding electron-electron interaction effects in these materials.

\begin{acknowledgments}

We wish to thank Arta Safari for useful discussion on the construction of heterobilayers moir\'e cells. All authors acknowledge funding from the EPSRC grant E/P77380. Via our membership of the UK's HEC Materials Chemistry Consortium, which is funded by EPSRC (EP/L000202, EP/R029431, EP/T022213), this work used the the ARCHER UK National Supercomputing Service. This work also used the Cirrus UK National Tier-2 HPC Service at EPCC (http://www.cirrus.ac.uk) funded by the University of Edinburgh and EPSRC (EP/P020267/1). Finally, we acknowledge the Imperial College London Research Computing Service (DOI:10.14469/hpc/2232) for the computational resources used in carrying out this work.

\end{acknowledgments}

\appendix
\section{Improvements on the Tight-binding Model}
\label{sec:appendix-tbimprovement}
\begin{figure}[t!]
    \centering
    \subcaptionbox{}{
        \includegraphics[width=0.22\textwidth,trim={10px 10px 0px 0px},clip]{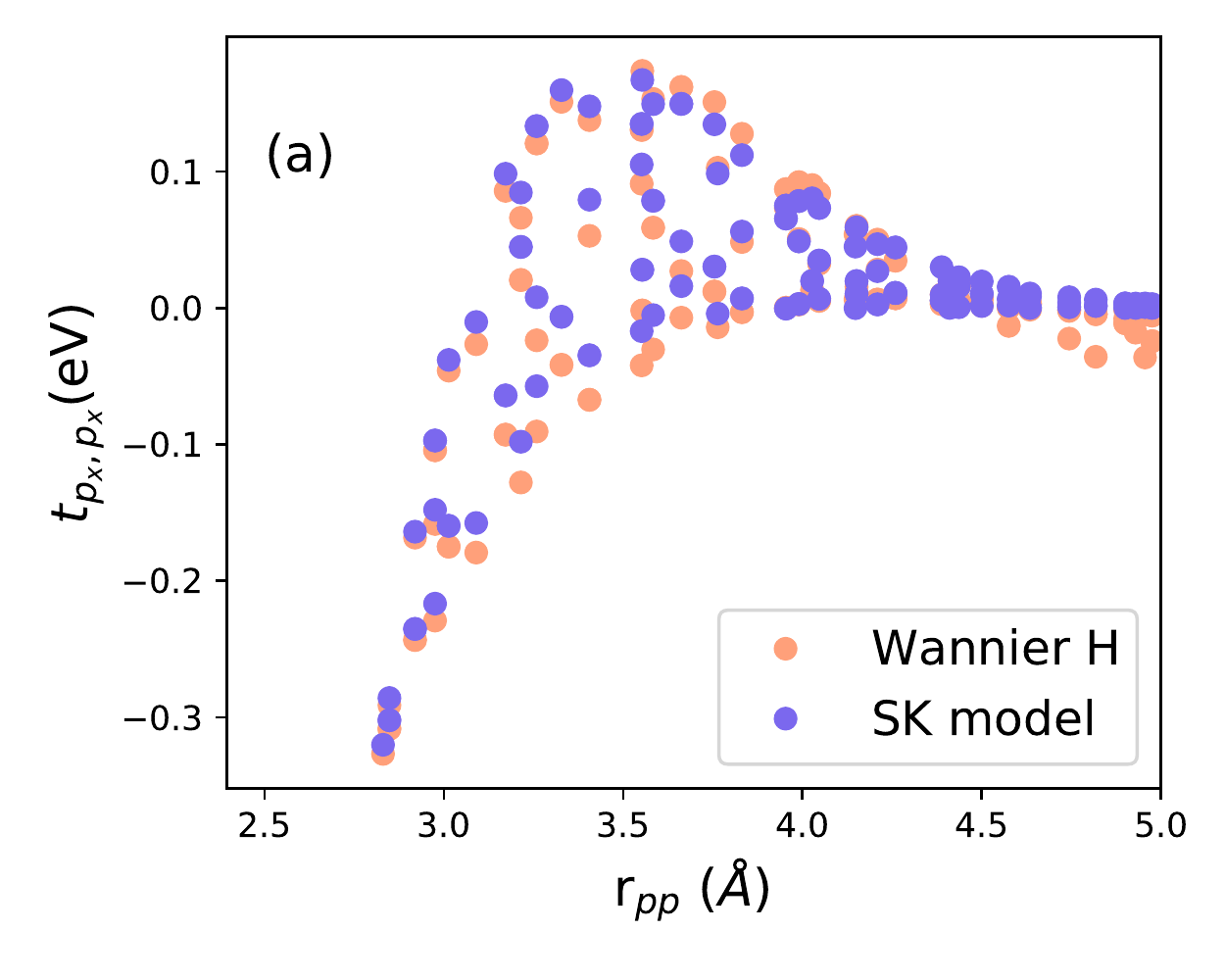}}
    \subcaptionbox{}{
        \includegraphics[width=0.22\textwidth,trim={10px 10px 0px 0px},clip]{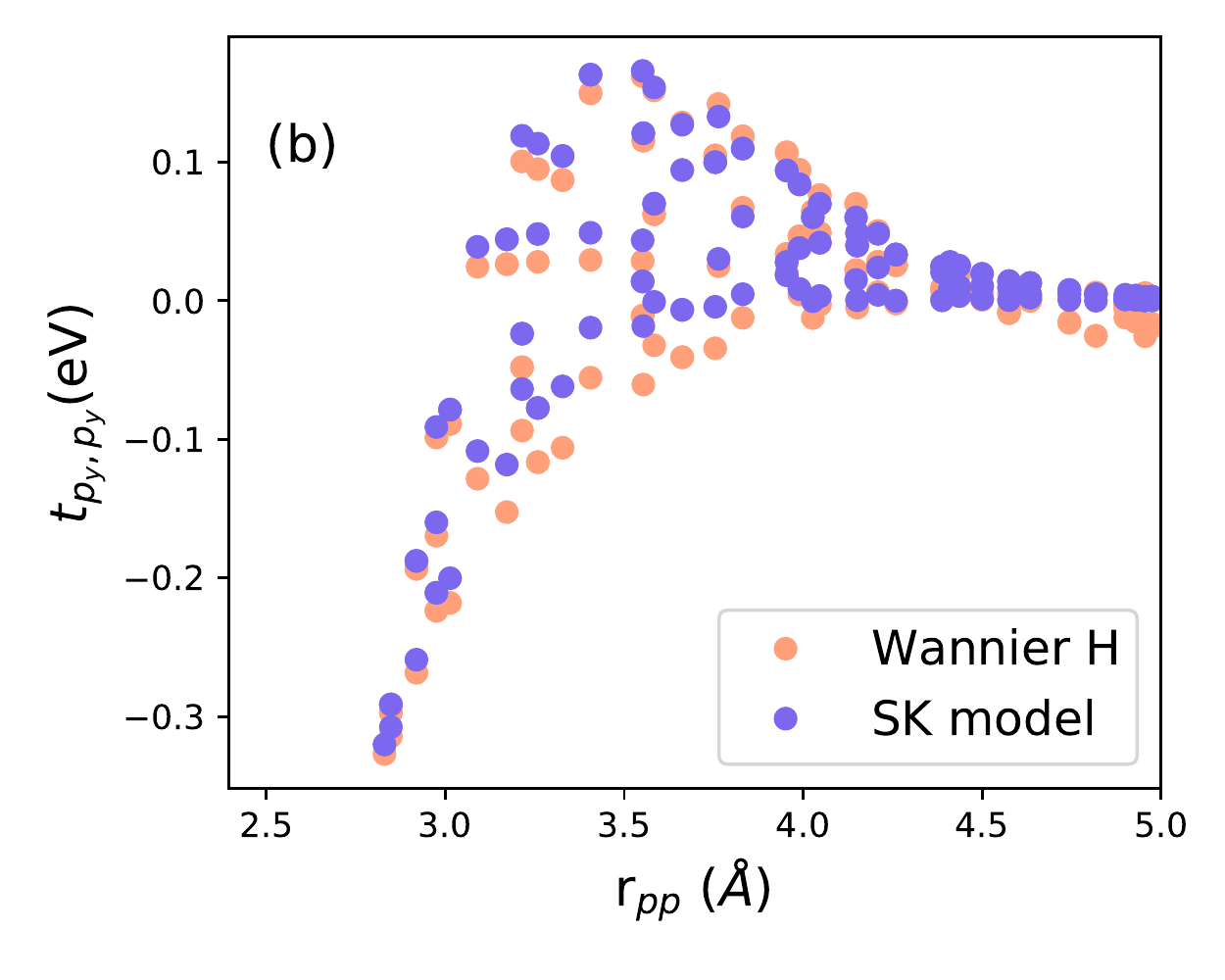}}\\
    \subcaptionbox{}{
        \includegraphics[width=0.22\textwidth,trim={10px 10px 0px 0px},clip]{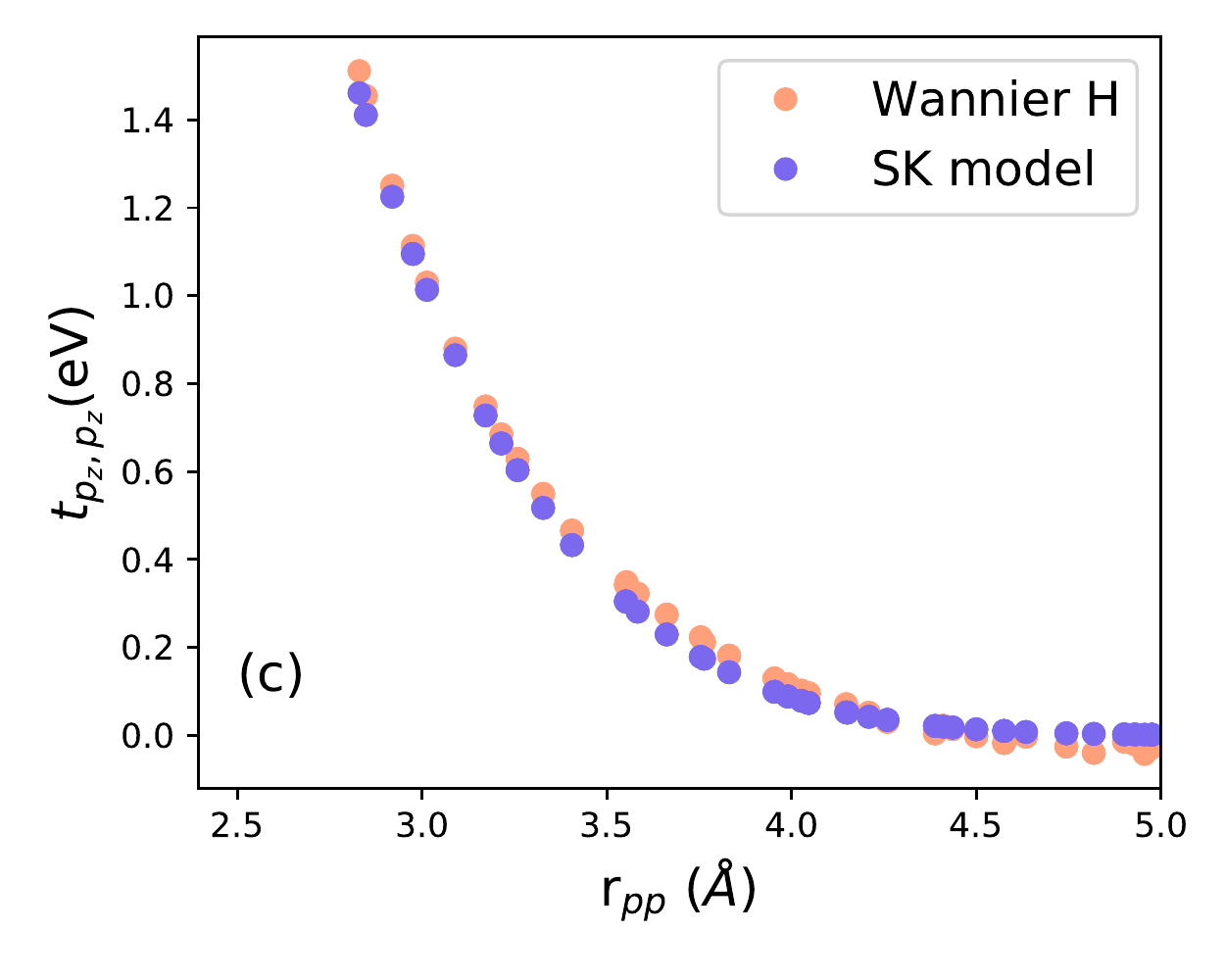}}
    \subcaptionbox{}{
        \includegraphics[width=0.22\textwidth,trim={10px 10px 0px 0px},clip]{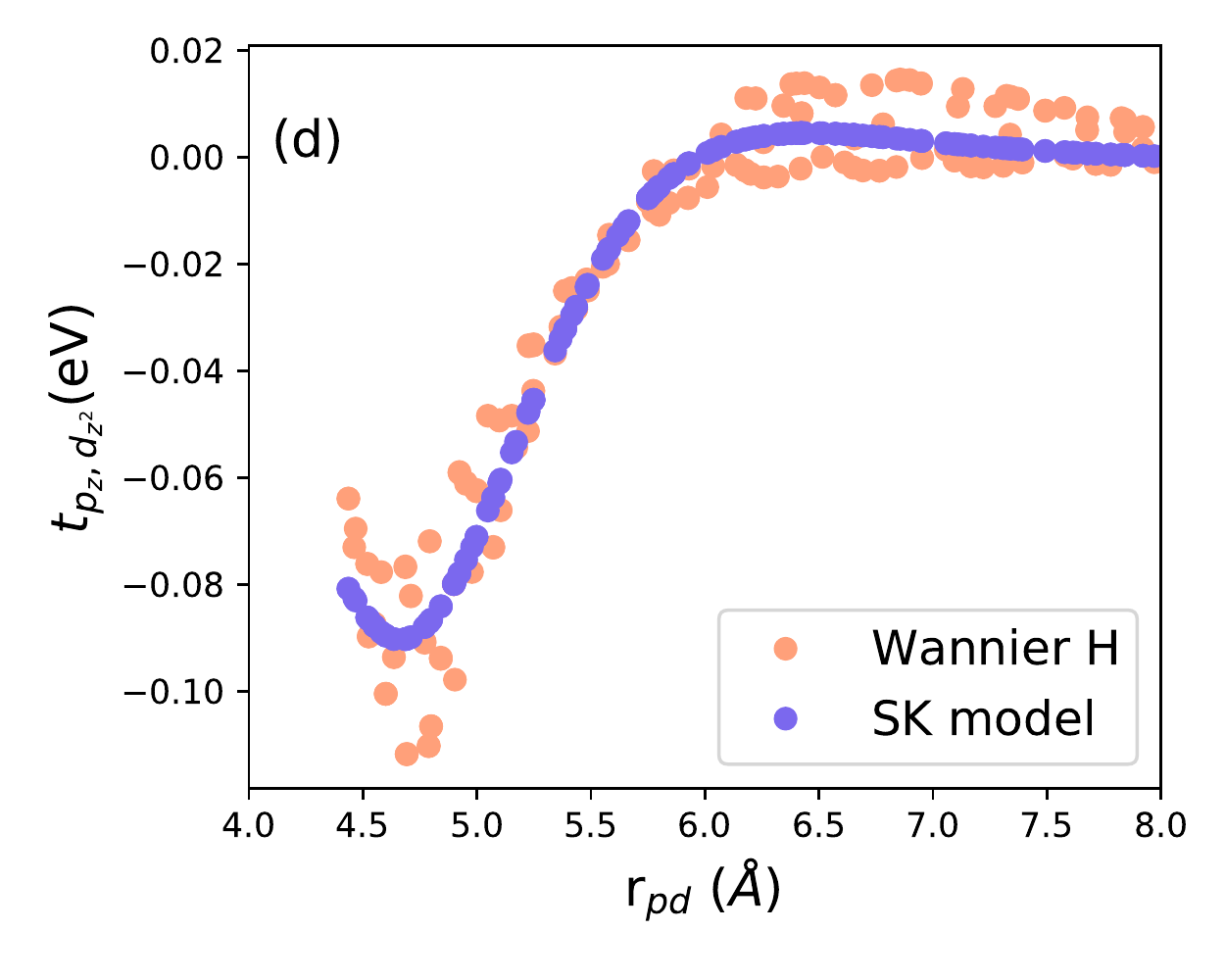}}
    \caption{Comparison of interlayer (a) p$_x$-p$_x$, (b) p$_y$-p$_y$, (c) p$_z$-p$_z$ and (d) p$_z$-$d_{z^2}$ hopping parameters obtained from a Wannierization of the DFT Hamiltonian to Slater-Koster (SK) model for untwisted MoS$_2$ bilayers.}.
    \label{fig:pz-dz2}
\end{figure}
In this Appendix, we describe the modifications that were required to generalize the tight-binding model for untwisted TMD bilayers developed by Fang and coworkers~\cite{Kaxiras_11bands} to twisted TMD bilayers. In particular, we improved the description of interlayer hoppings and parametrized these hoppings for all combinations of homo- and heterobilayers composed of \bilayer{1}, \bilayer{2}, \bilayer{3} and \bilayer{4} monolayers.
\begin{figure*}[t!]
    \centering
    \subcaptionbox{}{
        \includegraphics[width=0.31\textwidth]{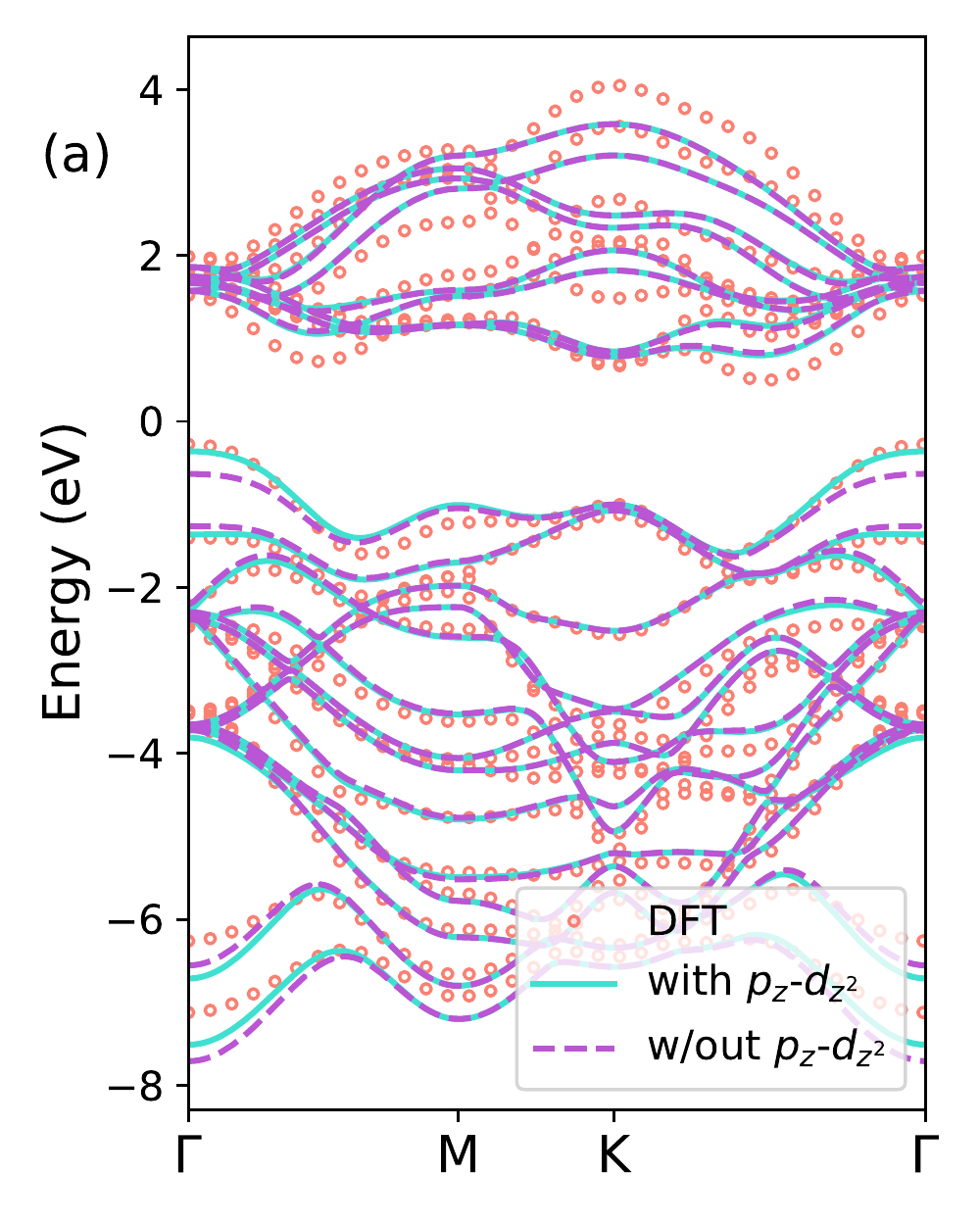}}
    \subcaptionbox{}{
        \includegraphics[width=0.31\textwidth]{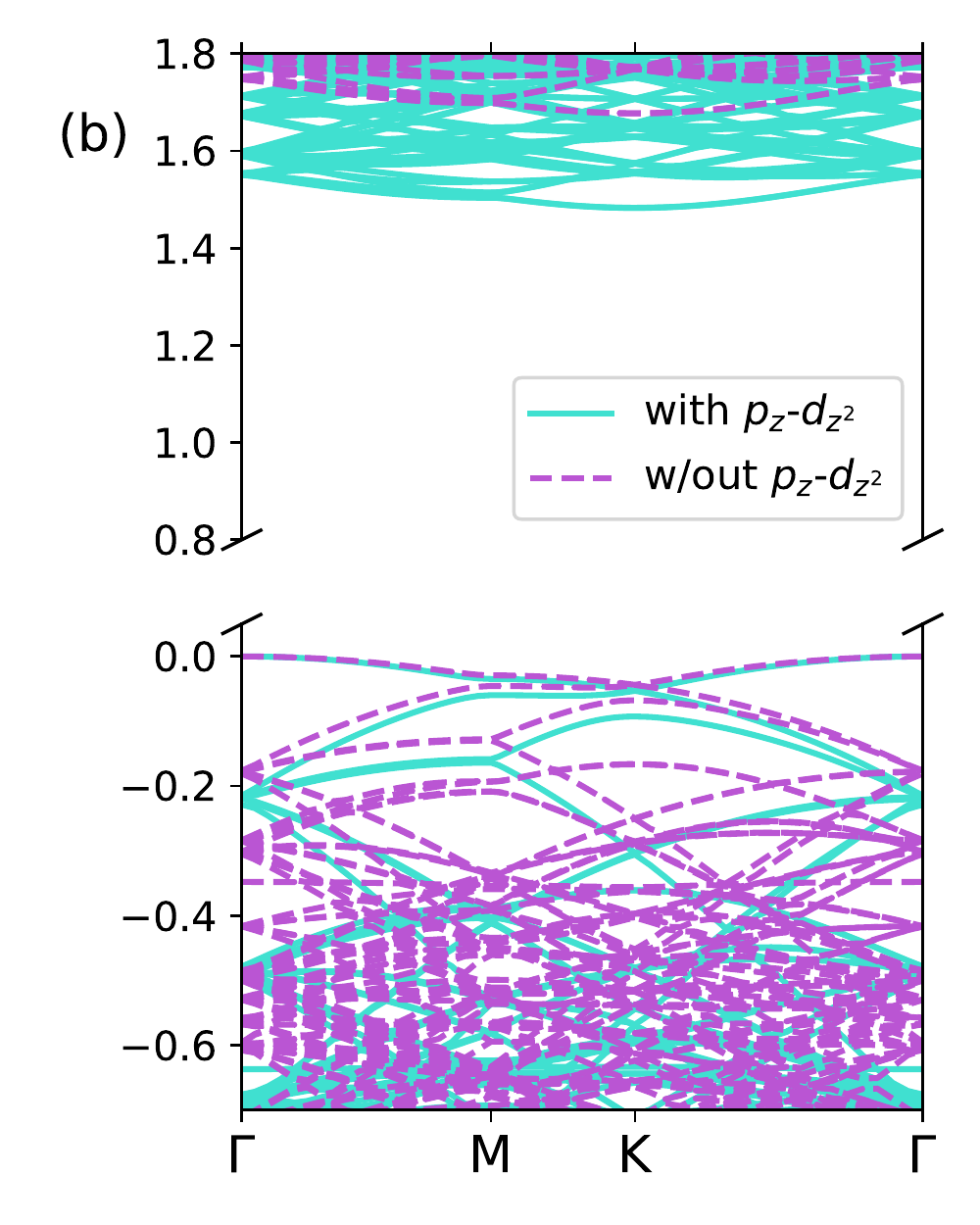}}
    \subcaptionbox{}{
        \includegraphics[width=0.31\textwidth]{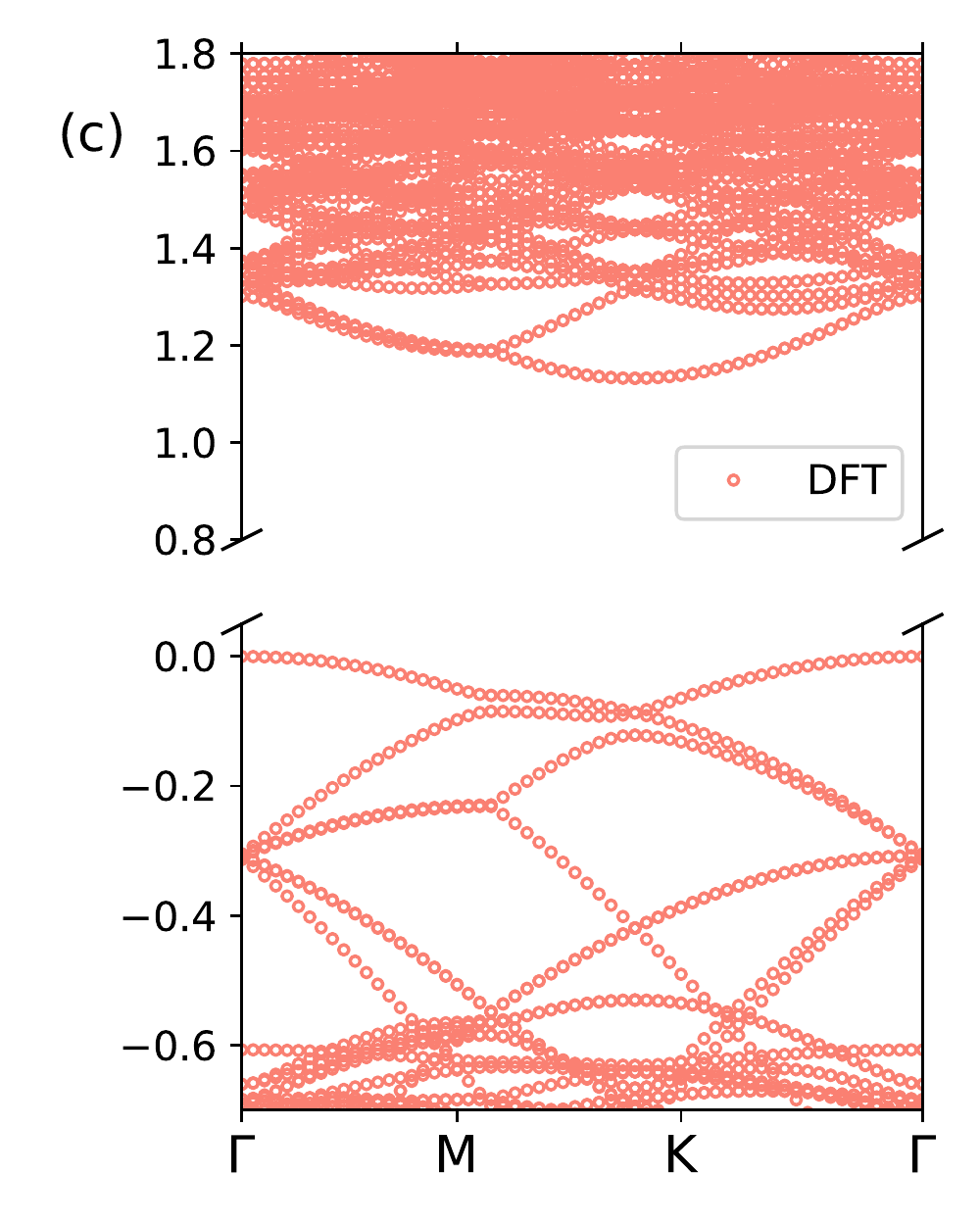}}
    \caption{Comparison of the tight-binding band structure with and without p$_z$-$d_{z^2}$ hopping to the first-principles DFT result for (a) untwisted AA-stacked bilayer MoS$_2$ and (b)  twisted bilayer MoS$_2$ at $\theta=7.3\si{\degree}$. (c) Corresponding DFT bandstructure. 
    }
    \label{fig:BS_with_pz-dz2}
\end{figure*}

To calculate the band structure of untwisted TMD bilayers, Fang and coworkers used Slater-Koster expressions for the interlayer hoppings between chalcogen p-orbitals. They also included a term for the interlayer hopping between chalcogen p$_z$-orbitals and transition metal $d_{z^2}$-orbitals, but did not describe this with a Slater-Koster expression. Instead, they only calculated the value of this hopping for the specific geometry of an untwisted 2H bilayer. 

To generalize the description of p$_z$ to $d_{z^2}$ hoppings to twisted bilayers, we used the Slater-Koster formula  
\begin{eqnarray}
t_{p_z,d_{z^2}}(\textbf{r}) &=& n \left[ n^2-\frac{1}{2}(l^2+m^2)\right] V_{pd\sigma}(\textbf{r}) \\ 
 & & +\sqrt{3}n(l^2+m^2)V_{pd\pi}(\textbf{r}),
\end{eqnarray}

\noindent where the directional cosines are defined as $l=r_x/r$, $m=r_y/r$ and $n=r_z/r$. 

To determine the functions $V_{pd\sigma}(\mathbf{r})$ and $V_{pd\pi}(\mathbf{r})$, we calculated $t_{p_z,d_{z^2}}$ and also $t_{p_z,d_{xz}}$ and $t_{p_z,d_{yz}}$ for a set of untwisted bilayers with different stacking configurations and different interlayer separations using a Wannier transformation of the DFT Hamiltonian. Next, a least square fitting process was used to extract $V_{pd\sigma}$ and $V_{pd\pi}$ at different interatomic distances and the results were fitted to functions of the type
\begin{equation}
    V_{pd}(\textbf{r})=V \left(\frac{r}{d}\right)^{\alpha} \cos \left(\beta \frac{r}{d} + \gamma\right)
\end{equation}
with $V,\alpha,\beta$ and $\gamma$ denoting fitting parameters, and $d=3.5$~\AA~ is an average interlayer distance. Fig.~\ref{fig:pz-dz2} (d) demonstrates that this yields an accurate description of the calculated $t_{p_z,d_{z^2}}$ hopping parameters. We have also tested the influence of other hoppings between chalcogen p-orbitals and transition metal d-orbitals, but found that the most important contribution arises from p$_z$ to d$_{z^2}$ hoppings.

The accuracy of the Slater-Koster approximation, in describing the orientation dependence of the interlayer hopping integrals for p-p and p-d orbitals, is demonstrated in Fig.~\ref{fig:pz-dz2} for hoppings extracted from displaced untwisted bilayers. The main error for p-p hoppings (see Fig.~\ref{fig:pz-dz2}a-c) arises from approximating the orthogonal Wannier basis as non-orthogonal atomic-like orbitals to make use of the Slater-Koster rules. This discrepancy, however, is very small, i.e. less than 20 meV on average for interlayer hopping matrix elements which suggests that it is an appropriate model to describe various configurations seen in twisted bilayers.

\begin{figure}[t!]
    \centering
    \includegraphics[width=0.33\textwidth]{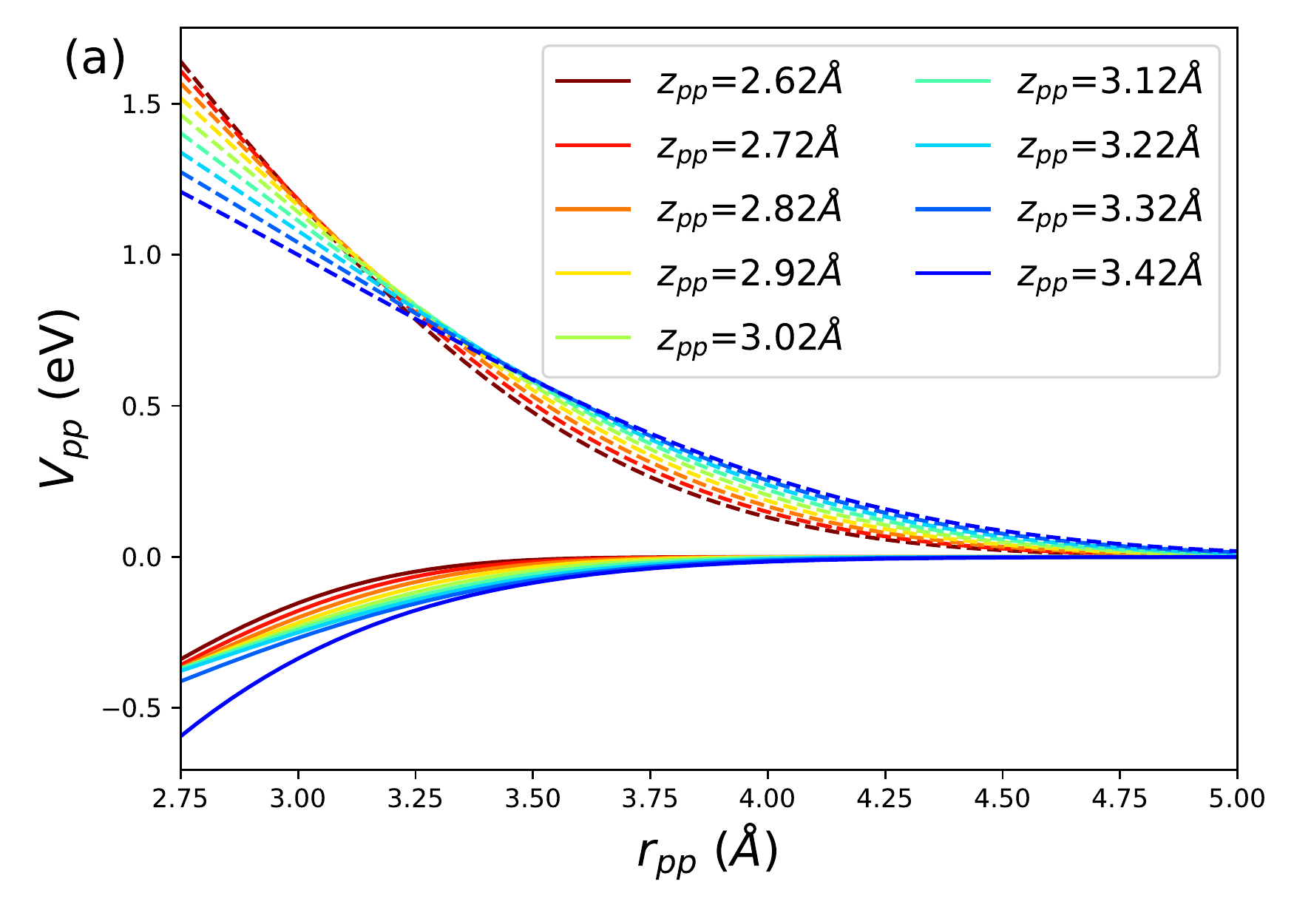}
    \includegraphics[width=0.33\textwidth]{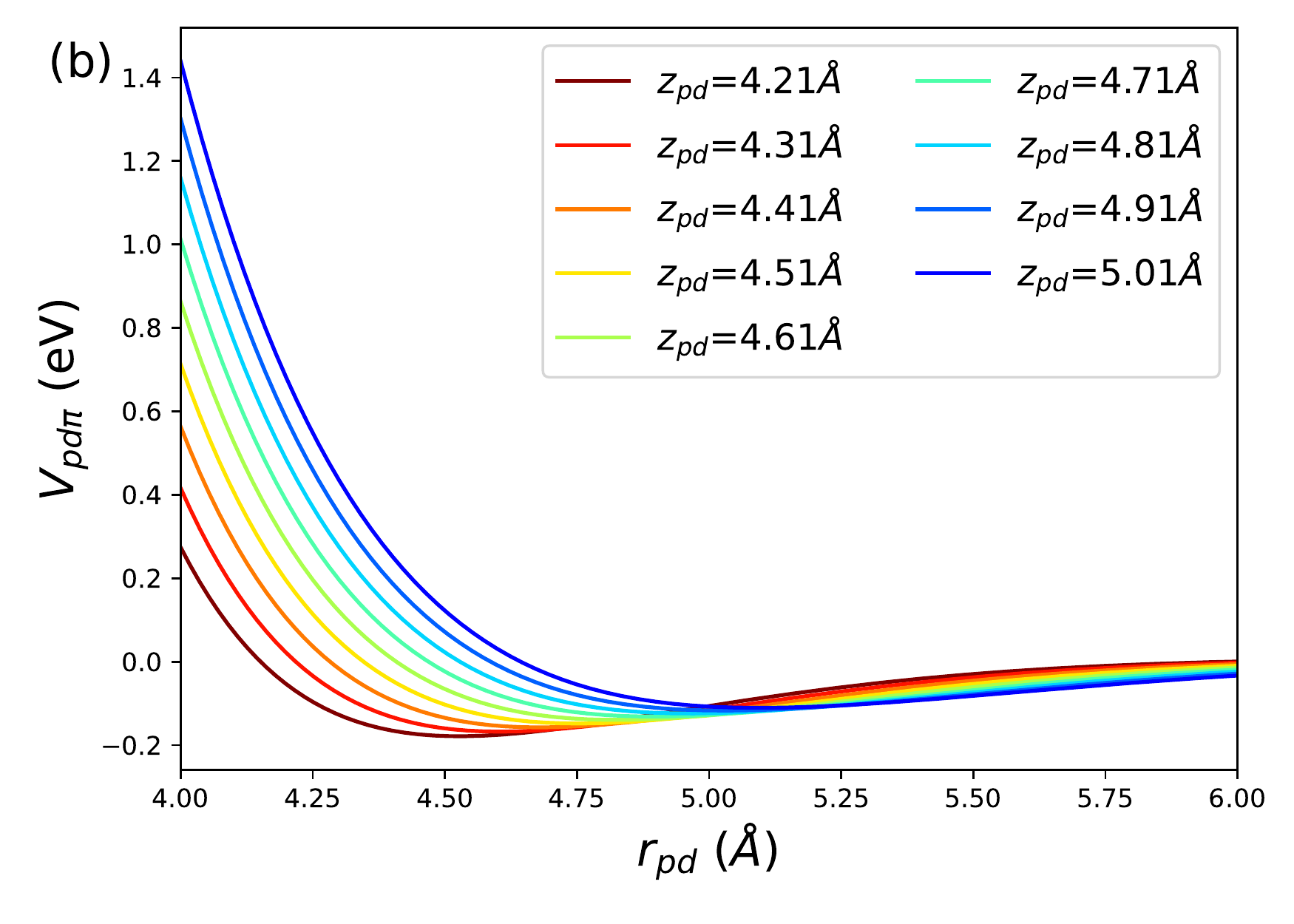}
    \caption{Distance dependence of the Slater-Koster parameters: (a) p-p interlayer hopping with dashed lines denoting $V_{pp\sigma}$ and solid lines denoting $V_{pp\pi}$ as function of interlayer separation and (b) p$_z$-d$_{z^2}$ interlayer hopping parameter $V_{pd\pi}$ as function of interlayer separation. 
    }
    \label{fig:interlayer-sepdependence}
\end{figure}

Figure~\ref{fig:BS_with_pz-dz2} compares the band structures of untwisted and twisted bilayer MoS$_2$ at a twist angle of $7.3$\si{\degree} from tight-binding with and without the p$_z$ to d$_{z^2}$ hopping to a first-principles density-functional theory result. We find that inclusion of p$_z$ to d$_{z^2}$ hoppings improves the agreement with the first principles result significantly. In particular, the valence band states near $\Gamma$ are pushed to higher energies which reduces the band gap by approximately 200 meV. A similar shift of the highest valence bands is found also in the twisted bilayers.

Aside from the inclusion of interlayer p$_z$ to d$_{z^2}$ hoppings, we also discovered that the description of interlayer hoppings between chalcogen p-orbitals developed by Fang and coworkers required improvements to obtain an accurate description of twisted bilayers. To parametrize the corresponding Slater-Koster expressions, Fang et al. carried out first-principles calculations of untwisted bilayers with different stacking configurations, but using a fixed interlayer separation. As we have demonstrated in the main section of the manuscript, the introduction of a twist results in significant atomic relaxations and concomitant variations in the interlayer separation. We have found that these changes in the interlayer separation are not well captured by the simple Slater-Koster expressions used by Fang and coworkers. Fig.~\ref{fig:interlayer-sepdependence}(a) shows the Slater-Koster parameters $V_{pp\sigma}$ and $V_{pp\pi}$ as function of the interatomic distance for different interlayer separations: for small interatomic distances, the Slater-Koster parameters depend sensitively on the interlayer separation.

To account for the dependence of the Slater-Koster parameters on the interlayer separation, the following procedure is used: for a given pair of atoms, we first calculate the interlayer separation as the difference of their z-coordinates as well as the interatomic distance. Then, the Slater-Koster parameters for the specific interlayer separation are used to obtain the desired hopping matrix element.

The DFT calculations for monolayers and untwisted bilayers were performed with Quantum Espresso\cite{QE} within the optB88 generalized gradient approximation for the exchange-correlation potential and a plane-wave cutoff value of 70 Ry ($\approx$950 eV). Monolayer calculations were performed with a $25\times25\times1$ Monkhorst-Pack k-point grid whereas a $12\times12\times1$ k-point grid was used for the untwisted bilayer calculations. The DFT Hamitonian was transformed into the basis of 11 (22) Wannier functions consisting of atomic-like p and d orbitals with average spreads around $2.2$~\AA~ for the monolayer (bilayer) calculations. For twisted bilayers DFT calculations have been carried out with \textsc{Onetep}\cite{ONETEP,PhysRevB.98.125123}, a linear-scaling DFT code. We use the Perdew-Burke-Ernzerhof exchange-correlation functional~\cite{PBE_PRL77} with projector-augmented-wave pseudopotentials~\cite{PAW_PRB50,JOLLET20141246}, generated from ultra-soft pseudopotentials\cite{GARRITY2014446}, and a kinetic energy cutoff of 800~eV. A basis consisting of 9 non-orthogonal generalized Wannier functions (NGWFs) for calcogen atoms and 13 NGWFs for metal atoms is employed. The NGWFs' radii are set to 9.0~a$_0$.

\bibliographystyle{unsrt}
\bibliography{biblio}

\end{document}


\thispagestyle{plain}
\begin{center}
\textbf{\Large{Supplementary information for  ``Chemical trends in flat band properties of twisted transition metal dichalcogenide homo- and heterobilayers''}}

\vspace{1cm}
Valerio Vitale$^1$, Kemal Atalar$^1$, Arash Mostofi$^1$ and Johannes Lischner$^1$ \\
\vspace{0.5cm}
$^1$ Departments of Materials and Physics and the Thomas Young Centre for Theory and Simulation of Materials, Imperial College London, London SW7 2AZ, UK\\
\date{}

\end{center}

\renewcommand{\thefigure}{S\arabic{figure}}
\renewcommand{\thetable}{S\arabic{table}}

\section*{S1 - Flat vs relaxed band structures}
\begin{figure}[t]
    \centering
    \includegraphics[width=0.9\columnwidth]{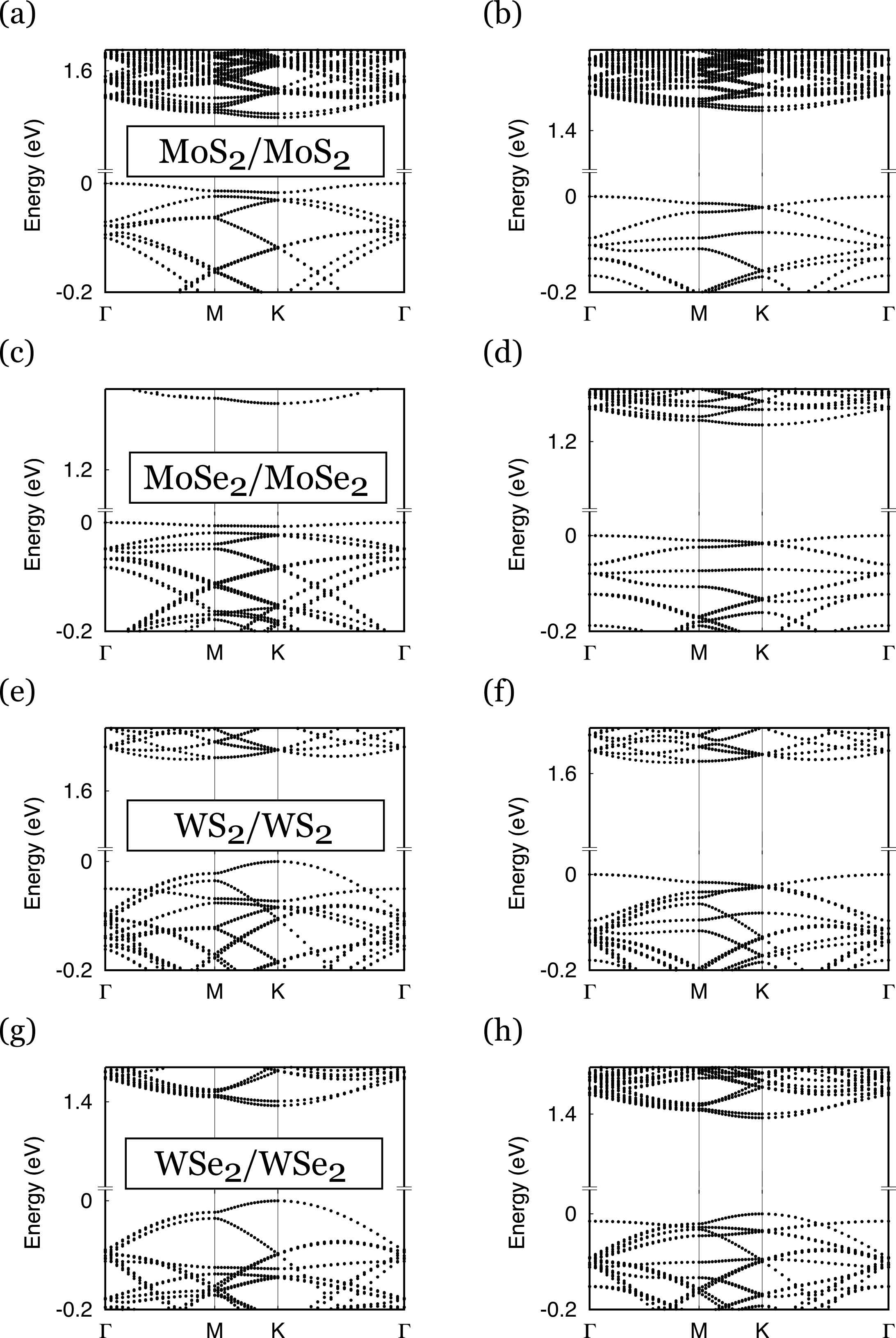}
    \caption{Comparison of flat (left column) vs relaxed (right column) band structures of (a)-(b) \protect\bilayer{1}, (c)-(d) \protect\bilayer{2}, (e)-(f)\protect\bilayer{3} and (g)-(h) \protect\bilayer{4} at $\theta=5.1^\circ$. For the flat systems only the interlayer separation has been relaxed.}
    \label{fig:all_flat_vs_relaxed_hbls}
\end{figure}

Fig.~\ref{fig:all_flat_vs_relaxed_hbls} shows the band structures of unrelaxed (flat) and relaxed homobilayers for $\theta=$\homoangle{5}. The main effect of relaxation is to shift up in energy the flat bands (compared to the $K$/$K$'-derived bands) and to close the gap at the $K$-point, giving rise to Dirac cones. Moreover, in \bilayer{3} and \bilayer{4} flat and $K$/$K$'-derived bands intersect. 

\begin{figure}[t]
    \centering
    \includegraphics[width=0.9\columnwidth]{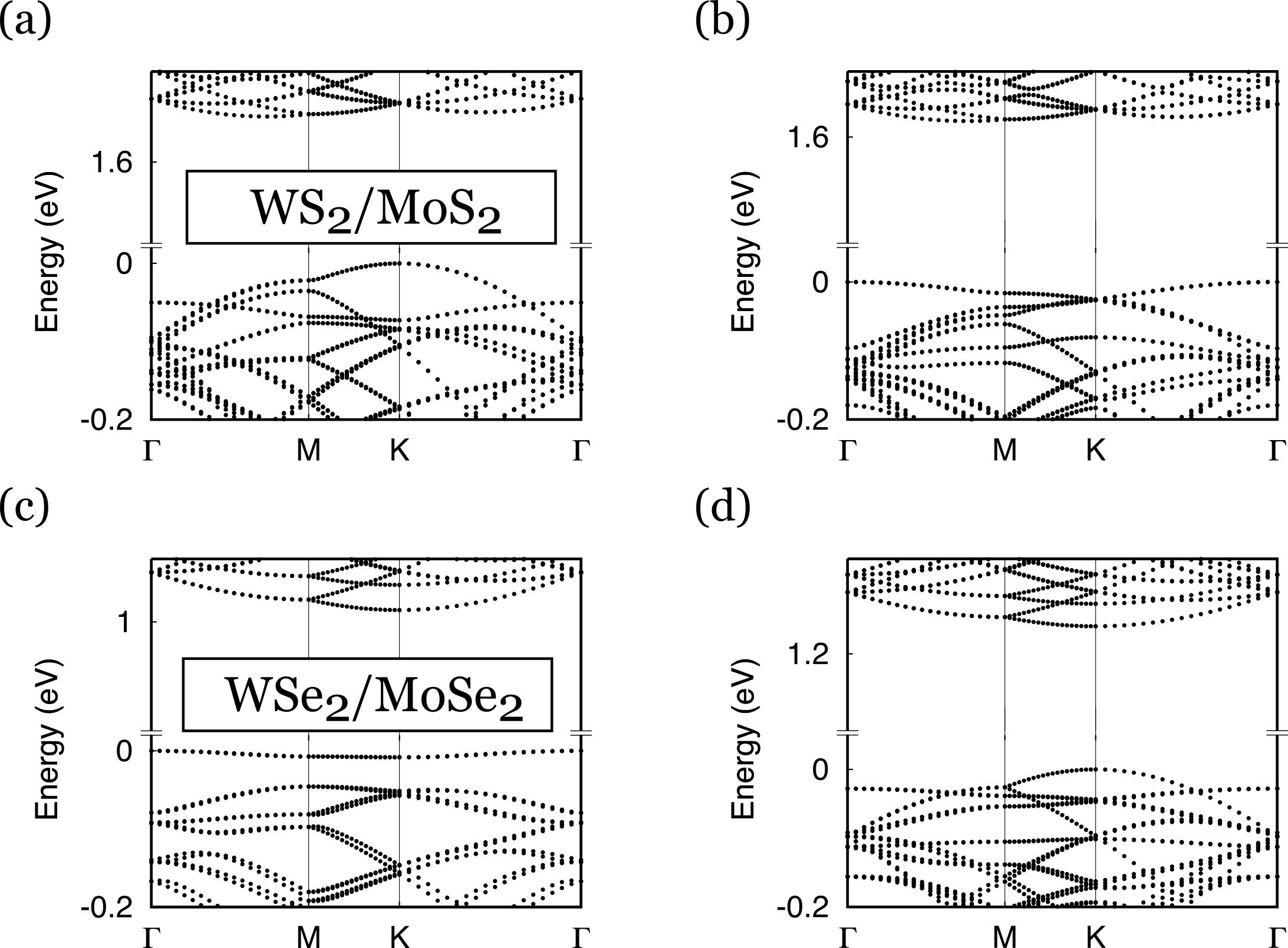}
    \caption{Comparison of flat (left column) vs relaxed (right column) band structures of (a)-(b) \protect\bilayer{5}, (c)-(d) \protect\bilayer{6}, at $\theta=5.1^\circ$. For the flat systems only the interlayer separation has been relaxed.}
    \label{fig:flat_vs_rel_hetero1}
\end{figure}

Similarly, unrelaxed (flat) and relaxed band structures for heterobilayers with same chalcogen species for $\theta=$\homoangle{5} are shown in Fig.~\ref{fig:flat_vs_rel_hetero1}. The ordering of $K$/$K$'-derived vs flat bands is reversed by relaxation. In particular, in unrelaxed \bilayer{5}, $K$/$K$'-derived bands and flat bands intersect and the valence band maximum (VBM) is at the $K$/$K$'-point, wherease in the relaxed structure the VBM is at the $\Gamma$-point. In \bilayer{6} the situation is reversed. In both cases, the gap between the flat bands at the $K$-point vanishes.

\begin{figure}[t]
    \centering
    \includegraphics[width=0.9\columnwidth]{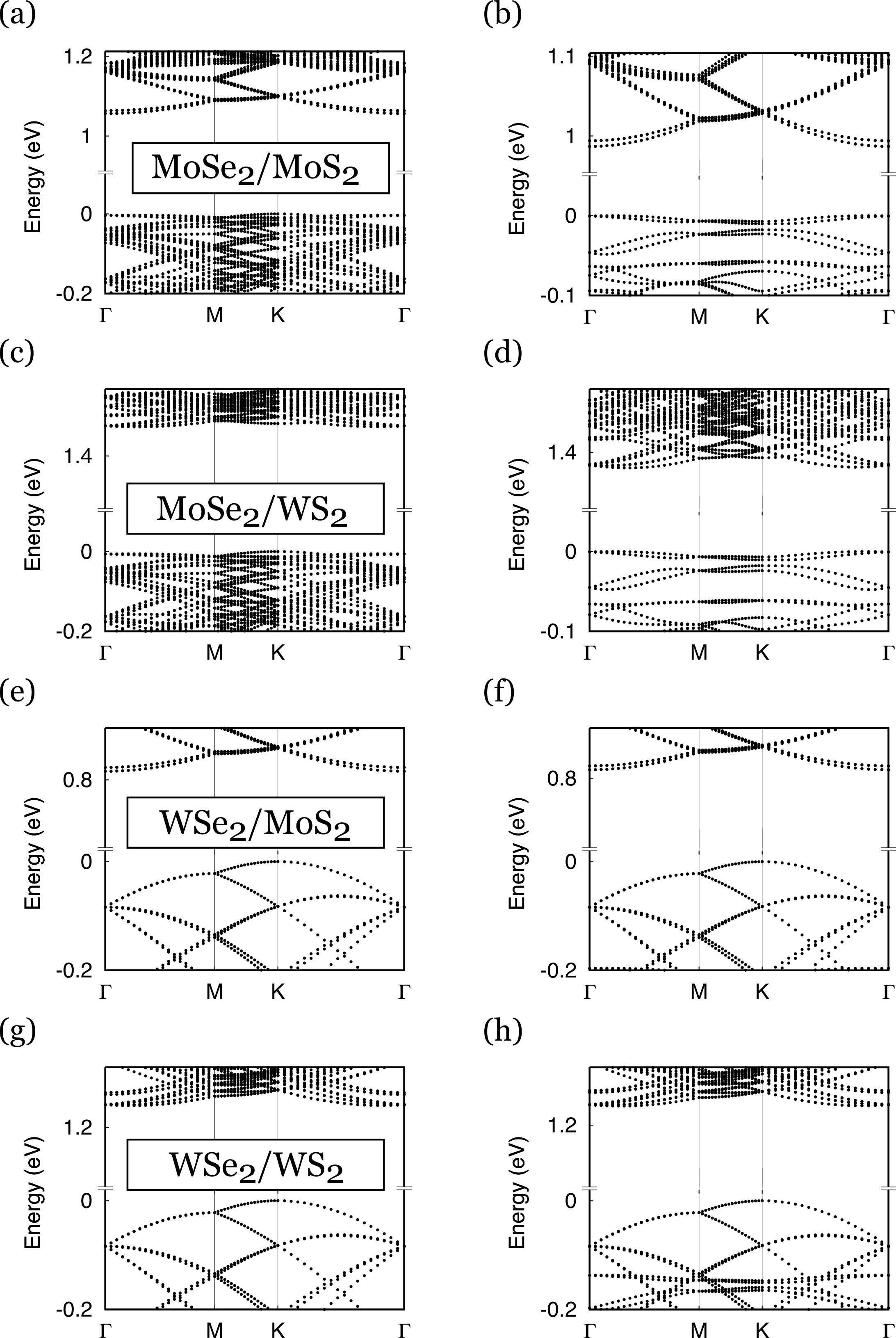}
    \caption{Comparison of flat (left column) vs relaxed (right column) band structures of (a)-(b) \protect\bilayer{7}, (c)-(d) \protect\bilayer{8}, (e)-(f) \protect\bilayer{9} and (g)-(h) \protect\bilayer{10} at $\theta=4.5^\circ$. For the flat systems only the interlayer separation has been relaxed.}
    \label{fig:flat_vs_rel_hetero2}
\end{figure}

Fig.~\ref{fig:flat_vs_rel_hetero2} shows the band structures for unrelaxed and relaxed heterobilayers with different chalcogen species for $\theta=$\heteroangle{1}. In \bilayer{8} and \bilayer{9} the effect of relaxation is to shift up in energy the flat bands and make them separated from all remote bands. For these systems, the gap at the $K$ point remains finite and each band is not doubly degenerate.
In \bilayer{9} and \bilayer{10} bilayers the $K$/$K$'-derived bands are always on top. Relaxation has a minor effect for these systems, as $K$/$K$'-derived bands and flat bands are separated by several hundreds meV.

\newpage

\section*{S2 - DFT vs TB band structures}
\begin{figure}[h!]
    \centering
    \includegraphics[width=0.9\columnwidth]{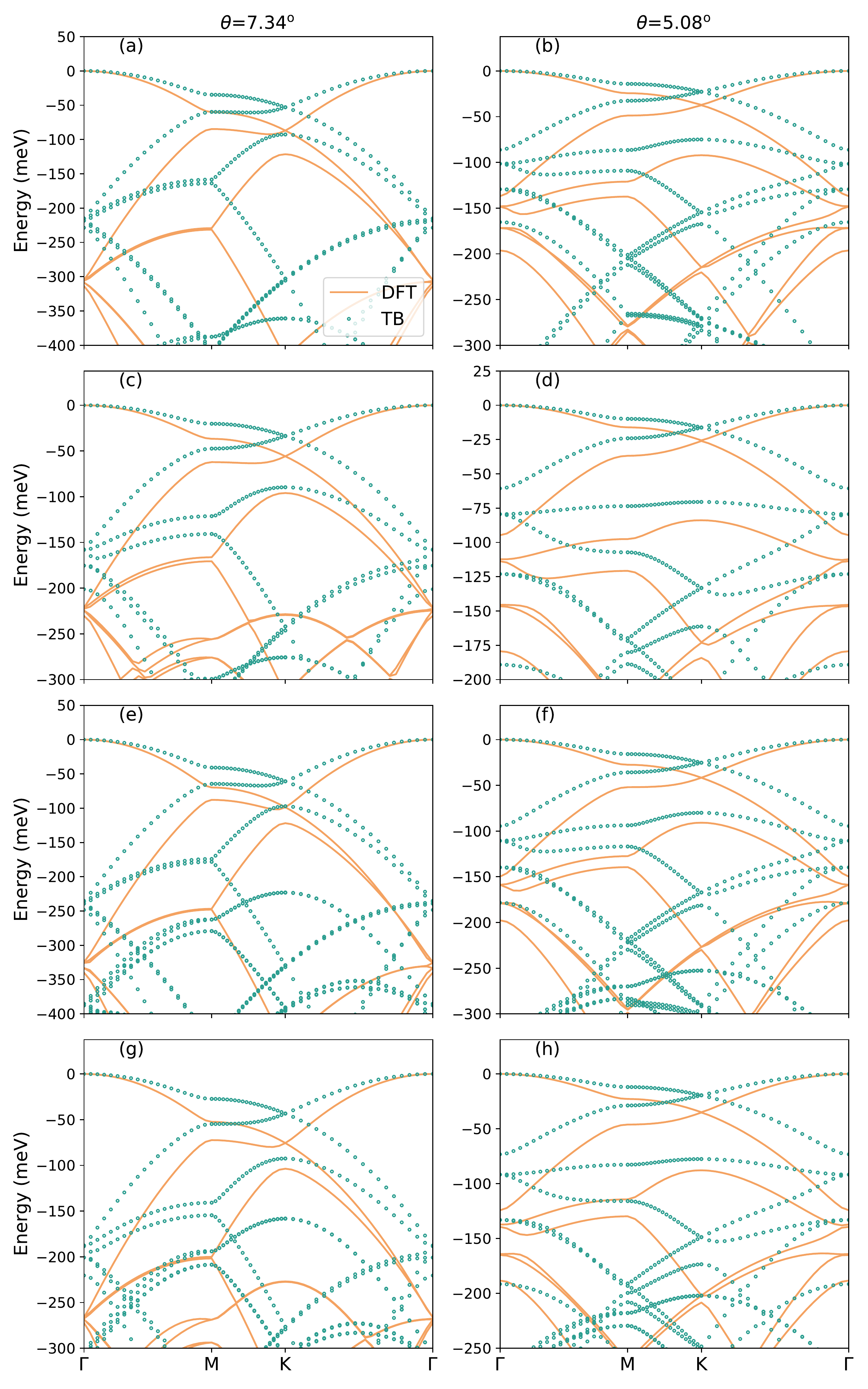}
    \caption{Comparison of relaxed DFT (solid orange) vs TB (green empty dots) bandstructures \textit{without SOC} at two different twist angles, $\theta=7.3$\si{\degree} (left column) and $\theta=5.1$\si{\degree} (right column) for a)-b) \protect\bilayer{1}, c)-d) \protect\bilayer{2}, e)-f) \protect\bilayer{3} and g)-h) \protect\bilayer{4}}\label{fig:DFT_vs_TB}
\end{figure}

Fig.~\ref{fig:DFT_vs_TB} shows the comparison between DFT and tight-binding (TB) band structures without spin-orbit coupling (SOC) for relaxed homobilayers at two twist angles, $\theta=$\homoangle{6} and $\theta=$\homoangle{5}, respectively. DFT band structure have been computed with \textsc{Onetep}\cite{ONETEP,PhysRevB.98.125123}, a linear-scaling DFT code. We use the Perdew-Burke-Ernzerhof exchange-correlation functional~\cite{PBE_PRL77} with projector-augmented-wave pseudopotentials~\cite{PAW_PRB50,JOLLET20141246}, generated from ultra-soft pseudopotentials\cite{GARRITY2014446}, and a kinetic energy cutoff of 800~eV. A basis consisting of 9 non-orthogonal generalized Wannier functions (NGWFs) for chalcogen atoms and 13 NGWFs for metal atoms is employed. We set the NGWFs' radii to 9.0~a$_0$. Agreement between DFT and TB band structures improves at smaller angles.

\section*{S3 - Projections of top valence bands onto atomic orbitals}

\begin{figure}[h]
    \centering
    \includegraphics[width=0.9\columnwidth]{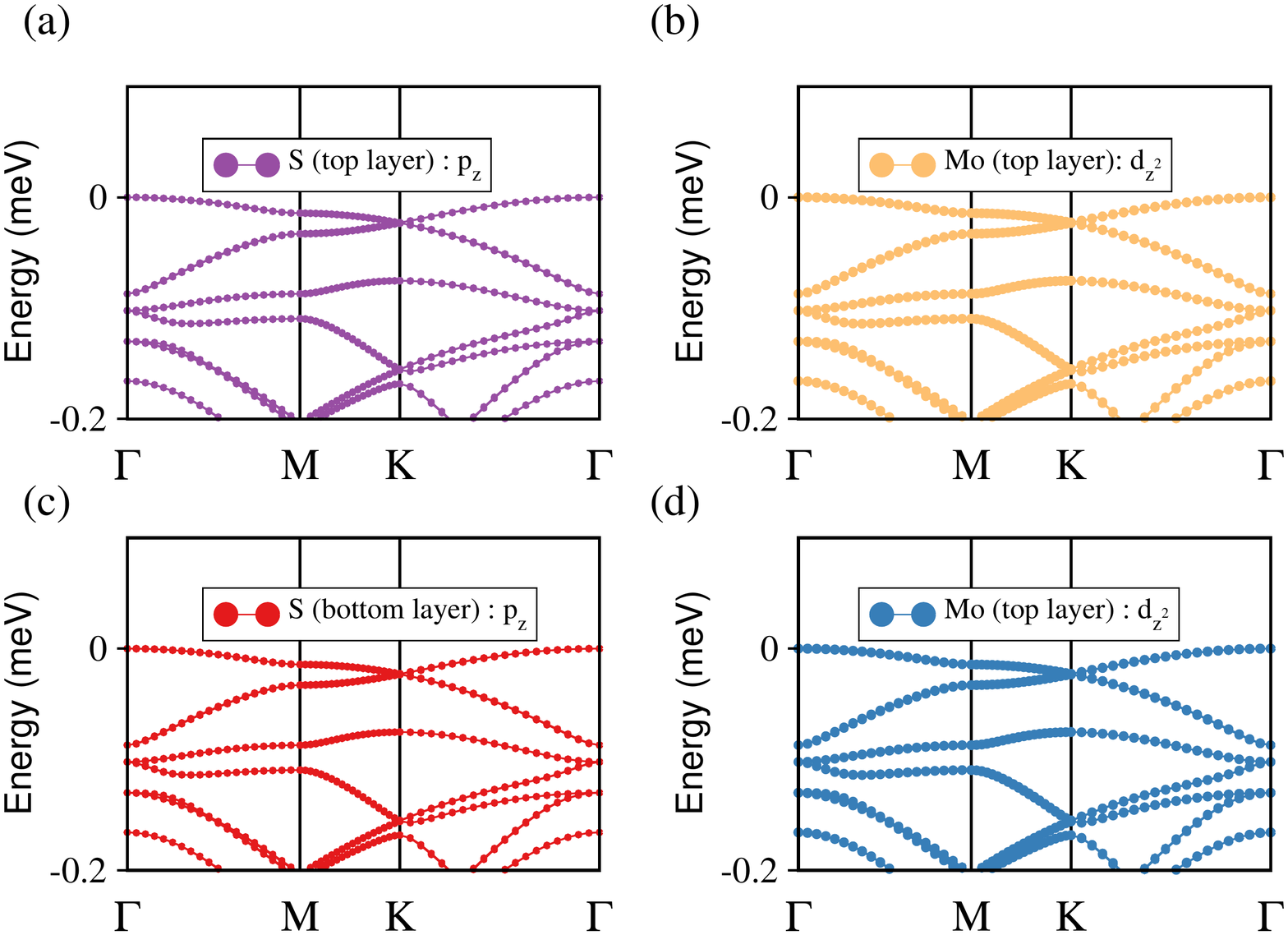}
    \caption{Projections of the highest valence states of twisted \protect\bilayer{1} for $\theta=$\protect\homoangle{5} onto (a) p$_z$-like orbitals of inner S atoms of top layer, (b) d$_{z^2}$-like orbitals of Mo atoms of top layer, (c) p$_z$-like orbitals of inner S atoms of bottom layer and (d) d$_{z^2}$-like orbitals of Mo atoms of bottom layer. In all panels, the size of circles is proportional to the magnitude of the projection.}
    \label{fig5:tBLWSe2}
\end{figure}

Fig.~\ref{fig5:tBLWSe2} shows the projections of the highest valence bands states onto Mo d$_z^2$-like orbitals and inner S p$_z$-like orbitals in \bilayer{1} for $\theta=$\homoangle{5}. Flat bands have large projections on these orbitals ($\sim 98\%$). These bands originate from $\Gamma$-states of the top valence band of \monolayer{1} monolayers, which also have large projections onto Mo d$_z^2$-like orbitals and S p$_z$-like orbitals. For homobilayers projections onto the two layers are symmetric.

\begin{figure}[h]
\centering
\includegraphics[width=0.9\columnwidth]{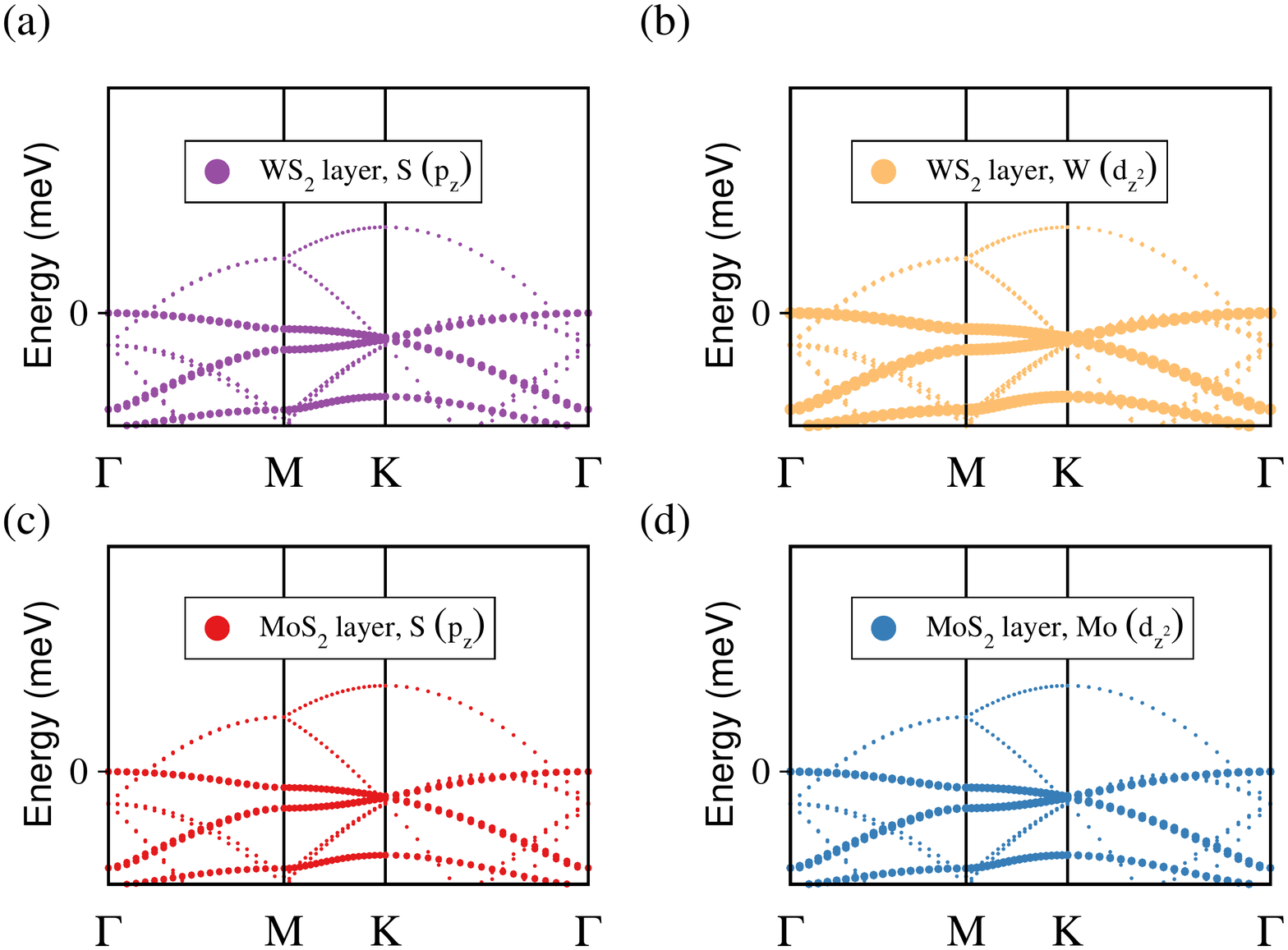}
\caption{Projections of the highest valence band states of \protect\bilayer{5} for $\theta=$\protect\homoangle{5} onto (a) p$_z$-like orbitals of inner S atoms of \protect\monolayer{3}; (b) d$_{z^2}$-like orbitals centred on W atoms of \protect\monolayer{3}; (c) p$_z$-like orbitals centred on inner S atoms of \protect\monolayer{1}; and (d) d$_{z^2}$-like orbitals centred on Mo atoms of \protect\monolayer{1}. In all panels, the size of circles is proportional to the magnitude of the projection.}\label{fig10:hetero_projection_on_orbitals}
\end{figure}

Fig.~\ref{fig10:hetero_projection_on_orbitals} shows the projections of the highest valence bands states onto Mo and W d$_z^2$-like orbitals and inner S p$_z$-like orbitals in \bilayer{5} for $\theta=$\homoangle{5}. As for homobilayers, flat bands bands have large projections on these orbitals and these bands originate from $\Gamma$-states of the corresponding monolayers. In contrast to homobilayers, projections on the two layers are not layer-symmetric, with projections onto \monolayer{3} larger than projections onto \monolayer{1}. 

\begin{figure}[h]
\centering
\includegraphics[width=0.9\columnwidth]{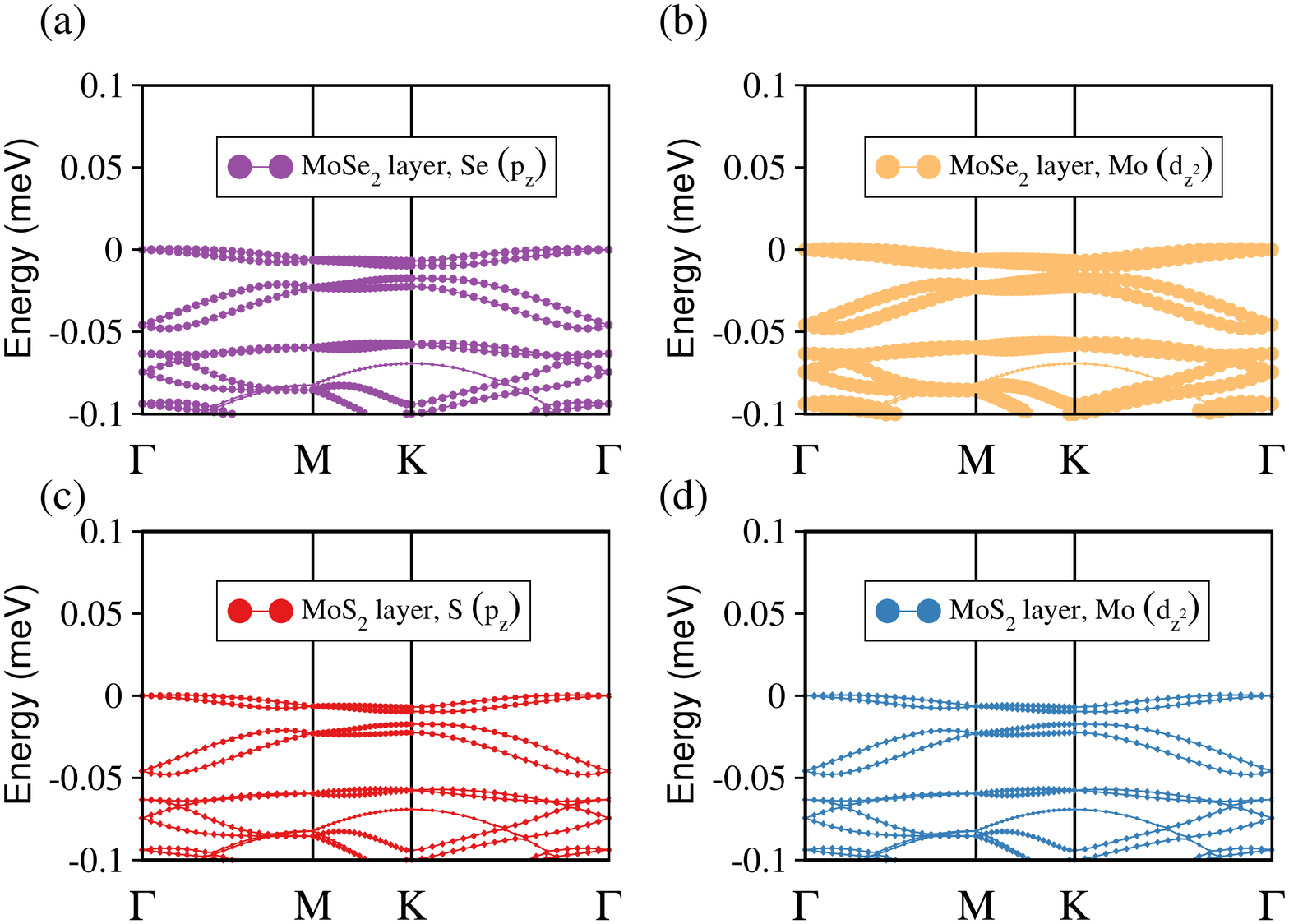}
\caption{Projections of the top valence states of \protect\bilayer{7} for $\theta=$\protect\heteroangle{1} onto (a) p$_z$ orbitals on inner Se atoms; (b) d$_{z^2}$ orbitals of Mo atoms in \protect\monolayer{2} layer; (c) p$_z$ orbitals on inner S atoms; and (d) d$_{z^2}$ orbitals on Mo atoms of \protect\monolayer{1}. In all panels, the size of the circles is proportional to the magnitude of the projection.}\label{fig13:hetero_projection_on_orbitals}
\end{figure}

Fig.~\ref{fig13:hetero_projection_on_orbitals} shows the projections of the highest valence bands states onto Mo d$_z^2$ orbitals and inner S and Se p$_z$ orbitals in \bilayer{7} at $\theta=$\heteroangle{1}. Similarly to other bilayers, flat bands bands have large projections on these orbitals and these bands originate from $\Gamma$-states of the corresponding monolayers. Projections on the two layers are not layer-symmetric, with projections onto d$_z^2$ orbitals centred on the Mo atoms of \monolayer{2} showing the largest contribution.

\begin{figure}[h]
\centering
\includegraphics[width=0.9\columnwidth]{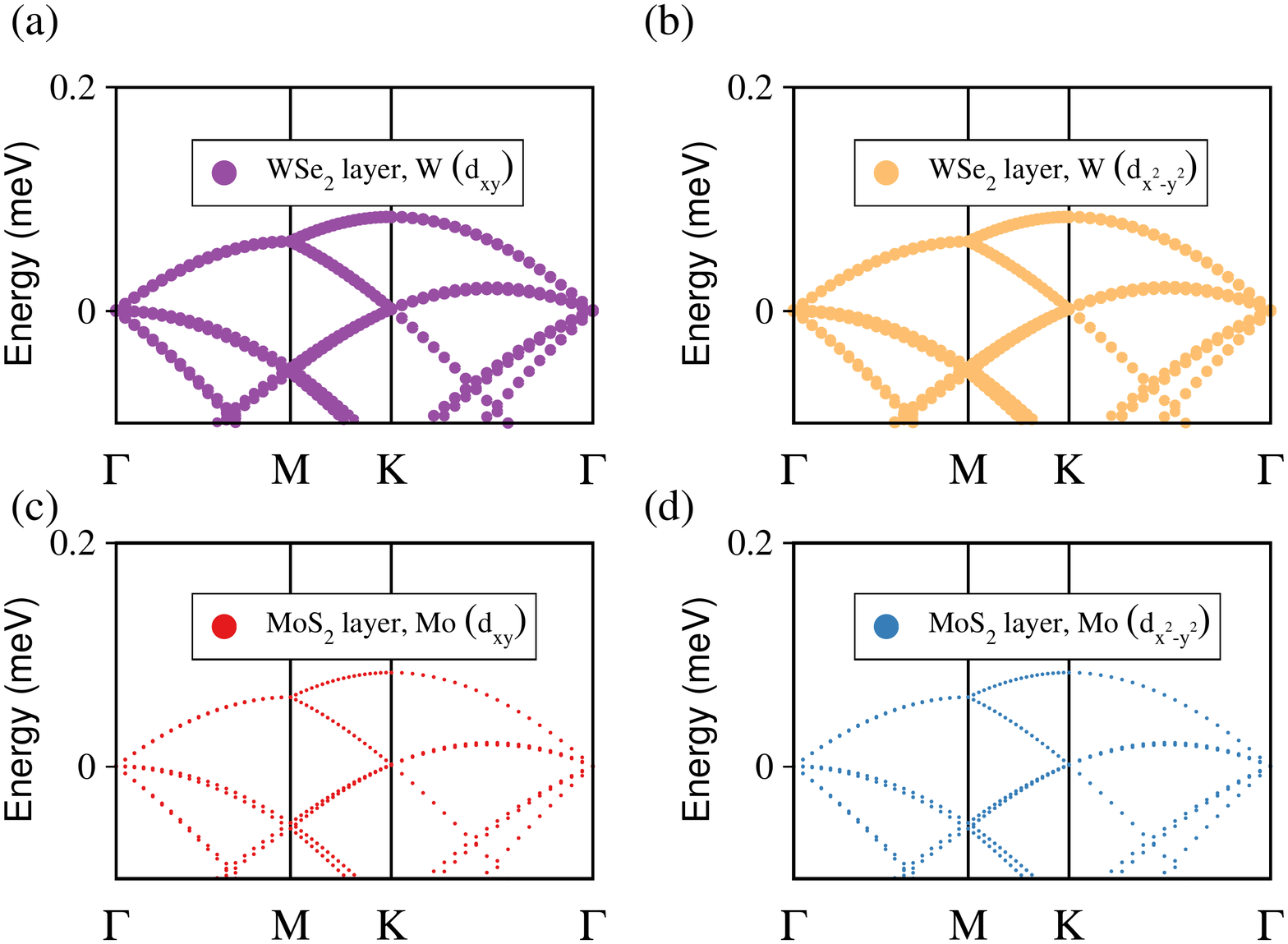}
\caption{Projections of the highest valence states of twisted \protect\bilayer{9} for $\theta=$\protect\heteroangle{1} onto (a) d$_{xy}$ and (b) d$_{x^2-y^2}$ orbitals of W atoms, (c) d$_{xy}$ and (d) d$_{x^2-y^2}$ orbitals on Mo atoms. In all these panels, the size of circles is proportional to the magnitude of the projection.}
\label{fig15:MoS2_WSe2_layer_projection}
\end{figure}

Finally, Fig.~\ref{fig15:MoS2_WSe2_layer_projection} shows the projections of the highest valence bands states onto Mo and W d$_{xy}$-like and d$_{x^2-y^2}$-like orbitals in \bilayer{9} for $\theta=$\heteroangle{1}. Contrary to all other bilayers,\bilayer{5}, \bilayer{9} and \bilayer{10} exhibit $K$/$K$'-derived bands at the valence band edge even at small twist angles. These bands have large projections onto d$_{xy}$ and d$_{x^2-y^2}$ orbitals centred on \monolayer{3} (for \bilayer{5}) and \monolayer{4} (for \bilayer{9} and \bilayer{10}) and originate from $K$/$K$' states of the top valence bands of \monolayer{3} monolayer (for \bilayer{5}), and \monolayer{4} monolayer ((for \bilayer{9} and \bilayer{10}). 

\section*{S4 - Wavefunction densities}
\begin{figure}[h]
        \centering
        \includegraphics[width=0.5\columnwidth]{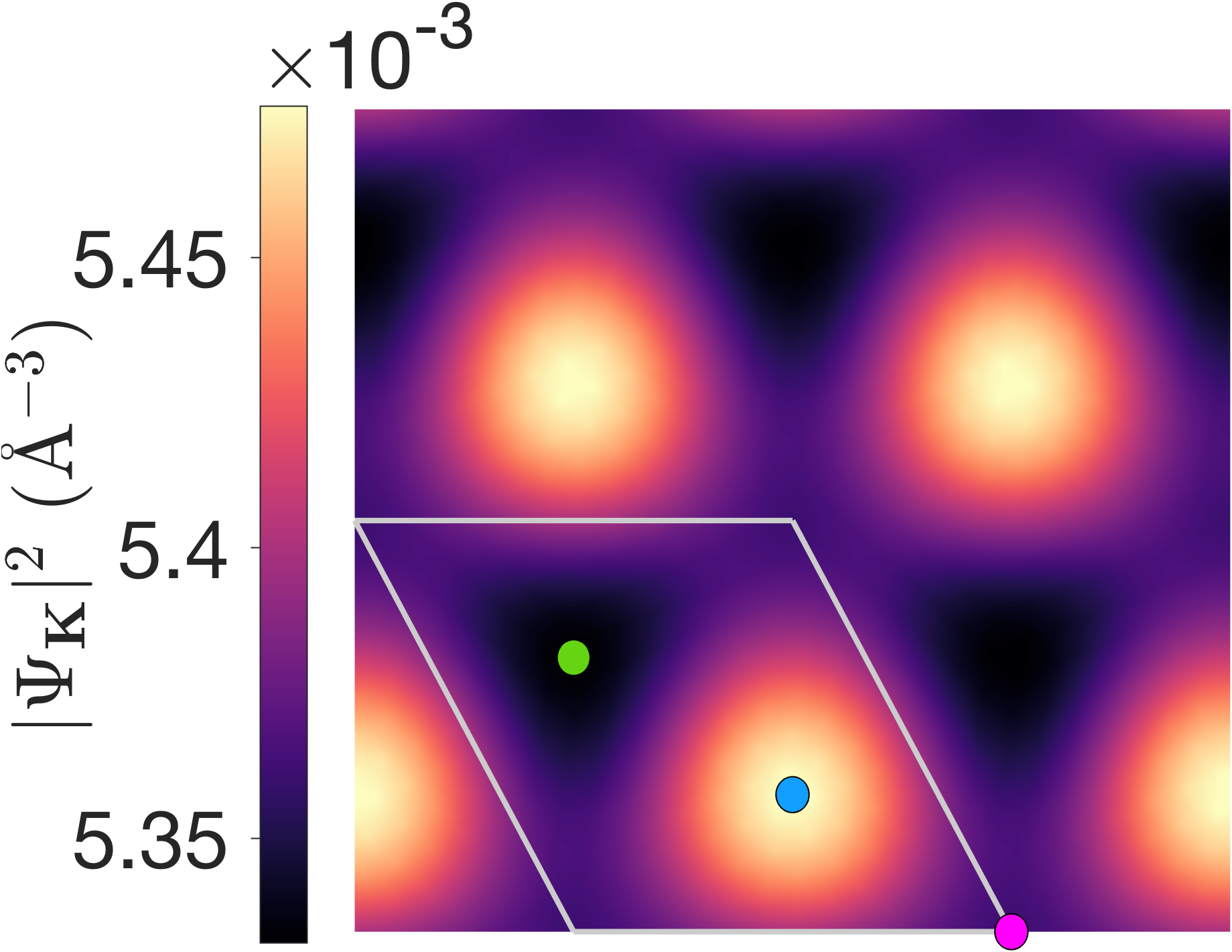}
            \caption{Top view of $|\psi_\mathbf{K}|^2$ of VBM in \protect\bilayer{9} at $\theta=$\protect\heteroangle{1}. Colored dots refer to different stacking regions as described in Fig.~1(a) of main text and moir\'e cell indicated by grey lines.}
        \label{fig15b:psi_K}
\end{figure}

Fig.~\ref{fig15b:psi_K} shows the square of the wavefunction of VBM at the K-point ($|\psi_\mathbf{K}|^2$) in \bilayer{9} for $\theta=$\heteroangle{1}. This state is localized mainly on \region{AB} regions, which form a triangular lattice, and is more diffuse than flat bands states.

\section{S7 Effect of interlayer distance on flat bands}
\begin{figure}[b]
\centering
\includegraphics[width=\columnwidth]{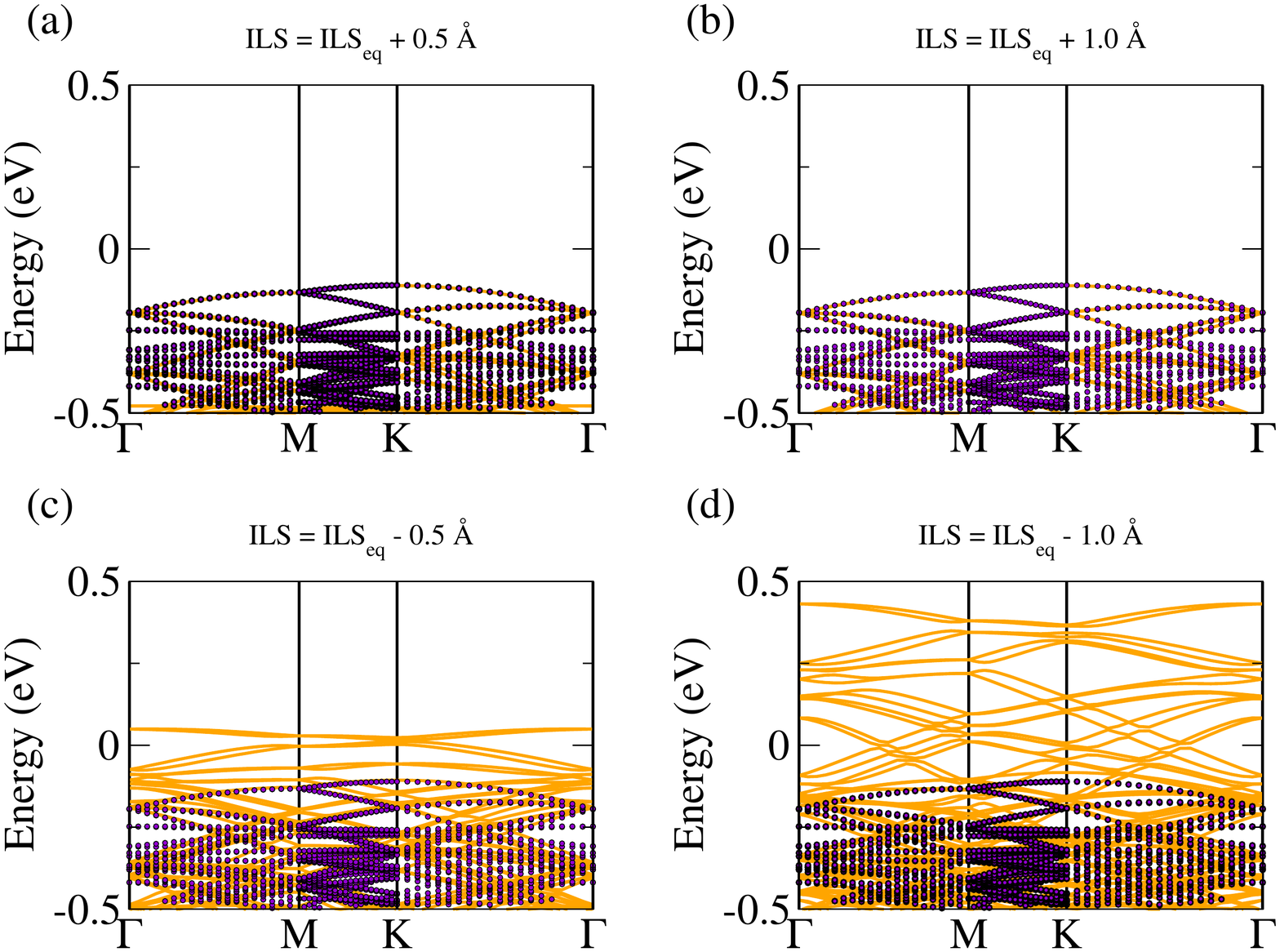}
\caption{Effect of interlayer separation on the ordering of flat vs $K$/$K$'-derived bands at the top valence manifold in \protect\bilayer{10} with $\theta=$\protect\heteroangle{1}. Starting from a relaxed structure, the ILS is increased/decreased by rigidly moving the layers without further relaxing the system. In panel (a) the ILS is increased by 0.5~\AA. In panel (b) the ILS is increased by 1.0~\AA. In panel (c) the ILS is decreased by 0.5~\AA and in panel (d) the ILS is decreased by 1.0~\AA. In all panels the band structure at equilibrium is shown with violet circles and that with modified ILS with solid orange lines. The ordering of flat vs $K$/$K$'-derived bands is reversed when the two layers are closer, compared to the equilibrium case (which exhibit $K$/$K$'-derived bands on top), as shown in bottom panels (which exhibit flat bands on top).}\label{fig18:effect_ILS}
\end{figure}

Fig.~\ref{fig18:effect_ILS} shows the effect of interlayer separation (ILS) on the ordering of flat bands over $K$/$K$'-derived bands in \bilayer{10} for $\theta=$\heteroangle{1}. To compute the band structures in Fig.~\ref{fig18:effect_ILS}, we start from a relaxed \bilayer{10} bilayer and rigidly translate the layers in the $z$-direction without further relaxing them. For ILS greater or equal to the equilibrium ILS we find $K$/$K$'-derived bands at the top of the valence band edge (top panels in Fig.~\ref{fig18:effect_ILS}). As the ILS is reduced (bottom panels in Fig.~\ref{fig18:effect_ILS}), flat bands emerge at the top of the valence band edge, which suggests that pressure could potentially be used to change the ordering between flat bands and $K$/$K$'-derived bands, and consequently, being able to modify flat-band properties in these systems.

\section*{S5 - Tables of tight-binding parameters}
In this section we present tables of tigh-binding hopping parameters, both intralayer and interlayer, for all homo- and hetero-bilayers in our work. 
\setlength{\tabcolsep}{12pt}
\begin{table}[h]
\sisetup{round-mode=places}
\centering
\caption{Tight-binding independent parameters in units of eV for MoS$_2$, MoSe$_2$, WS$_2$ and WSe$_2$ from Wannierised DFT results.}
\resizebox{0.55\columnwidth}{!}{%
\begin{tabular}{@{}l *{4}{S[round-precision=3]}}
\toprule
 & \multicolumn{1}{c}{MoS$_2$} & \multicolumn{1}{c}{MoSe$_2$} & \multicolumn{1}{c}{WS$_2$} & \multicolumn{1}{c}{WSe$_2$} \\
 \midrule
$\varepsilon_1 = \varepsilon_2$ &  0.0342 & 1.0371 & 0.6787 & 1.6145 \\ 
$\varepsilon_3$ &  -1.8961 & -0.4479 & -1.8681 & -0.4095 \\ 
$\varepsilon_4 = \varepsilon_5$ &  -2.3242 & -0.9647 & -2.2318 & -0.8712 \\ 
$\varepsilon_6$ &  -1.1249 & 0.0232 & -0.6951 & 0.3974 \\ 
$\varepsilon_7 = \varepsilon_8$ &  -0.9159 & 0.2490 & -0.4702 & 0.6436 \\ 
$\varepsilon_9$ &  -3.8919 & -2.6737 & -4.0282 & -2.8150 \\ 
$\varepsilon_{10} = \varepsilon_{11}$ &  -2.9073 & -1.5956 & -2.9916 & -1.6980 \\ 
$t_{1,1}^{(1)}$ &  -0.1826 & -0.1515 & -0.1828 & -0.1531 \\ 
$t_{2,2}^{(1)}$ &  0.0312 & 0.0186 & 0.0272 & 0.0146 \\ 
$t_{3,3}^{(1)}$ &  -0.1746 & -0.2063 & -0.1758 & -0.2097 \\ 
$t_{4,4}^{(1)}$ &  0.8666 & 0.9528 & 0.8770 & 0.9644 \\ 
$t_{5,5}^{(1)}$ &  -0.1873 & -0.1805 & -0.2158 & -0.2061 \\ 
$t_{6,6}^{(1)}$ &  -0.3063 & -0.2647 & -0.3805 & -0.3314 \\ 
$t_{7,7}^{(1)}$ &  0.2841 & 0.2493 & 0.3588 & 0.3129 \\ 
$t_{8,8}^{(1)}$ &  -0.5604 & -0.4747 & -0.6848 & -0.5771 \\ 
$t_{9,9}^{(1)}$ &  -0.2067 & -0.2200 & -0.2252 & -0.2395 \\ 
$t_{10,10}^{(1)}$ &  0.9160 & 0.9891 & 0.9718 & 1.0481 \\ 
$t_{11,11}^{(1)}$ &  0.0018 & -0.0033 & 0.0097 & 0.0020 \\ 
$t_{3,5}^{(1)}$ &  -0.0686 & -0.0724 & -0.0814 & -0.0881 \\ 
$t_{6,8}^{(1)}$ &  0.4092 & 0.3578 & 0.4875 & 0.4290 \\ 
$t_{9,11}^{(1)}$ &  0.0003 & -0.0015 & -0.0385 & -0.0430 \\ 
$t_{1,2}^{(1)}$ &  -0.2533 & -0.1963 & -0.3077 & -0.2386 \\ 
$t_{3,4}^{(1)}$ &  -0.1066 & -0.0893 & -0.1153 & -0.0940 \\ 
$t_{4,5}^{(1)}$ &  -0.0687 & -0.0671 & -0.0963 & -0.0919 \\ 
$t_{6,7}^{(1)}$ &  -0.1103 & -0.1023 & -0.1437 & -0.1342 \\ 
$t_{7,8}^{(1)}$ &  -0.2413 & -0.2058 & -0.2964 & -0.2504 \\ 
$t_{9,10}^{(1)}$ &  0.1047 & 0.1182 & 0.1610 & 0.1815 \\ 
$t_{10,11}^{(1)}$ &  -0.0388 & -0.0436 & -0.1000 & -0.1095 \\ 
$t_{4,1}^{(5)}$ &  -0.7767 & -0.7074 & -0.8719 & -0.7913 \\ 
$t_{3,2}^{(5)}$ &  -1.3644 & -1.3220 & -1.4291 & -1.3958 \\ 
$t_{5,2}^{(5)}$ &  2.1017 & 1.9706 & 2.2684 & 2.1307 \\ 
$t_{9,6}^{(5)}$ &  -0.8447 & -0.7825 & -0.9795 & -0.9151 \\ 
$t_{11,6}^{(5)}$ &  -0.9398 & -0.8698 & -0.9909 & -0.8953 \\ 
$t_{10,7}^{(5)}$ &  1.3644 & 1.2795 & 1.5223 & 1.4220 \\ 
$t_{9,8}^{(5)}$ &  -0.9409 & -0.8700 & -0.9441 & -0.8759 \\ 
$t_{11,8}^{(5)}$ &  0.6175 & 0.5702 & 0.6462 & 0.5975 \\ 
$t_{9,6}^{(6)}$ &  -0.0691 & -0.0670 & -0.0645 & -0.0638 \\ 
$t_{11,6}^{(6)}$ &  -0.1533 & -0.1506 & -0.1570 & -0.1521 \\ 
$t_{9,8}^{(6)}$ &  -0.2288 & -0.2211 & -0.2710 & -0.2547 \\ 
$t_{11,8}^{(6)}$ &  -0.2378 & -0.2142 & -0.2699 & -0.2434 \\
\bottomrule
\end{tabular}
}
\end{table}

\begin{table}[t]
\centering
\sisetup{round-mode=places}
\caption{p$_z$-p$_z$ interlayer hopping parameters for MoS$_2$/MoS$_2$}
\resizebox{\columnwidth}{!}{%
\begin{tabular}{*{7}{S[round-precision=3]}}
 \toprule
  \multicolumn{1}{c}{$\Delta z_{pp}$ [\AA]} & \multicolumn{2}{c}{$\nu$ [eV]} & \multicolumn{2}{c}{R [\AA]} & \multicolumn{2}{c}{$\eta$} \\
  \cline{2-3} \cline{4-5} \cline{6-7}\\
    & \multicolumn{1}{c}{$\sigma$} & \multicolumn{1}{c}{$\pi$} & \multicolumn{1}{c}{$\sigma$} & \multicolumn{1}{c}{$\pi$} & \multicolumn{1}{c}{$\sigma$} & \multicolumn{1}{c}{$\pi$} \\
   \midrule        
 2.6170 &      4.07660 & -1.27092 & 2.82371 & 2.61189 & 3.54713 & 5.41905 \\
 2.7170 &      3.72336 & -1.11170 & 2.88775 & 2.68782 & 3.59004 & 5.49683 \\
 2.8170 &      3.39650 & -0.98342 & 2.95225 & 2.75873 & 3.62986 & 5.52781 \\
 2.9170 &     3.09684 & -0.88484 & 3.01667 & 2.82178 & 3.66561 & 5.49774 \\
 3.0170 &      2.82420 & -0.81786 & 3.08049 & 2.87201 & 3.69634 & 5.38604 \\
 3.1170 &      2.57738 & -0.79236 & 3.14336 & 2.89917 & 3.72152 & 5.15810 \\
 3.2170 &      2.35482 & -0.84288 & 3.20489 & 2.87939 & 3.74068 & 4.75358 \\
 3.3170 &      2.15466 & -1.13777 & 3.26482 & 2.73898 & 3.75364 & 4.04510 \\
 3.4170 &      1.97483 & -6.14163 & 3.32293 & 1.95866 & 3.76049 & 2.50090 \\
 \bottomrule
\end{tabular}
}
\end{table}

\begin{table}[h]
\centering
\sisetup{round-mode=places}
\caption{p$_z$-d$_{z^2}$ interlayer hopping parameters for MoS$_2$/MoS$_2$}
\resizebox{\columnwidth}{!}{%
\begin{tabular}{*{9}{S[round-precision=3]}}
 \toprule
 \multicolumn{1}{c}{$\Delta z_{pd}$ [\AA]} & \multicolumn{2}{c}{$V$ [eV] } & \multicolumn{2}{c}{$\alpha$} & \multicolumn{2}{c}{$\beta$} & \multicolumn{2}{c}{$\gamma$} \\
 \cline{2-3} \cline{4-5} \cline{6-7} \cline{8-9}\\
  & \multicolumn{1}{c}{$\sigma$} & \multicolumn{1}{c}{$\pi$} & \multicolumn{1}{c}{$\sigma$} & \multicolumn{1}{c}{$\pi$} & \multicolumn{1}{c}{$\sigma$} & \multicolumn{1}{c}{$\pi$} & \multicolumn{1}{c}{$\sigma$} & \multicolumn{1}{c}{$\pi$} \\
 \midrule
4.208 & -0.400 & -4.256 & -5.843 & -10.332 & 2.386 & 1.710 & -4.784 & -2.388 \\
4.309 & -0.370 & -4.508 & -5.455 & -10.186 & 2.245 & 1.689 & -4.247 & -2.412 \\
4.409 & -0.388 & -4.751 & -5.465 & -10.050 & 2.171 & 1.666 & -4.045 & -2.423 \\
4.508 & -0.424 & -4.988 & -5.588 & -9.925 & 2.120 & 1.642 & -3.959 & -2.422 \\
4.609 & -0.472 & -5.222 & -5.760 & -9.813 & 2.081 & 1.616 & -3.933 & -2.410 \\
4.708 & -0.531 & -5.461 & -5.957 & -9.713 & 2.049 & 1.589 & -3.943 & -2.387 \\
4.809 & -0.603 & -5.705 & -6.167 & -9.627 & 2.022 & 1.561 & -3.979 & -2.357 \\
4.909 & -0.689 & -5.949 & -6.382 & -9.549 & 1.998 & 1.533 & -4.029 & -2.318 \\
5.008 & -0.789 & -6.197 & -6.598 & -9.482 & 1.976 & 1.504 & -4.092 & -2.272 \\
  \bottomrule
  \end{tabular}
  }
  \end{table}

\newpage

\begin{table}[t]
\centering
\sisetup{round-mode=places}
\caption{p$_z$-p$_z$ interlayer hopping parameters for MoSe$_2$/MoSe$_2$}
\resizebox{\columnwidth}{!}{%
\begin{tabular}{*{7}{S[round-precision=3]}}
 \toprule
  \multicolumn{1}{c}{$\Delta z_{pp}$ [\AA]} & \multicolumn{2}{c}{$\nu$ [eV]} & \multicolumn{2}{c}{R [\AA]} & \multicolumn{2}{c}{$\eta$} \\
  \cline{2-3} \cline{4-5} \cline{6-7}\\
    & \multicolumn{1}{c}{$\sigma$} & \multicolumn{1}{c}{$\pi$} & \multicolumn{1}{c}{$\sigma$} & \multicolumn{1}{c}{$\pi$} & \multicolumn{1}{c}{$\sigma$} & \multicolumn{1}{c}{$\pi$} \\
   \midrule
2.4850 &       4.64287 & -1.97036 & 2.88111 & 2.51311 & 3.64215 & 5.04578 \\
 2.5850 &       4.30810 & -1.74472 & 2.93987 & 2.58816 & 3.68439 & 5.15746 \\
 2.6850 &       3.98402 & -1.55447 & 3.00067 & 2.66023 & 3.72644 & 5.23449 \\
  2.7850 &      3.67379 & -1.39773 & 3.06319 & 2.72775 & 3.76824 & 5.26873 \\
  2.8850 &      3.38136 & -1.27678 & 3.12674 & 2.78749 & 3.80830 & 5.24518 \\
  2.9850 &      3.10875 & -1.19704 & 3.19075 & 2.83428 & 3.84547 & 5.14403 \\
  3.0850 &      2.85655 & -1.17622 & 3.25482 & 2.85730 & 3.87906 & 4.93086 \\
  3.1850 &      2.62487 & -1.27587 & 3.31852 & 2.82962 & 3.90822 & 4.53871 \\
   3.2850 &     2.41308 & -1.84610 & 3.38148 & 2.65430 & 3.93244 & 3.79729 \\
 \bottomrule
\end{tabular}
}
\end{table}

\begin{table}[h]
\sisetup{round-mode=places}
\centering
\caption{p$_z$-d$_{z^2}$ interlayer hopping parameters for MoSe$_2$/MoSe$_2$}
\resizebox{\columnwidth}{!}{%
\begin{tabular}{*{9}{S[round-precision=3]}}
 \toprule
 \multicolumn{1}{c}{$\Delta z_{pd}$ [\AA]} & \multicolumn{2}{c}{$V$ [eV] } & \multicolumn{2}{c}{$\alpha$} & \multicolumn{2}{c}{$\beta$} & \multicolumn{2}{c}{$\gamma$} \\
 \cline{2-3} \cline{4-5} \cline{6-7} \cline{8-9}\\
  & \multicolumn{1}{c}{$\sigma$} & \multicolumn{1}{c}{$\pi$} & \multicolumn{1}{c}{$\sigma$} & \multicolumn{1}{c}{$\pi$} & \multicolumn{1}{c}{$\sigma$} & \multicolumn{1}{c}{$\pi$} & \multicolumn{1}{c}{$\sigma$} & \multicolumn{1}{c}{$\pi$} \\
 \midrule
4.143 & -1.431 & -5.458 & -9.656 & -10.558 & 2.770 & 1.605 & -6.696 & -1.924 \\
4.242 & -0.515 & -5.773 & -6.015 & -10.379 & 2.340 & 1.601 & -4.768 & -2.015 \\
4.343 & -0.407 & -6.035 & -5.159 & -10.195 & 2.141 & 1.594 & -3.899 & -2.090 \\
4.442 & -0.422 & -6.258 & -5.124 & -10.015 & 2.068 & 1.583 & -3.675 & -2.144 \\
4.543 & -0.459 & -6.451 & -5.229 & -9.843 & 2.022 & 1.569 & -3.594 & -2.178 \\
4.643 & -0.511 & -6.626 & -5.387 & -9.682 & 1.987 & 1.551 & -3.577 & -2.193 \\
4.742 & -0.575 & -6.791 & -5.571 & -9.535 & 1.960 & 1.531 & -3.598 & -2.192 \\
4.843 & -0.653 & -6.954 & -5.769 & -9.402 & 1.938 & 1.508 & -3.643 & -2.177 \\
4.942 & -0.745 & -7.117 & -5.973 & -9.283 & 1.918 & 1.484 & -3.706 & -2.148 \\
 \bottomrule
\end{tabular}
}
\end{table}

\newpage

\begin{table}[t]
\sisetup{round-mode=places}
\centering
\caption{p$_z$-p$_z$ interlayer hopping parameters for WS$_2$/WS$_2$}
\resizebox{\columnwidth}{!}{%
\begin{tabular}{*{7}{S[round-precision=3]}}
 \toprule
  \multicolumn{1}{c}{$\Delta z_{pp}$ [\AA]} & \multicolumn{2}{c}{$\nu$ [eV]} & \multicolumn{2}{c}{R [\AA]} & \multicolumn{2}{c}{$\eta$} \\
  \cline{2-3} \cline{4-5} \cline{6-7}\\
    & \multicolumn{1}{c}{$\sigma$} & \multicolumn{1}{c}{$\pi$} & \multicolumn{1}{c}{$\sigma$} & \multicolumn{1}{c}{$\pi$} & \multicolumn{1}{c}{$\sigma$} & \multicolumn{1}{c}{$\pi$} \\
   \midrule 
2.6180 &       3.90795 & -1.02097 & 2.84655 & 2.70446 & 3.54628 & 6.13100 \\
 2.7180 &       3.55262 & -0.85858 & 2.91511 & 2.79938 & 3.59832 & 6.34796 \\
2.8180 &       3.25312 & -0.83053 & 2.97958 & 2.83687 & 3.63551 & 6.09974 \\
  2.9180 &       2.98662 & -0.71419 & 3.03971 & 2.92167 & 3.65795 & 6.18028 \\
 3.0180 &       2.71910 & -0.70953 & 3.10544 & 2.94457 & 3.69106 & 5.81570 \\
 3.1180 &       2.46881 & -0.70544 & 3.17253 & 2.96179 & 3.72388 & 5.45537 \\
  3.2180 &      2.24472 & -0.80992 & 3.23794 & 2.90569 & 3.74936 & 4.81419 \\
 3.3180   &    2.04267 & -1.53765 & 3.30210 & 2.59232 & 3.76926 & 3.61969 \\
  \bottomrule
\end{tabular}
}
\end{table}

\begin{table}[h]
\sisetup{round-mode=places}
\centering
\caption{p$_z$-d$_{z^2}$ interlayer hopping parameters for WS$_2$/WS$_2$}
\resizebox{\columnwidth}{!}{%
\begin{tabular}{*{9}{S[round-precision=3]}}
 \toprule
 \multicolumn{1}{c}{$\Delta z_{pd}$ [\AA]} & \multicolumn{2}{c}{$V$ [eV] } & \multicolumn{2}{c}{$\alpha$} & \multicolumn{2}{c}{$\beta$} & \multicolumn{2}{c}{$\gamma$} \\
 \cline{2-3} \cline{4-5} \cline{6-7} \cline{8-9}\\
  & \multicolumn{1}{c}{$\sigma$} & \multicolumn{1}{c}{$\pi$} & \multicolumn{1}{c}{$\sigma$} & \multicolumn{1}{c}{$\pi$} & \multicolumn{1}{c}{$\sigma$} & \multicolumn{1}{c}{$\pi$} & \multicolumn{1}{c}{$\sigma$} & \multicolumn{1}{c}{$\pi$} \\
 \midrule
4.209 & -0.212 & -3.798 & -3.744 & -9.761 & 2.153 & 1.645 & -3.584 & -2.128 \\
4.309 & -0.235 & -3.664 & -3.928 & -9.369 & 2.115 & 1.638 & -3.559 & -2.194 \\
4.409 & -0.247 & -3.975 & -3.923 & -9.324 & 2.052 & 1.630 & -3.399 & -2.268 \\
4.509 & -0.282 & -4.238 & -4.198 & -9.241 & 2.045 & 1.620 & -3.532 & -2.329 \\
4.609 & -0.319 & -4.156 & -4.435 & -8.990 & 2.029 & 1.599 & -3.626 & -2.321 \\
4.709 & -0.363 & -4.209 & -4.695 & -8.836 & 2.019 & 1.575 & -3.752 & -2.306 \\
4.809 & -0.416 & -4.261 & -4.955 & -8.701 & 2.010 & 1.550 & -3.888 & -2.274 \\
4.909 & -0.476 & -4.342 & -5.210 & -8.595 & 2.003 & 1.523 & -4.030 & -2.234 \\
 \bottomrule
\end{tabular}
}
\end{table}

\newpage

\begin{table}[t]
\sisetup{round-mode=places}
\centering
\caption{p$_z$-p$_z$ interlayer hopping parameters for WSe$_2$/WSe$_2$}
\resizebox{\columnwidth}{!}{%
\begin{tabular}{*{7}{S[round-precision=3]}}
 \toprule
\multicolumn{1}{c}{$\Delta z_{pp}$ [\AA]} & \multicolumn{2}{c}{$\nu$ [eV]} & \multicolumn{2}{c}{R [\AA]} & \multicolumn{2}{c}{$\eta$} \\
  \cline{2-3} \cline{4-5} \cline{6-7}\\
    & \multicolumn{1}{c}{$\sigma$} & \multicolumn{1}{c}{$\pi$} & \multicolumn{1}{c}{$\sigma$} & \multicolumn{1}{c}{$\pi$} & \multicolumn{1}{c}{$\sigma$} & \multicolumn{1}{c}{$\pi$} \\
   \midrule 
2.4855 &       3.77532 & -1.04153 & 3.02031 & 2.75682 & 3.94867 & 7.53891 \\
 2.5855 &       4.26945 & -1.01179 & 2.92522 & 2.79872 & 3.55666 & 7.16000 \\
 2.6855 &       3.86290 & -0.83403 & 3.00973 & 2.91045 & 3.66829 & 7.48761 \\
 2.7855 &       3.80744 & -0.77887 & 3.02406 & 2.96963 & 3.59039 & 7.27066 \\
 2.8855 &       3.47484 & -1.01937 & 3.09585 & 2.89059 & 3.64621 & 5.94986 \\
 2.9855 &       2.94072 & -0.83752 & 3.22501 & 2.99760 & 3.83990 & 6.14819 \\
 3.0855 &       2.69241 & -0.73659 & 3.29391 & 3.07555 & 3.88085 & 6.10354 \\
 3.1855 &       2.44160 & -0.70889 & 3.36929 & 3.10991 & 3.93751 & 5.76339 \\
 3.2855 &       2.24240 & -0.71238 & -3.43370 & 3.11388 & 3.96233 & 5.27715 \\
 \bottomrule
\end{tabular}
}
\end{table}

\begin{table}[h]
\sisetup{round-mode=places}
\centering
\caption{p$_z$-d$_{z^2}$ interlayer hopping parameters for WSe$_2$/WSe$_2$}
\resizebox{\columnwidth}{!}{%
\begin{tabular}{*{9}{S[round-precision=3]}}
 \toprule
 \multicolumn{1}{c}{$\Delta z_{pd}$ [\AA]} & \multicolumn{2}{c}{$V$ [eV] } & \multicolumn{2}{c}{$\alpha$} & \multicolumn{2}{c}{$\beta$} & \multicolumn{2}{c}{$\gamma$} \\
 \cline{2-3} \cline{4-5} \cline{6-7} \cline{8-9}\\
    & \multicolumn{1}{c}{$\sigma$} & \multicolumn{1}{c}{$\pi$} & \multicolumn{1}{c}{$\sigma$} & \multicolumn{1}{c}{$\pi$} & \multicolumn{1}{c}{$\sigma$} & \multicolumn{1}{c}{$\pi$} & \multicolumn{1}{c}{$\sigma$} & \multicolumn{1}{c}{$\pi$} \\
 \midrule
4.143 & -0.111 & -2.013 & -2.049 & -7.743 & 1.852 & 1.466 & -2.087 & -1.343 \\
4.243 & -0.831 & -4.070 & -7.560 & -9.161 & 2.654 & 1.586 & -6.409 & -1.981 \\
4.343 & -0.170 & -2.574 & -2.789 & -7.867 & 1.906 & 1.462 & -2.672 & -1.476 \\
4.443 & -0.259 & -3.402 & -3.824 & -8.275 & 1.984 & 1.528 & -3.244 & -1.889 \\
4.543 & -0.304 & -3.443 & -4.150 & -8.101 & 1.983 & 1.520 & -3.395 & -1.932 \\
4.643 & -0.326 & -4.038 & -4.151 & -8.286 & 1.981 & 1.529 & -3.552 & -2.086 \\
4.743 & -0.407 & -3.766 & -4.621 & -7.973 & 2.003 & 1.501 & -3.849 & -2.009 \\
4.843 & -0.469 & -3.683 & -4.850 & -7.782 & 1.996 & 1.479 & -3.980 & -1.985 \\
4.943 & -0.534 & -4.192 & -5.061 & -7.913 & 1.991 & 1.477 & -4.132 & -2.089 \\
 \bottomrule
\end{tabular}
}
\end{table}

\newpage

\begin{table}[t]
\sisetup{round-mode=places}
\centering
\caption{p$_z$-p$_z$ interlayer hopping parameters for MoSe$_2$/MoS$_2$}
\resizebox{\columnwidth}{!}{%
\begin{tabular}{*{7}{S[round-precision=3]}}
 \toprule
  \multicolumn{1}{c}{$\Delta z_{pp}$ [\AA]} & \multicolumn{2}{c}{$\nu$ [eV]} & \multicolumn{2}{c}{R [\AA]} & \multicolumn{2}{c}{$\eta$} \\
  \cline{2-3} \cline{4-5} \cline{6-7}\\
    & \multicolumn{1}{c}{$\sigma$} & \multicolumn{1}{c}{$\pi$} & \multicolumn{1}{c}{$\sigma$} & \multicolumn{1}{c}{$\pi$} & \multicolumn{1}{c}{$\sigma$} & \multicolumn{1}{c}{$\pi$} \\
   \midrule 
2.6170 &       3.98374 & -1.39320 & 2.93078 & 2.64108 & 3.83778 & 5.78948 \\
 2.7170 &       3.65139 & -1.22388 & 2.99599 & 2.71836 & 3.88590 & 5.88571 \\
 2.8170 &       3.33959 & -1.08019 & 3.06245 & 2.79356 & 3.93278 & 5.94871 \\
 2.9170 &       3.04770 & -0.96824 & 3.13013 & 2.86171 & 3.97854 & 5.94317 \\
 3.0170 &       2.78078 & -0.88325 & 3.19769 & 2.92202 & 4.01989 & 5.87291 \\
 3.1170 &       2.53809 & -0.82947 & 3.26472 & 2.96877 & 4.05585 & 5.71308 \\
 3.2170 &       2.31797 & -0.82565 & 3.33103 & 2.98767 & 4.08605 & 5.40888 \\
 3.3170 &       2.11763 & -0.92931 & 3.39678 & 2.94595 & 4.11172 & 4.87775 \\
 3.4170 &       1.93624 & -1.57638 & 3.46157 & 2.70227 & 4.13200 & 3.87194 \\
 \bottomrule
\end{tabular}
}
\end{table}

\begin{table}[h]
\sisetup{round-mode=places}
\centering
\caption{p$_z$-d$_{z^2}$ and $d_{z^2}$-$p_z$ interlayer hopping parameters for MoSe$_2$/MoS$_2$}
\resizebox{\columnwidth}{!}{%
\begin{tabular}{*{9}{S[round-precision=3]}}
 \toprule
 \multicolumn{1}{c}{$\Delta z_{pd}$ [\AA]} & \multicolumn{2}{c}{$V$ [eV] } & \multicolumn{2}{c}{$\alpha$} & \multicolumn{2}{c}{$\beta$} & \multicolumn{2}{c}{$\gamma$} \\
 \cline{2-3} \cline{4-5} \cline{6-7} \cline{8-9}\\
    & \multicolumn{1}{c}{$\sigma$} & \multicolumn{1}{c}{$\pi$} & \multicolumn{1}{c}{$\sigma$} & \multicolumn{1}{c}{$\pi$} & \multicolumn{1}{c}{$\sigma$} & \multicolumn{1}{c}{$\pi$} & \multicolumn{1}{c}{$\sigma$} & \multicolumn{1}{c}{$\pi$} \\
 \midrule
4.208 & -0.407 & -4.392 & -5.649 & -10.118 & 2.397 & 1.720 & -4.866 & -2.451 \\
4.309 & -0.375 & -4.551 & -5.250 & -9.916 & 2.257 & 1.707 & -4.333 & -2.510 \\
4.409 & -0.395 & -4.692 & -5.272 & -9.725 & 2.190 & 1.695 & -4.169 & -2.568 \\
4.508 & -0.432 & -4.869 & -5.399 & -9.578 & 2.144 & 1.675 & -4.109 & -2.591 \\
4.609 & -0.481 & -5.066 & -5.569 & -9.458 & 2.110 & 1.654 & -4.106 & -2.602 \\
4.708 & -0.542 & -5.288 & -5.764 & -9.362 & 2.081 & 1.630 & -4.137 & -2.598 \\
4.809 & -0.615 & -5.523 & -5.972 & -9.282 & 2.058 & 1.605 & -4.193 & -2.584 \\
4.909 & -0.701 & -5.781 & -6.184 & -9.219 & 2.037 & 1.579 & -4.262 & -2.564 \\
5.008 & -0.803 & -6.051 & -6.397 & -9.166 & 2.018 & 1.553 & -4.343 & -2.536 \\
\midrule
\multicolumn{1}{c}{$\Delta z_{dp}$ [\AA]} &  &  &  &  &  &  &  & \\ 
\midrule
4.208 & 0.647 &  2.918 & -6.762 & -8.926 & 2.269 & 1.662 & -4.238 & -1.906 \\
4.309 & 0.668 &  3.045 & -6.636 & -8.779 & 2.180 & 1.625 & -3.960 & -1.839 \\
4.409 & 0.716 &  3.173 & -6.637 & -8.647 & 2.109 & 1.588 & -3.763 & -1.765 \\
4.508 & 0.792 &  3.373 & -6.727 & -8.581 & 2.054 & 1.558 & -3.645 & -1.725 \\
4.609 & 0.884 &  3.610 & -6.842 & -8.547 & 2.005 & 1.528 & -3.551 & -1.684 \\
4.708 & 0.996 &  3.919 & -6.979 & -8.553 & 1.962 & 1.501 & -3.486 & -1.662 \\
4.809 & 1.131 &  4.318 & -7.129 & -8.599 & 1.924 & 1.479 & -3.444 & -1.664 \\
4.909 & 1.289 &  4.812 & -7.287 & -8.674 & 1.891 & 1.461 & -3.422 & -1.687 \\
5.008 & 1.475 &  5.424 & -7.448 & -8.774 & 1.860 & 1.446 & -3.415 & -1.732 \\
 \bottomrule
\end{tabular}
}
\end{table}

\newpage

\begin{table}[t]
\sisetup{round-mode=places}
\centering
\caption{p$_z$-p$_z$ interlayer hopping parameters for WS$_2$/MoS$_2$}
\resizebox{\columnwidth}{!}{%
\begin{tabular}{*{7}{S[round-precision=3]}}
 \toprule
  \multicolumn{1}{c}{$\Delta z_{pp}$ [\AA]} & \multicolumn{2}{c}{$\nu$ [eV]} & \multicolumn{2}{c}{R [\AA]} & \multicolumn{2}{c}{$\eta$} \\
  \cline{2-3} \cline{4-5} \cline{6-7}\\
    & \multicolumn{1}{c}{$\sigma$} & \multicolumn{1}{c}{$\pi$} & \multicolumn{1}{c}{$\sigma$} & \multicolumn{1}{c}{$\pi$} & \multicolumn{1}{c}{$\sigma$} & \multicolumn{1}{c}{$\pi$} \\
   \midrule 
2.6170 &       4.06380 & -1.18176 & 2.82369 & 2.64417 & 3.52230 & 5.67773 \\
 2.7170 &       3.70887 & -1.03778 & 2.88870 & 2.71923 & 3.56647 & 5.74120 \\
 2.8170 &       3.38690 & -0.92649 & 2.95272 & 2.78696 & 3.60350 & 5.73562 \\
 2.9170 &       3.08863 & -0.84453 & 3.01715 & 2.84485 & 3.63759 & 5.65153 \\
 3.0170 &       2.81252 & -0.79403 & 3.08213 & 2.88763 & 3.66961 & 5.47194 \\
  3.1170 &      2.56293 & -0.79742 & 3.14598 & 2.89765 & 3.69566 & 5.13566 \\
 3.2170 &       2.33596 & -0.91543 & 3.20914 & 2.83965 & 3.71731 & 4.57475 \\
 3.3170 &      2.13429 & -1.58087 & 3.26988 & 2.57589 & 3.73137 & 3.60418 \\
  \bottomrule
\end{tabular}
}
\end{table}

\begin{table}[h]
\sisetup{round-mode=places}
\centering
\caption{p$_z$-d$_{z^2}$ and $d_{z^2}$-$p_z$ interlayer hopping parameters for WS$_2$/MoS$_2$}
\resizebox{\columnwidth}{!}{%
\begin{tabular}{*{9}{S[round-precision=3]}}
 \toprule
 \multicolumn{1}{c}{$\Delta z_{pd}$ [\AA]} & \multicolumn{2}{c}{$V$ [eV] } & \multicolumn{2}{c}{$\alpha$} & \multicolumn{2}{c}{$\beta$} & \multicolumn{2}{c}{$\gamma$} \\
 \cline{2-3} \cline{4-5} \cline{6-7} \cline{8-9}\\
    & \multicolumn{1}{c}{$\sigma$} & \multicolumn{1}{c}{$\pi$} & \multicolumn{1}{c}{$\sigma$} & \multicolumn{1}{c}{$\pi$} & \multicolumn{1}{c}{$\sigma$} & \multicolumn{1}{c}{$\pi$} & \multicolumn{1}{c}{$\sigma$} & \multicolumn{1}{c}{$\pi$} \\
 \midrule
4.208 & -0.280 & -3.950 & -4.704 & -10.109 & 2.204 & 1.662 & -3.834 & -2.171 \\
4.309 & -0.302 & -4.144 & -4.812 & -9.944 & 2.148 & 1.651 & -3.719 & -2.231 \\
4.409 & -0.335 & -4.309 & -5.003 & -9.778 & 2.111 & 1.638 & -3.710 & -2.280 \\
4.508 & -0.377 & -4.494 & -5.221 & -9.645 & 2.082 & 1.620 & -3.740 & -2.308 \\
4.609 & -0.426 & -4.692 & -5.450 & -9.534 & 2.058 & 1.601 & -3.798 & -2.326 \\
4.708 & -0.484 & -4.929 & -5.684 & -9.454 & 2.039 & 1.583 & -3.878 & -2.343 \\
4.809 & -0.552 & -5.161 & -5.919 & -9.383 & 2.021 & 1.559 & -3.963 & -2.332 \\
4.909 & -0.628 & -5.326 & -6.141 & -9.291 & 2.003 & 1.532 & -4.055 & -2.301 \\
\midrule
\multicolumn{1}{c}{$\Delta z_{dp}$ [\AA]} &  &  &  &  &  &  &  & \\ 
\midrule
4.208 & 0.488 &  2.815 & -6.322 & -9.057 & 2.416 & 1.755 & -5.080 & -2.470 \\
4.309 & 0.450 &  2.966 & -5.899 & -8.929 & 2.290 & 1.719 & -4.613 & -2.426 \\
4.409 & 0.468 &  2.975 & -5.849 & -8.691 & 2.209 & 1.682 & -4.366 & -2.357 \\
4.508 & 0.510 &  3.127 & -5.934 & -8.598 & 2.147 & 1.651 & -4.216 & -2.324 \\
4.609 & 0.564 &  3.243 & -6.064 & -8.492 & 2.095 & 1.616 & -4.109 & -2.262 \\
4.708 & 0.634 &  3.477 & -6.224 & -8.477 & 2.049 & 1.591 & -4.034 & -2.251 \\
4.809 & 0.718 &  3.772 & -6.403 & -8.497 & 2.009 & 1.567 & -3.988 & -2.251 \\
4.909 & 0.812 &  4.171 & -6.572 & -8.561 & 1.978 & 1.555 & -3.988 & -2.309 \\
 \bottomrule
\end{tabular}
}
\end{table}

\newpage

\begin{table}[t]
\sisetup{round-mode=places}
\centering
\caption{p$_z$-p$_z$ interlayer hopping parameters for WSe$_2$/MoS$_2$}
\resizebox{\columnwidth}{!}{%
\begin{tabular}{*{7}{S[round-precision=3]}}
 \toprule
  \multicolumn{1}{c}{$\Delta z_{pp}$ [\AA]} & \multicolumn{2}{c}{$\nu$ [eV]} & \multicolumn{2}{c}{R [\AA]} & \multicolumn{2}{c}{$\eta$} \\
  \cline{2-3} \cline{4-5} \cline{6-7}\\
    & \multicolumn{1}{c}{$\sigma$} & \multicolumn{1}{c}{$\pi$} & \multicolumn{1}{c}{$\sigma$} & \multicolumn{1}{c}{$\pi$} & \multicolumn{1}{c}{$\sigma$} & \multicolumn{1}{c}{$\pi$} \\
   \midrule 
2.6180 &       3.91409 & -1.35682 & 2.93643 & 2.64582 & 3.79269 & 5.78831 \\
 2.7180 &       3.57585 & -1.19066 & 3.00534 & 2.72403 & 3.84974 & 5.87604 \\
 2.8180 &       3.23939 & -1.06245 & 3.08019 & 2.79461 & 3.91958 & 5.88730 \\
 2.9180 &       2.97585 & -0.97945 & 3.14392 & 2.85207 & 3.95293 & 5.80485 \\
3.0180 &       2.71780 & -0.93666 & 3.21223 & 2.89085 & 4.00005 & 5.59080 \\
  3.1180 &       2.48808 & -0.92244 & 3.27770 & 2.91545 & 4.03170 & 5.32233 \\
 3.2180 &       2.26266 & -1.02033 & 3.34786 & 2.88116 & 4.07324 & 4.81768 \\
 3.3180 &      2.05294 & -1.59811 & 3.41947 & 2.67207 & 4.11589 & 3.89001 \\
   \bottomrule
\end{tabular}
}
\end{table}

\begin{table}[h]
\centering
\caption{p$_z$-d$_{z^2}$ and $d_{z^2}$-$p_z$ interlayer hopping parameters for WSe$_2$/MoS$_2$}
\resizebox{\columnwidth}{!}{%
\begin{tabular}{*{9}{S[round-precision=3]}}
 \toprule
 \multicolumn{1}{c}{$\Delta z_{pd}$ [\AA]} & \multicolumn{2}{c}{$V$ [eV] } & \multicolumn{2}{c}{$\alpha$} & \multicolumn{2}{c}{$\beta$} & \multicolumn{2}{c}{$\gamma$} \\
 \cline{2-3} \cline{4-5} \cline{6-7} \cline{8-9}\\
    & \multicolumn{1}{c}{$\sigma$} & \multicolumn{1}{c}{$\pi$} & \multicolumn{1}{c}{$\sigma$} & \multicolumn{1}{c}{$\pi$} & \multicolumn{1}{c}{$\sigma$} & \multicolumn{1}{c}{$\pi$} & \multicolumn{1}{c}{$\sigma$} & \multicolumn{1}{c}{$\pi$} \\
 \midrule
4.209 & -0.305 & -3.495 & -4.791 & -9.414 & 2.282 & 1.690 & -4.285 & -2.315 \\
4.309 & -0.324 & -3.789 & -4.841 & -9.375 & 2.222 & 1.687 & -4.159 & -2.417 \\
4.409 & -0.361 & -4.076 & -5.028 & -9.321 & 2.184 & 1.686 & -4.144 & -2.527 \\
4.509 & -0.401 & -4.193 & -5.216 & -9.165 & 2.152 & 1.671 & -4.164 & -2.571 \\
4.609 & -0.449 & -4.404 & -5.401 & -9.087 & 2.123 & 1.659 & -4.194 & -2.626 \\
4.709 & -0.506 & -4.689 & -5.605 & -9.053 & 2.101 & 1.635 & -4.264 & -2.627 \\
4.809 & -0.574 & -4.928 & -5.817 & -8.997 & 2.082 & 1.616 & -4.344 & -2.648 \\
4.909 & -0.652 & -5.245 & -6.026 & -8.978 & 2.067 & 1.600 & -4.452 & -2.685 \\
\midrule
\multicolumn{1}{c}{$\Delta z_{dp}$ [\AA]} &  &  &  &  &  &  &  & \\ 
\midrule
4.209 & 0.326 &  2.397 & -4.374 & -8.266 & 2.044 & 1.610 & -3.097 & -1.656 \\
4.309 & 0.371 &  1.955 & -4.615 & -7.472 & 2.015 & 1.496 & -3.080 & -1.156 \\
4.409 & 0.456 &  2.164 & -5.152 & -7.497 & 1.999 & 1.456 & -3.158 & -1.039 \\
4.509 & 0.535 &  2.915 & -5.444 & -8.022 & 1.973 & 1.524 & -3.177 & -1.556 \\
4.609 & 0.620 &  3.048 & -5.685 & -7.946 & 1.930 & 1.494 & -3.123 & -1.503 \\
4.709 & 0.701 &  3.308 & -5.830 & -7.953 & 1.910 & 1.473 & -3.168 & -1.501 \\
4.809 & 0.817 &  3.819 & -6.074 & -8.104 & 1.892 & 1.475 & -3.227 & -1.642 \\
4.909 & 0.953 &  3.603 & -6.314 & -7.815 & 1.880 & 1.399 & -3.327 & -1.289 \\
 \bottomrule
\end{tabular}
}
\end{table}

\newpage

\begin{table}[t]
\sisetup{round-mode=places}
\centering
\caption{p$_z$-p$_z$ interlayer hopping parameters for WS$_2$/MoS$_2$}
\resizebox{\columnwidth}{!}{%
\begin{tabular}{*{7}{S[round-precision=3]}}
 \toprule
  \multicolumn{1}{c}{$\Delta z_{pp}$ [\AA]} & \multicolumn{2}{c}{$\nu$ [eV]} & \multicolumn{2}{c}{R [\AA]} & \multicolumn{2}{c}{$\eta$} \\
  \cline{2-3} \cline{4-5} \cline{6-7}\\
    & \multicolumn{1}{c}{$\sigma$} & \multicolumn{1}{c}{$\pi$} & \multicolumn{1}{c}{$\sigma$} & \multicolumn{1}{c}{$\pi$} & \multicolumn{1}{c}{$\sigma$} & \multicolumn{1}{c}{$\pi$} \\
   \midrule
2.6180 &        3.91791 & -1.03906 & 2.93822 & 2.75304 & 3.82989 & 6.80133 \\
2.7180 &        3.50152 & -0.96067 & 3.02366 & 2.81212 & 3.93860 & 6.68807 \\
2.8180 &        3.20957 & -0.82222 & 3.09033 & 2.90183 & 3.98619 & 6.85931 \\
2.9180 &        2.92665 & -0.72550 & 3.16029 & 2.97926 & 4.03748 & 6.89229 \\
3.0180 &        2.66852 & -0.65330 & 3.22964 & 3.04756 & 4.08174 & 6.80656 \\
3.1180 &        2.43023 & -0.61616 & 3.29922 & 3.09516 & 4.12223 & 6.53321 \\
3.2180 &        2.20681 & -0.60285 & 3.37025 & 3.12354 & 4.16368 & 6.14019 \\
3.3180 &        2.00326 & -0.63959 & 3.44085 & 3.10983 & 4.20112 & 5.55576 \\
3.4180 &      1.81836 & -0.95112 & 3.51122 & 2.92416 & 4.23544 & 4.43524 \\
   \bottomrule
\end{tabular}
}
\end{table}

\begin{table}[h]
\sisetup{round-mode=places}
\centering
\caption{p$_z$-d$_{z^2}$ and $d_{z^2}$-$p_z$ interlayer hopping parameters for WS$_2$/MoS$_2$}
\resizebox{\columnwidth}{!}{%
\begin{tabular}{*{9}{S[round-precision=3]}}
 \toprule
 \multicolumn{1}{c}{$\Delta z_{pd}$ [\AA]} & \multicolumn{2}{c}{$V$ [eV] } & \multicolumn{2}{c}{$\alpha$} & \multicolumn{2}{c}{$\beta$} & \multicolumn{2}{c}{$\gamma$} \\
 \cline{2-3} \cline{4-5} \cline{6-7} \cline{8-9}\\
    & \multicolumn{1}{c}{$\sigma$} & \multicolumn{1}{c}{$\pi$} & \multicolumn{1}{c}{$\sigma$} & \multicolumn{1}{c}{$\pi$} & \multicolumn{1}{c}{$\sigma$} & \multicolumn{1}{c}{$\pi$} & \multicolumn{1}{c}{$\sigma$} & \multicolumn{1}{c}{$\pi$} \\
 \midrule
4.209 & -0.315 & -3.892 & -4.788 & -9.546 & 2.375 & 1.735 & -4.816 & -2.556 \\
4.309 & -0.282 & -3.957 & -4.326 & -9.305 & 2.224 & 1.723 & -4.201 & -2.618 \\
4.409 & -0.289 & -4.180 & -4.260 & -9.187 & 2.145 & 1.709 & -3.957 & -2.667 \\
4.509 & -0.326 & -4.373 & -4.472 & -9.066 & 2.117 & 1.694 & -3.985 & -2.708 \\
4.609 & -0.372 & -4.616 & -4.708 & -8.986 & 2.096 & 1.676 & -4.048 & -2.736 \\
4.709 & -0.425 & -4.913 & -4.947 & -8.941 & 2.077 & 1.655 & -4.125 & -2.748 \\
4.809 & -0.488 & -5.205 & -5.190 & -8.901 & 2.059 & 1.629 & -4.207 & -2.729 \\
4.909 & -0.561 & -5.421 & -5.432 & -8.834 & 2.044 & 1.601 & -4.304 & -2.695 \\
5.009 & -0.645 & -5.599 & -5.669 & -8.760 & 2.030 & 1.575 & -4.412 & -2.662 \\
\midrule
\multicolumn{1}{c}{$\Delta z_{dp}$ [\AA]} &  &  &  &  &  &  &  & \\ 
\midrule 
4.209 & 0.475 &  2.182 & -5.789 & -8.165 & 2.158 & 1.545 & -3.619 & -1.266 \\
4.309 & 0.513 &  2.211 & -5.821 & -7.958 & 2.089 & 1.497 & -3.433 & -1.122 \\
4.409 & 0.586 &  2.326 & -6.018 & -7.860 & 2.068 & 1.465 & -3.496 & -1.061 \\
4.509 & 0.652 &  2.477 & -6.126 & -7.817 & 2.025 & 1.435 & -3.438 & -1.006 \\
4.609 & 0.735 &  2.735 & -6.281 & -7.871 & 1.989 & 1.419 & -3.423 & -1.041 \\
4.709 & 0.830 &  3.038 & -6.436 & -7.945 & 1.959 & 1.406 & -3.433 & -1.086 \\
4.809 & 0.941 &  3.374 & -6.601 & -8.024 & 1.932 & 1.393 & -3.459 & -1.127 \\
4.909 & 1.073 &  3.795 & -6.783 & -8.137 & 1.909 & 1.384 & -3.498 & -1.200 \\
5.009 & 1.207 &  5.020 & -6.930 & -8.592 & 1.857 & 1.431 & -3.373 & -1.628 \\
 \bottomrule
\end{tabular}
}
\end{table}

\newpage

\begin{table}[t]
\sisetup{round-mode=places}
\centering
\caption{p$_z$-p$_z$ interlayer hopping parameters for WSe$_2$/WS$_2$}
\resizebox{\columnwidth}{!}{%
\begin{tabular}{*{7}{S[round-precision=3]}}
 \toprule
  \multicolumn{1}{c}{$\Delta z_{pp}$ [\AA]} & \multicolumn{2}{c}{$\nu$ [eV]} & \multicolumn{2}{c}{R [\AA]} & \multicolumn{2}{c}{$\eta$} \\
  \cline{2-3} \cline{4-5} \cline{6-7}\\
    & \multicolumn{1}{c}{$\sigma$} & \multicolumn{1}{c}{$\pi$} & \multicolumn{1}{c}{$\sigma$} & \multicolumn{1}{c}{$\pi$} & \multicolumn{1}{c}{$\sigma$} & \multicolumn{1}{c}{$\pi$} \\
   \midrule
2.6180 &        3.92000 & -1.57602 & 2.93276 & 2.57963 & 3.74489 & 5.38711 \\
2.7180 &        3.48936 & -1.22990 & 3.02359 & 2.70567 & 3.86727 & 5.74245 \\
2.8180 &        3.15243 & -1.33089 & 3.10122 & 2.69872 & 3.94262 & 5.29563 \\
2.9180 &        2.81221 & -1.39387 & 3.18922 & 2.70165 & 4.04606 & 4.95728 \\
3.0180 &        2.51569 & -1.43231 & 3.27344 & 2.70799 & 4.13632 & 4.68777 \\
3.1180 &        2.25836 & -1.52720 & 3.35505 & 2.69566 & 4.21617 & 4.39146 \\
3.2180 &        2.00990 & -1.67156 & 3.44402 & 2.66578 & 4.30962 & 4.06663 \\
3.3180 &       1.79743 & -2.49663 & 3.52780 & 2.48173 & 4.39094 & 3.44422 \\
   \bottomrule
\end{tabular}
}
\end{table}

\begin{table}[h]
\sisetup{round-mode=places}
\centering
\caption{p$_z$-d$_{z^2}$ and $d_{z^2}$-$p_z$ interlayer hopping parameters for WSe$_2$/WS$_2$}
\resizebox{\columnwidth}{!}{%
\begin{tabular}{*{9}{S[round-precision=3]}}
 \toprule
 \multicolumn{1}{c}{$\Delta z_{pd}$ [\AA]} & \multicolumn{2}{c}{$V$ [eV] } & \multicolumn{2}{c}{$\alpha$} & \multicolumn{2}{c}{$\beta$} & \multicolumn{2}{c}{$\gamma$} \\
 \cline{2-3} \cline{4-5} \cline{6-7} \cline{8-9}\\
    & \multicolumn{1}{c}{$\sigma$} & \multicolumn{1}{c}{$\pi$} & \multicolumn{1}{c}{$\sigma$} & \multicolumn{1}{c}{$\pi$} & \multicolumn{1}{c}{$\sigma$} & \multicolumn{1}{c}{$\pi$} & \multicolumn{1}{c}{$\sigma$} & \multicolumn{1}{c}{$\pi$} \\
 \midrule
4.209 & -0.386 & -4.075 & -5.301 & -9.433 & 2.511 & 1.789 & -5.565 & -2.827 \\
4.309 & -0.312 & -4.108 & -4.551 & -9.151 & 2.352 & 1.793 & -4.939 & -2.978 \\
4.409 & -0.324 & -3.870 & -4.544 & -8.739 & 2.300 & 1.793 & -4.861 & -3.097 \\
4.509 & -0.342 & -3.880 & -4.590 & -8.534 & 2.263 & 1.783 & -4.857 & -3.180 \\
4.609 & -0.370 & -4.156 & -4.716 & -8.507 & 2.238 & 1.780 & -4.920 & -3.310 \\
4.709 & -0.400 & -4.026 & -4.835 & -8.272 & 2.215 & 1.759 & -4.996 & -3.332 \\
4.809 & -0.437 & -4.264 & -4.972 & -8.253 & 2.198 & 1.747 & -5.099 & -3.408 \\
4.909 & -0.481 & -4.248 & -5.133 & -8.115 & 2.182 & 1.725 & -5.212 & -3.421 \\
\midrule
\multicolumn{1}{c}{$\Delta z_{dp}$ [\AA]} &  &  &  &  &  &  &  & \\ 
\midrule
4.209 & 0.412 &  1.737 & -5.286 & -7.459 & 2.119 & 1.521 & -3.536 & -1.153 \\
4.309 & 0.386 &  1.111 & -4.810 & -6.068 & 1.985 & 1.285 & -2.979 & 0.070 \\
4.409 & 0.419 &  1.270 & -4.878 & -6.216 & 1.949 & 1.268 & -2.941 & 0.077 \\
4.509 & 0.488 &  1.429 & -5.157 & -6.333 & 1.954 & 1.246 & -3.124 & 0.107 \\
4.609 & 0.534 &  1.599 & -5.256 & -6.443 & 1.933 & 1.221 & -3.156 & 0.152 \\
4.709 & 0.583 &  1.914 & -5.325 & -6.744 & 1.916 & 1.264 & -3.223 & -0.211 \\
4.809 & 0.677 &  2.086 & -5.590 & -6.769 & 1.907 & 1.206 & -3.326 & 0.035 \\
4.909 & 0.774 &  2.405 & -5.798 & -6.957 & 1.888 & 1.220 & -3.387 & -0.148 \\
 \bottomrule
\end{tabular}
}
\end{table}

\newpage

\begin{table}[t]
\sisetup{round-mode=places}
\centering
\caption{p$_z$-p$_z$ interlayer hopping parameters for WSe$_2$/MoSe$_2$}
\resizebox{\columnwidth}{!}{%
\begin{tabular}{*{7}{S[round-precision=3]}}
 \toprule
  \multicolumn{1}{c}{$\Delta z_{pp}$ [\AA]} & \multicolumn{2}{c}{$\nu$ [eV]} & \multicolumn{2}{c}{R [\AA]} & \multicolumn{2}{c}{$\eta$} \\
  \cline{2-3} \cline{4-5} \cline{6-7}\\
    & \multicolumn{1}{c}{$\sigma$} & \multicolumn{1}{c}{$\pi$} & \multicolumn{1}{c}{$\sigma$} & \multicolumn{1}{c}{$\pi$} & \multicolumn{1}{c}{$\sigma$} & \multicolumn{1}{c}{$\pi$} \\
   \midrule
2.4850 &        4.65311 & -1.79133 & 2.87302 & 2.55163 & 3.57415 & 5.30139 \\
2.5850 &        4.33977 & -1.46749 & 2.93041 & 2.66289 & 3.61438 & 5.64587 \\
2.6850 &        3.98009 & -1.26448 & 2.99818 & 2.74881 & 3.67592 & 5.79404 \\
2.7850 &        3.64490 & -1.11078 & 3.06592 & 2.82628 & 3.73220 & 5.86999 \\
2.8850 &        3.35169 & -0.97766 & 3.13048 & 2.90311 & 3.77453 & 5.91835 \\
2.9850 &        3.08740 & -0.88039 & 3.19271 & 2.96857 & 3.80643 & 5.86950 \\
3.0850 &        2.82712 & -0.82439 & 3.25957 & 3.01411 & 3.84737 & 5.67251 \\
3.1850 &        2.58675 & -0.84085 & 3.32638 & 3.01450 & 3.88484 & 5.23662 \\
3.2850 &      2.38127 & -1.50940 & 3.38847 & 2.71856 & 3.90453 & 3.86883 \\
   \bottomrule
\end{tabular}
}
\end{table}

\begin{table}[h]
\sisetup{round-mode=places}
\centering
\caption{p$_z$-d$_{z^2}$ and $d_{z^2}$-$p_z$ interlayer hopping parameters for WSe$_2$/WS$_2$}
\resizebox{\columnwidth}{!}{%
\begin{tabular}{*{9}{S[round-precision=3]}}
 \toprule
 \multicolumn{1}{c}{$\Delta z_{pd}$ [\AA]} & \multicolumn{2}{c}{$V$ [eV] } & \multicolumn{2}{c}{$\alpha$} & \multicolumn{2}{c}{$\beta$} & \multicolumn{2}{c}{$\gamma$} \\
 \cline{2-3} \cline{4-5} \cline{6-7} \cline{8-9}\\
    & \multicolumn{1}{c}{$\sigma$} & \multicolumn{1}{c}{$\pi$} & \multicolumn{1}{c}{$\sigma$} & \multicolumn{1}{c}{$\pi$} & \multicolumn{1}{c}{$\sigma$} & \multicolumn{1}{c}{$\pi$} & \multicolumn{1}{c}{$\sigma$} & \multicolumn{1}{c}{$\pi$} \\
 \midrule
4.143 & -1.204 & -5.668 & -9.104 & -10.639 & 2.731 & 1.529 & -6.500 & -1.600 \\
4.242 & -0.289 & -5.324 & -4.234 & -10.151 & 2.061 & 1.557 & -3.288 & -1.810 \\
4.343 & -0.311 & -5.557 & -4.338 & -9.970 & 2.011 & 1.565 & -3.183 & -1.946 \\
4.442 & -0.339 & -5.626 & -4.467 & -9.733 & 1.980 & 1.573 & -3.183 & -2.086 \\
4.543 & -0.381 & -5.741 & -4.683 & -9.544 & 1.963 & 1.559 & -3.254 & -2.115 \\
4.643 & -0.430 & -5.783 & -4.897 & -9.343 & 1.949 & 1.541 & -3.347 & -2.127 \\
4.742 & -0.489 & -6.062 & -5.125 & -9.258 & 1.938 & 1.528 & -3.454 & -2.164 \\
4.843 & -0.555 & -6.179 & -5.337 & -9.120 & 1.931 & 1.516 & -3.590 & -2.203 \\
4.942 & -0.641 & -6.277 & -5.593 & -9.003 & 1.912 & 1.481 & -3.653 & -2.116 \\
\midrule
\multicolumn{1}{c}{$\Delta z_{dp}$ [\AA]} &  &  &  &  &  &  &  & \\ 
\midrule
4.143 & 0.943 &  3.390 & -8.137 & -9.183 & 2.460 & 1.680 & -5.449 & -2.177 \\
4.242 & 1.084 &  2.872 & -8.270 & -8.422 & 2.502 & 1.621 & -5.781 & -1.985 \\
4.343 & 0.585 &  3.307 & -6.179 & -8.534 & 2.282 & 1.614 & -4.803 & -2.089 \\
4.442 & 0.677 &  3.512 & -6.436 & -8.464 & 2.252 & 1.592 & -4.817 & -2.087 \\
4.543 & 0.712 &  3.496 & -6.379 & -8.233 & 2.197 & 1.565 & -4.683 & -2.035 \\
4.643 & 0.827 &  3.120 & -6.620 & -7.776 & 2.134 & 1.527 & -4.522 & -1.908 \\
4.742 & 0.897 &  2.937 & -6.663 & -7.488 & 2.077 & 1.468 & -4.371 & -1.676 \\
4.843 & 0.912 &  3.100 & -6.535 & -7.478 & 2.012 & 1.438 & -4.164 & -1.623 \\
4.942 & 1.047 &  3.664 & -6.695 & -7.676 & 1.962 & 1.450 & -4.053 & -1.805 \\
 \bottomrule
\end{tabular}
}
\end{table}

\bibliographystyle{unsrt}
\bibliography{SI}